\input harvmac
\input epsf
\input tables
\def\ra{\rightarrow}
\def\gsim{{~\raise.15em\hbox{$>$}\kern-.85em
          \lower.35em\hbox{$\sim$}~}}
\def\lsim{{~\raise.15em\hbox{$<$}\kern-.85em
          \lower.35em\hbox{$\sim$}~}} 
\def\Re{{\cal R}e}
\def\Im{{\cal I}m}
\def\epsK{{\varepsilon}_K}
\def\epspK{{\varepsilon}^\prime_K}
\def\epe{\varepsilon^\prime/\varepsilon}
\def\H{{\cal H}}
\def\O{{\cal O}}
\def\L{{\cal L}}
\def\U{{\cal U}}
\def\CP{{\rm CP}}
\def\ACF{A_{\rm CF}}

\def\YGTitle#1#2{\nopagenumbers\abstractfont\hsize=\hstitle\rightline{#1}%
\vskip .4in\centerline{\titlefont #2}\abstractfont\vskip .3in\pageno=0}
\YGTitle{\vbox{
\hbox{IASSNS-HEP-99-96}
\hbox{hep-ph/9911321}}}
{\vbox{\centerline{CP Violation In and Beyond the Standard Model}}}
\smallskip
\centerline{Yosef Nir}
\smallskip
\centerline{\it School of Natural Sciences,
 Institute for Advanced Study, Princeton NJ 08540, USA\foot{
 Address for the academic year 1999-2000; {\rm nir@sns.ias.edu}}}
\smallskip
\centerline{and}
\smallskip
\centerline{\it Department of Particle Physics,
 Weizmann Institute of Science, Rehovot 76100, Israel}
\bigskip
\baselineskip 18pt
\noindent
The special features of CP violation in the Standard Model are presented.
The significance of measuring CP violation in $B$, $K$ and $D$ decays
is explained. The predictions of the Standard Model for 
CP asymmetries in $B$ decays are analyzed in detail. Then, four frameworks of 
new physics are reviewed: (i) Supersymmetry provides an excellent 
demonstration of the power of CP violation as a probe of new physics. 
(ii) Left-right symmetric models are discussed as an example of an extension 
of the gauge sector. CP violation suggests that the scale of LRS breaking is 
low. (iii) The variety of extensions of the scalar sector are presented and
their unique CP violating signatures are emphasized.
(iv) Vector-like down quarks are presented as an example of an extension
of the fermion sector. Their implications for CP asymmetries in $B$ decays
are highly interesting. 
\bigskip
\bigskip
\baselineskip 15pt
\centerline{Lectures given in the}
\centerline{\bf XXVII SLAC Summer Institute on Particle Physics}
\centerline{\it CP Violation In and Beyond the Standard Model}
\centerline{July 7 $-$ 16, 1999}
\bigskip
\baselineskip 18pt
\leftskip=0cm\rightskip=0cm
 
\Date{}

\newsec{Introduction}
CP violation arises naturally in the three generation Standard Model.
The CP violation that has been measured in neutral $K$-meson decays
($\epsK$ and $\epspK$) is accommodated in the Standard Model in a simple way
\ref\KoMa{M. Kobayashi and T. Maskawa, Prog. Theo. Phys. 49 (1973) 652.}. 
Yet, CP violation is one of the least tested aspects of the 
Standard Model. The value of the $\epsK$ parameter
\ref\CCFT{J.H. Christenson, J.W. Cronin, V.L. Fitch and R. Turlay,
 Phys. Rev. Lett. 13 (1964) 138.}\ 
as well as bounds 
on other CP violating parameters (most noticeably, the electric
dipole moments of the neutron, $d_N$, and of the electron, $d_e$)
can be accounted for in models where CP violation has features
that are very different from the Standard Model ones. 

It is unlikely that the Standard Model provides the complete description
of CP violation in nature. First, it is quite clear that there exists
New Physics beyond the Standard Model. Almost any extension of the Standard
Model has additional sources of CP violating effects (or effects that
change the relationship of the measurable quantitites to the CP violating
parameters of the Standard Model). In addition there is a great puzzle
in cosmology that relates to CP violation, and that is the baryon
asymmetry of the universe 
\ref\Sakh{A.D. Sakharov, ZhETF Pis. Red. 5 (1967) 32;
 JETP Lett. 5 (1967) 24.}.
Theories that explain the observed asymmetry
must include new sources of CP violation
\ref\CKN{For a review see, A.G. Cohen, D.B. Kaplan and
A.E. Nelson,  Ann. Rev. Nucl. Part. Sci. 43 (1993) 27.}: 
the Standard Model cannot
generate a large enough matter-antimatter imbalance to produce the
baryon number to entropy ratio observed in the universe today
\nref\FS{G.R. Farrar and M.E. Shaposhnikov,
 Phys. Rev. D50 (1994) 774, hep-ph/9305275.}%
\nref\Gave{M.B. Gavela {\it et al.},
 Nucl. Phys. B430 (1994) 382, hep-ph/9406288.}%
\nref\HuSa{P. Huet and E. Sather, Phys. Rev. D51 (1995) 379, hep-ph/9404302.}%
\refs{\FS-\HuSa}.

In the near future, significant new information on CP violation
will be provided by various experiments. The main source of information
will be measurements of CP violation in various $B$ decays, particularly
neutral $B$ decays into final CP eigenstates
\nref\BCP{A.B. Carter and A.I. Sanda, Phys. Rev. Lett. 45 (1980) 952;
 Phys. Rev. D23 (1981) 1567.}%
\nref\BiSa{I.I. Bigi and A.I. Sanda, Nucl. Phys. B193 (1981) 85; 
 Nucl. Phys. B281 (1987) 41.}%
\nref\DuRo{
I. Dunietz and J. Rosner, Phys. Rev. D34 (1986) 1404.}%
\refs{\BCP-\DuRo}. First attempts have already been reported
\nref\OPALpre{K. Ackerstaff {\it et al.}, OPAL collaboration,
 Eur. Phys. J. C5 (1998) 379, hep-ex/9801022.}%
\nref\CDFpre{F. Abe {\it et al.}, CDF Collaboration,
 Phys. Rev. Lett. 81 (1998) 5513, hep-ex/9806025.}%
\nref\CDFBpK{T. Affolder {\it et al.}, CDF Collaboration, hep-ex/9909003.}%
\nref\BEN{G. Barenboim, G. Eyal and Y. Nir,
 Phys. Rev. Lett. 83 (1999) 4486, hep-ph/9905397.}%
\nref\ENcp{G. Eyal and Y. Nir, JHEP 9909 (1999) 013, hep-ph/9908296.}%
\refs{\OPALpre-\CDFBpK}\ and interpreted in the framework of various
models of new physics \refs{\BEN-\ENcp}. Another piece of valuable information 
might come from a measurement of the $K_L\ra\pi^0\nu\bar\nu$ decay
\nref\Litt{L.S. Littenberg, Phys. Rev. D39 (1989) 3322.}%
\nref\BuBu{G. Buchalla and A.J. Buras, Nucl. Phys. B400 (1993) 225.}%
\nref\Bupnn{A.J. Buras, Phys. Lett. B333 (1994) 476, hep-ph/9405368.}%
\nref\GrNi{Y. Grossman and Y. Nir,
 Phys. Lett. B398 (1997) 163, hep-ph/9701313.}%
\refs{\Litt-\GrNi}. For the first time, the
pattern of CP violation that is predicted by the Standard Model
will be tested. Basic questions such as whether CP is an approximate
symmetry in nature will be answered.

It could be that the scale where new CP violating sources appear
is too high above the Standard Model scale (e.g. the GUT scale)
to give any observable deviations from the Standard Model predictions.
In such a case, the outcome of the experiments will be a (frustratingly)
successful test of the Standard Model and a significant improvement in
our knowledge of the CKM matrix.

A much more interesting situation will arise if the new sources of 
CP violation appear at a scale that is not too high above the electroweak 
scale. Then they might be discovered in the forthcoming experiments. 
Once enough independent observations of CP violating effects are made, 
we will find that there is no single choice of CKM parameters that is 
consistent with all measurements. There may even be enough information 
in the pattern of the inconsistencies to tell us something about the nature 
of the new physics contributions
\nref\NiSi{Y. Nir and D. Silverman, Nucl. Phys. B345 (1990) 301.}%
\nref\DLN{C.O. Dib, D. London and Y. Nir, Int. J. Mod. Phys. A6 (1991) 1253.}%
\nref\NiQuR{Y. Nir and H.R. Quinn, Ann. Rev. Nucl. Part. Sci. 42 (1992) 211.}%
\refs{\NiSi-\NiQuR}.

The aim of these lectures is to explain the theoretical tools
with which we will analyze new information about CP violation.
The first part, chapters 2-6, deal with the Standard Model while
the second, chapters 7-10, discuss physics beyond the Standard Model.
In chapter 2, we present the Standard Model picture of CP violation.
We emphasize the features that are unique to the Standard Model.
In chapter 3, we  give a brief, model-independent discussion
of CP violating observables in $B$ meson decays. In chapter 4, we discuss
CP violation in the $K$ system (particularly, the $\epsK$ and $\epspK$
parameters) in a model independent way and in the framework of
the Standard Model. We also describe CP violation in $K\ra\pi\nu\bar\nu$.
In chapter 5, we discuss CP violation in $D\ra K\pi$ decays. In chapter 6, we
present in detail the Standard Model predictions for CP asymmetries
in $B$ decays. In chapter 7, the power of CP violation as a probe
of new physics is explained. Then, we discuss specific frameworks
of new physics: Supersymmetry (chapter 8), Left-Right symmetry as
an example of extensions of the gauge sector (chapter 9), extensions
of the scalar sector (chapter 10), and extra down singlet-quarks as
an example of extensions of the fermion sector (chapter 11).
Finally, we summarize our main points in chapter 12.

\newsec{Theory of CP Violation in the Standard Model}
\subsec{Yukawa Interactions Are the Source of CP Violation}
A model of the basic interactions between elementary particles
is defined by the following three ingredients:
\item{(i)} The symmetries of the Lagrangian;
\item{(ii)} The representations of fermions and scalars;
\item{(iii)} The pattern of spontaneous symmetry breaking.

The Standard Model (SM) is defined as follows:

(i) The gauge symmetry is
\eqn\SMgg{
G_{\rm SM}=SU(3)_{\rm C}\times SU(2)_{\rm L}\times U(1)_{\rm Y}.}

(ii) There are three fermion generations, each consisting of
five representations:
\eqn\SMrep{Q_{Li}^I(3,2)_{+1/6},\ \ u_{Ri}^I(3,1)_{+2/3},\ \ 
d_{Ri}^I(3,1)_{-1/3},\ \ L_{Li}^I(1,2)_{-1/2},\ \ \ell_{Ri}^I(1,1)_{-1}.}
Our notations mean that, for example, the left-handed quarks, $Q^I_L$, 
are in a triplet (3) of the $SU(3)_{\rm C}$ group, a doublet (2) of 
$SU(2)_{\rm L}$ and carry hypercharge $Y=Q_{\rm EM}-T_3=+1/6$. 
The index $I$ denotes interaction eigenstates. The 
index $i=1,2,3$ is the flavor (or generation) index.  
There is a single scalar multiplet:
\eqn\SMrepH{\phi(1,2)_{+1/2}.}

(iii) The $\phi$ scalar assumes a VEV,
\eqn\phivev{\vev{\phi}=\pmatrix{0\cr {v\over\sqrt2}\cr},}
so that the gauge group is spontaneously broken:
\eqn\SSB{G_{\rm SM}\rightarrow SU(3)_{\rm C}\times U(1)_{\rm EM}.}

The Standard Model Lagrangian, $\L_{\rm SM}$, is the most general 
renormalizable Lagrangian that is consistent with the gauge symmetry 
$G_{\rm SM}$ of eq. \SMgg. It can be divided to three parts:
\eqn\LagSM{\L_{\rm SM}=\L_{\rm kinetic}+\L_{\rm Higgs}+\L_{\rm Yukawa}.}

As concerns the kinetic terms, to maintain gauge invariance, one has 
to replace the derivative with a covariant derivative:
\eqn\SMDmu{D^\mu=\partial^\mu+ig_s G^\mu_a L_a+ig W^\mu_b T_b
+ig^\prime B^\mu Y.}
Here $G^\mu_a$ are the eight gluon fields, $W^\mu_b$ the three
weak interaction bosons and $B^\mu$ the single hypercharge boson.
The $L_a$'s are $SU(3)_{\rm C}$ generators (the $3\times3$
Gell-Mann matrices ${1\over2}\lambda_a$ for triplets, $0$ for singlets),
the $T_b$'s are $SU(2)_{\rm L}$ generators (the $2\times2$
Pauli matrices ${1\over2}\tau_b$ for doublets, $0$ for singlets),
and $Y$ are the $U(1)_{\rm Y}$ charges. For example, for the
left-handed quarks $Q_L^I$, we have
\eqn\DmuQL{\eqalign{
{\cal L}_{\rm kinetic}(Q_L)=&\ 
i{\overline{Q_{Li}^I}}\gamma^\mu D_\mu Q_{Li}^I,\cr
D^\mu Q_{Li}^I=&\ \left(\partial^\mu+{i\over2}g_s G^\mu_a\lambda_a
+{i\over2}g W^\mu_b\tau_b+{i\over6}g^\prime B^\mu\right)Q_{Li}^I.\cr}}
This part of the interaction Lagrangian is always CP conserving.

The Higgs potential, which describes the scalar self interactions, is given by:
\eqn\HiPo{\L_{\rm Higgs}=\mu^2\phi^\dagger\phi-\lambda(\phi^\dagger\phi)^2.}
For the Standard Model scalar sector, where there is a single doublet,
this part of the Lagrangian is also CP conserving.  For extended scalar
sector, such as that of a two Higgs doublet model, $\L_{\rm Higgs}$ can
be CP violating. Even in case that it is CP symmetric, it may lead
to spontaneous CP violation.

The Yukawa interactions are given by
\eqn\Hint{-{\cal L}_{\rm Yukawa}=
Y^d_{ij}{\overline {Q^I_{Li}}}\phi d^I_{Rj}
+Y^u_{ij}{\overline {Q^I_{Li}}}\tilde\phi u^I_{Rj}
+Y^\ell_{ij}{\overline {L^I_{Li}}}\phi\ell^I_{Rj}+{\rm h.c.}.}
This part of the Lagrangian is, in general, CP violating.
More precisely, CP is violated if and only if
\ref\Jarl{C. Jarlskog, Phys. Rev. Lett. 55 (1985) 1039.}
\eqn\JarCon{\Im\left\{\det[Y^d Y^{d\dagger},Y^u Y^{u\dagger}]\right\}\neq0.}

An intuitive explanation of why CP violation is related to {\it complex} 
Yukawa couplings goes as follows. The hermiticity of the Lagrangian implies
that ${\cal L}_{\rm Yukawa}$ has its terms in pairs of the form
\eqn\Yukpairs{Y_{ij}\overline{\psi_{Li}}\phi\psi_{Rj}
+Y_{ij}^*\overline{\psi_{Rj}}\phi^\dagger\psi_{Li}.}
A CP transformation exchanges the operators 
\eqn\CPoper{\overline{\psi_{Li}}\phi\psi_{Rj}\leftrightarrow
\overline{\psi_{Rj}}\phi^\dagger\psi_{Li},} 
but leaves their coefficients, $Y_{ij}$ and $Y_{ij}^*$, unchanged. 
This means that CP is a symmetry of ${\cal L}_{\rm Yukawa}$ 
if $Y_{ij}=Y_{ij}^*$.

How many independent CP violating parameters are there in 
${\cal L}_{\rm Yukawa}$? Each of the three Yukawa matrices $Y^f$ is
$3\times3$ and complex. Consequently, there are 27 real and 27 imaginary 
parameters in these matrices. Not all of them are, however, physical.
If we switch off the Yukawa matrices, there is a global
symmetry added to the Standard Model, 
\eqn\Gglob{G_{\rm global}^{\rm SM}(Y^f=0)=U(3)_Q\times U(3)_{\bar d}
\times U(3)_{\bar u}\times U(3)_L\times U(3)_{\bar\ell}.}
A unitary rotation of the three generations for each of the 
five representations in \SMrep\ would leave the Standard Model 
Lagrangian invariant. This means that the physics described by 
a given set of Yukawa matrices $(Y^d,Y^u,Y^\ell)$, and the physics 
described by another set,
\eqn\rotY{\tilde Y^d=V_Q^\dagger Y^d V_{\bar d},\ \ \ 
\tilde Y^u=V_Q^\dagger Y^u V_{\bar u},\ \ \
\tilde Y^\ell=V_L^\dagger Y^\ell V_{\bar\ell},}
where $V$ are all unitary matrices, is the same.
One can use this freedom to remove, at most, 15 real and 30 
imaginary parameters (the number of parameters in five $3\times3$ 
unitary matrices). However, the fact that the Standard Model with
the Yukawa matrices switched on has still a global symmetry of
\eqn\SMglobal{G_{\rm global}^{\rm SM}=U(1)_B\times U(1)_e\times U(1)_\mu
\times U(1)_\tau}
means that only 26 imaginary parameters can be removed.
We conclude that there are 13 flavor parameters: 12 real ones
and a single phase. This single phase is the source of CP violation.

\subsec{Quark Mixing is the (Only!) Source of CP Violation}
Upon the replacement $\Re(\phi^0)\rightarrow(v+H^0)/\sqrt2$ (see 
eq. \phivev), the Yukawa interactions \Hint\ give rise to mass terms:
\eqn\Hint{-{\cal L}_M=
(M_d)_{ij}{\overline {d^I_{Li}}} d^I_{Rj}
+(M_u)_{ij}{\overline {u^I_{Li}}} u^I_{Rj}
+(M_\ell)_{ij}{\overline {\ell^I_{Li}}}\ell^I_{Rj}+{\rm h.c.},}
where
\eqn\YtoM{M_f={v\over\sqrt2}Y^f,}
and we decomposed the $SU(2)_{\rm L}$ doublets into their components:
\eqn\doublets{Q^I_{Li}=\pmatrix{u^I_{Li}\cr d^I_{Li}\cr},\ \ \ 
L^I_{Li}=\pmatrix{\nu^I_{Li}\cr \ell^I_{Li}\cr}.}
Since the Standard Model neutrinos have no Yukawa interactions, 
they are predicted to be massless.

The mass basis corresponds, by definition, to diagonal mass matrices.
We can always find unitary matrices $V_{fL}$ and $V_{fR}$ such that
\eqn\diagM{V_{fL}M_f V_{fR}^\dagger=M_f^{\rm diag},}
with $M_f^{\rm diag}$ diagonal and real. The mass eigenstates
are then identified as
\eqn\masses{\eqalign{
d_{Li}=(V_{dL})_{ij}d_{Lj}^I&,\ \ \ d_{Ri}=(V_{dR})_{ij}d_{Rj}^I,\cr
u_{Li}=(V_{uL})_{ij}u_{Lj}^I&,\ \ \ u_{Ri}=(V_{uR})_{ij}u_{Rj}^I,\cr
\ell_{Li}=(V_{\ell L})_{ij}\ell_{Lj}^I&,\ \ \ 
\ell_{Ri}=(V_{\ell R})_{ij}\ell_{Rj}^I,\cr
\nu_{Li}=(V_{\nu L})_{ij}\nu_{Lj}^I&.\cr}}
Since the Standard Model neutrinos are massless, $V_{\nu L}$ is arbitrary.

The charged current interactions (that is the interactions of the 
charged $SU(2)_{\rm L}$ gauge bosons $W^\pm_\mu={1\over\sqrt2}
(W^1_\mu\mp iW_\mu^2)$) for quarks, which in the interaction basis are 
described by \DmuQL, have a complicated form in the mass basis:
\eqn\Wmas{-{\cal L}_{W^\pm}=
{g\over\sqrt2}{\overline {u_{Li}}}\gamma^\mu(V_{uL}V_{dL}^\dagger)_{ij}
d_{Lj} W_\mu^++{\rm h.c.}.}
The unitary $3\times3$ matrix,
\eqn\VCKM{V_{\rm CKM}=V_{uL}V_{dL}^\dagger,\ \ \ 
(V_{\rm CKM}V_{\rm CKM}^\dagger=1),} 
is the Cabibbo-Kobayashi-Maskawa (CKM) {\it mixing matrix} for quarks
\nref\Cabi{N. Cabibbo, Phys. Rev. Lett. 10 (1963) 531.}%
\refs{\Cabi,\KoMa}. A unitary $3\times3$ matrix depends on nine parameters: 
three real angles and six phases. 

The form of the matrix is not unique. Usually, the following two  
conventions are employed:

$(i)$ There is freedom in defining $V_{\rm CKM}$ in that
we can permute between the various generations. This
freedom is fixed by ordering the up quarks and the down quarks
by their masses, i.e. $m_{u_1}< m_{u_2}<m_{u_3}$ and 
$m_{d_1}<m_{d_2}<m_{d_3}$. Usually, we call $(u_1,u_2,u_3)\rightarrow
(u,c,t)$ and $(d_1,d_2,d_3)\rightarrow(d,s,b)$, and the elements
of $V_{\rm CKM}$ are written as follows:
\eqn\defVij{V_{\rm CKM}=\pmatrix{V_{ud}&V_{us}&V_{ub}\cr
V_{cd}&V_{cs}&V_{cb}\cr V_{td}&V_{ts}&V_{tb}\cr}.} 

$(ii)$ There is further freedom in the phase structure
of $V_{\rm CKM}$. Let us define $P_f$ ($f=u,d,\ell$) to be
diagonal unitary (phase) matrices. Then, if instead of using
$V_{fL}$ and $V_{fR}$ for the rotation \masses\ to the mass basis
we use $\tilde V_{fL}$ and $\tilde V_{fR}$, defined by
$\tilde V_{fL}=P_f V_{fL}$ and $\tilde V_{fR}=P_f V_{fR}$,
we still maintain a legitimate mass basis since $M_f^{\rm diag}$ remains
unchanged by such transformations. However, $V_{\rm CKM}$ does change:
\eqn\eqphase{V_{\rm CKM}\rightarrow P_u V_{\rm CKM}P_d^*.} 
This freedom is fixed by
demanding that $V_{\rm CKM}$ will have the minimal number of
phases. In the three generation case $V_{\rm CKM}$ has a single
phase. (There are five phase differences between the elements of
$P_u$ and $P_d$ and, therefore, five of the six phases in the
CKM matrix can be removed.) This is the Kobayashi-Maskawa phase 
$\delta_{\rm KM}$ which is the single source of {\it CP violation} in 
the Standard Model \KoMa. For example, the elemnts of the CKM matrix
can be written as follows (this is the standard parametrization
\nref\ChKe{L.L. Chau and W.Y. Keung, Phys. Rev. Lett. 53 (1984) 1802.}%
\nref\PDG{C. Caso {\it et al.}, Particle Data Group,
 Eur. Phys. J. C3 (1998) 1.}%
\refs{\ChKe,\PDG}): 
\eqn\stapar{V_{\rm CKM}=\pmatrix{c_{12}c_{13}&s_{12}c_{13}&
s_{13}e^{-i\delta_{\rm KM}}\cr 
-s_{12}c_{23}-c_{12}s_{23}s_{13}e^{i\delta_{\rm KM}}&
c_{12}c_{23}-s_{12}s_{23}s_{13}e^{i\delta_{\rm KM}}&s_{23}c_{13}\cr
s_{12}s_{23}-c_{12}c_{23}s_{13}e^{i\delta_{\rm KM}}&
-c_{12}s_{23}-s_{12}c_{23}s_{13}e^{i\delta_{\rm KM}}&c_{23}c_{13}\cr},}
where $c_{ij}\equiv\cos\theta_{ij}$ and $s_{ij}\equiv\sin\theta_{ij}$.
The three $\sin\theta_{ij}$ are the three real mixing parameters.

As a result of the fact that $V_{\rm CKM}$ is not diagonal,
the $W^\pm$ gauge bosons couple to quark (mass eigenstates)
of different generations. Within the Standard Model, this is
the only source of {\it flavor changing} interactions.
In principle, there could be additional sources of flavor
mixing (and of CP violation) in the lepton sector and in $Z^0$ interactions.
We now explain why, within the Standard Model, this does not happen.

{\it Mixing in the lepton sector}: An analysis similar to the above
applies also to the left-handed leptons. 
The mixing matrix is 
\ref\MNS{Z. Maki, M. Nakagawa and S. Sakata,
 Prog. Theo. Phys. 28 (1962) 247.}\
$V_{\rm MNS}=V_{\nu L}V_{\ell L}^\dagger$. However, 
we can use the arbitrariness of $V_{\nu L}$ (related to the 
masslessness of neutrinos) to choose $V_{\nu L}=V_{\ell L}$, 
and the mixing matrix becomes a unit matrix. We conclude that 
the masslessness of neutrinos (if true) implies that there is 
no mixing in the lepton sector. If neutrinos have masses then 
the leptonic charged current interactions will exhibit mixing 
and CP violation.

{\it Mixing in neutral current interactions}: We study the interactions
of the neutral $Z$-boson, $Z^\mu=\cos\theta_W W_3^\mu-\sin\theta_W B^\mu$
($\tan\theta_W\equiv g^\prime/g$) with, for example, left-handed
down quarks. The $W_3$-interactions are given in \DmuQL, while the $B$ 
interactions are given by
\eqn\Bint{-{\cal L}_B(Q_L)=-{g^\prime\over6}
{\overline{Q^I_{Li}}}\gamma^\mu Q^I_{Li}B_\mu.}
In the mass basis, we have then
\eqn\Zmas{\eqalign{-{\cal L}_Z=&{g\over\cos\theta_W}\left(-{1\over2}
+{1\over3}\sin^2\theta_W\right){\overline{d_{Li}}}
\gamma^\mu (V_{dL}^\dagger V_{dL})_{ij}d_{Lj}Z_\mu\cr
=&{g\over\cos\theta_W}\left(-{1\over2}
+{1\over3}\sin^2\theta_W\right){\overline{d_{Li}}}
\gamma^\mu d_{Li}Z_\mu.\cr}}
We learn that the neutral current interactions remain universal
in the mass basis and there are no additional flavor parameters
in their description. This situation goes beyond the Standard Model
to all models where all left-handed quarks are in $SU(2)_{\rm L}$
doublets and all right-handed ones in singlets. The $Z$-boson does
have flavor changing couplings in models where this is not the case. 

Examining the mass basis one can easily identify the flavor parameters.
In the quark sector, we have six quark masses, three mixing angles
(the number of real parameters in $V_{\rm CKM}$) and the single phase
$\delta_{\rm KM}$ mentioned above. In the lepton sector, we have
the three charged lepton masses. 

We have also learnt now some of the special features of CP violation
in the Standard Model:
\item{(i)} CP is explicitly broken.
\item{(ii)} There is a single source of CP violation, that is $\delta_{\rm KM}$.
\item{(iii)} CP violation appears only in the charged current interactions 
of quarks.
\item{(iv)} CP violation is closely related to flavor changing interactions.
 
\subsec{The CKM Matrix and the Unitarity Triangles}
In the mass basis, CP violation is related to the CKM matrix.
The fact that there are only three real and one imaginary physical
parameters in $V_{\rm CKM}$ can be made manifest by choosing an
explicit parametrization. One example was given above, in eq. \stapar,
with the four parameters $(s_{12},s_{23},s_{13},\delta_{\rm KM})$.
Another, very useful, example is the Wolfenstein parametrization
of $V_{\rm CKM}$, where the four mixing parameters are 
$(\lambda,A,\rho,\eta)$ with
$\lambda=|V_{us}|=0.22$ playing the role of an expansion parameter
and $\eta$ representing the CP violating phase
\ref\WOL{L. Wolfenstein, Phys. Rev. Lett. 51 (1983) 1945.}:
\eqn\WCKM{
V=\pmatrix{1-{\lambda^2\over2}&\lambda&A\lambda^3(\rho-i\eta)\cr
-\lambda&1-{\lambda^2\over2}&A\lambda^2\cr
A\lambda^3(1-\rho-i\eta)&-A\lambda^2&1\cr}+\O(\lambda^4).}

Various parametrizations differ in the way that the freedom of phase 
rotation, eq. \eqphase, is used to leave a single phase in $V_{\rm CKM}$.
One can define, however, a CP violating quantity in $V_{\rm CKM}$
that is independent of the parametrization \Jarl. This quantity is called
$J$ and defined through 
\eqn\defJ{\Im[V_{ij}V_{kl}V_{il}^*V_{kj}^*]=
J\sum_{m,n=1}^3\epsilon_{ikm}\epsilon_{jln},\ \ \ (i,j,k,l=1,2,3).}
CP is violated in the Standard Model only if $J\neq0$.
 
The usefulness of $J$ may not be clear from its formal definition in \defJ,
but does give useful insights once the {\it unitarity triangles} are
introduced. The unitarity of the CKM matrix leads to various relations among
the matrix elements, {\it e.g.}
\eqn\Unitds{
V_{ud}V_{us}^*+V_{cd}V_{cs}^*+V_{td}V_{ts}^*=0,}
\eqn\Unitsb{
V_{us}V_{ub}^*+V_{cs}V_{cb}^*+V_{ts}V_{tb}^*=0,}
\eqn\Unitdb{
V_{ud}V_{ub}^*+V_{cd}V_{cb}^*+V_{td}V_{tb}^*=0.}
Each of the three relations \Unitds-\Unitdb\ requires 
the sum of three complex quantities to vanish and so can be geometrically
represented in the complex plane as a triangle. These are
``the unitarity triangles", though the term ``unitarity triangle"
is usually reserved for the relation \Unitdb\ only. It is a surprising
feature of the CKM matrix that all unitarity triangles are equal in area:
the area of each unitarity triangle equals $|J|/2$ while the sign of $J$ 
gives the direction of the complex vectors around the triangles. The relation 
between Jarlskog's measure of CP violation $J$
and the Wolfenstein parameters is given by
\eqn\JAre{J\simeq \lambda^6 A^2\eta.}

The rescaled unitarity triangle  is derived from \Unitdb\
by (a) choosing a phase convention such that $(V_{cd}V_{cb}^*)$
is real, and (b) dividing the lengths of all sides by $|V_{cd}V_{cb}^*|$.
Step (a) aligns one side of the triangle with the real axis, and
step (b) makes the length of this side 1. The form of the triangle
is unchanged. Two vertices of the rescaled unitarity triangle are
thus fixed at (0,0) and (1,0). The coordinates of the remaining
vertex correspond to the Wolfenstein parameters $(\rho,\eta)$.
The area of the rescaled unitarity triangle is $|\eta|/2$.

Depicting the rescaled unitarity triangle in the
$(\rho,\eta)$ plane, the lengths of the two complex sides are
\eqn\RbRt{R_u\equiv\sqrt{\rho^2+\eta^2}={1\over\lambda}
\left|{V_{ub}\over V_{cb}}\right|,\ \ \
R_t\equiv\sqrt{(1-\rho)^2+\eta^2}={1\over\lambda}
\left|{V_{td}\over V_{cb}}\right|.}
The three angles of the  unitarity triangle are denoted
by $\alpha,\beta$ and $\gamma$ 
\ref\DDGN{
C.O. Dib, I. Dunietz, F.J. Gilman and Y. Nir, Phys. Rev. D41 (1990) 1522.}:
\eqn\abcangles{
\alpha\equiv\arg\left[-{V_{td}V_{tb}^*\over V_{ud}V_{ub}^*}\right],\ \ \
\beta\equiv\arg\left[-{V_{cd}V_{cb}^*\over V_{td}V_{tb}^*}\right],\ \ \
\gamma\equiv\arg\left[-{V_{ud}V_{ub}^*\over V_{cd}V_{cb}^*}\right].}
They are physical quantities and, we will soon see, can be
independently measured by CP asymmetries in $B$ decays.
It is also useful to define the two small angles of the unitarity triangles
\Unitsb\ and \Unitds:
\eqn\bbangles{
\beta_s\equiv\arg\left[-{V_{ts}V_{tb}^*\over V_{cs}V_{cb}^*}\right],\ \ \
\beta_K\equiv\arg\left[-{V_{cs}V_{cd}^*\over V_{us}V_{ud}^*}\right].}

To make predictions for future measurements of CP violating observables,
we need to find the allowed ranges for the CKM phases. There are three
ways to determine the CKM parameters (see {\it e.g.} 
\ref\HarNir{H. Harari and Y. Nir, Phys. Lett. 195B (1987) 586.}):
\item{(i)} {\bf Direct measurements} are related to SM tree level 
processes. At present,
we have direct measurements of $|V_{ud}|$, $|V_{us}|$, $|V_{ub}|$,
$|V_{cd}|$, $|V_{cs}|$, $|V_{cb}|$ and $|V_{tb}|$. 
\item{(ii)} {\bf CKM Unitarity}  ($V_{\rm CKM}^\dagger
V_{\rm CKM}={\bf 1}$) relates the various matrix elements. At present, 
these relations are useful to constrain
$|V_{td}|$, $|V_{ts}|$, $|V_{tb}|$ and $|V_{cs}|$. 
\item{(iii)} {\bf Indirect measurements} are related to SM loop processes. 
At present, we constrain in this way $|V_{tb}V_{td}|$ (from $\Delta m_B$ 
and  $\Delta m_{B_s}$) and $\delta_{\rm KM}$ or, equivalently, $\eta$ 
(from $\epsK$).

When all available data is taken into account, we find 
\nref\NirCern{Y. Nir, in {\it proceedings of the 1998 European School of 
High Energy Physics} (University of St. Andrews, 1998), hep-ph/9810520.}%
\nref\PlSc{S. Plaszczynski and M.-H. Schune, BaBar note 340.}%
\nref\Plas{S. Plaszczynski, hep-ph/9804330.}%
\nref\GNPS{Y. Grossman, Y. Nir, S. Plaszczynski and M.-H. Schune,
Nucl. Phys. B511 (1998) 69, hep-ph/9709288.}%
\nref\BaBar{The BaBar physics book, eds. P. Harrison and H.R. Quinn,
 SLAC-R-504 (1998).}%
\nref\BuLL{A.J. Buras, hep-ph/9905437.}%
\refs{\NirCern-\BuLL}:
\eqn\lacon{\lambda=0.2205\pm0.0018,\ \ \ A=0.826\pm0.041,}
\eqn\recon{-0.15\leq\rho\leq+0.35,\ \ \ +0.20\leq\eta\leq+0.45,}
\eqn\abccon{0.4\leq\sin2\beta\leq0.8,\ \ \ -0.9\leq\sin2\alpha\leq1.0,\ \ \ 
0.23\leq\sin^2\gamma\leq1.0.}
Of course, there are correlations between the various parameters.
The full information can be described by allowed regions in the 
$(\rho,\eta)$ or the $(\sin2\alpha,\sin2\beta)$ planes (see e.g. \BaBar).
(Recently, it has been shown that $B\ra\pi K$ decays can provide
bounds on the angle $\gamma$ of the unitarity triangle
\nref\FlMa{R. Fleischer and T. Mannel,
 Phys. Rev. D57 (1998) 2752, hep-ph/9704423.}%
\nref\NeRo{M. Neubert and J.L. Rosner,
 Phys. Lett. B441 (1998) 403, hep-ph/9808493.}%
\nref\BuFlg{A.J. Buras and R. Fleischer, hep-ph/9810260.}%
\nref\Neug{M. Neubert, JHEP 02 (1999) 014, hep-ph/9812396; hep-ph/9904321.}%
\nref\FlMat{R. Fleischer and J. Matias, hep-ph/9906274.}%
\refs{\FlMa-\FlMat}. A bound of $\cos\gamma\leq0.32$ was derived in \Neug.
We did not incorporate these bounds into our analysis.)

Eqs. \recon\ and \abccon\ show yet another
important feature of CP violation in the Standard Model. The fact that
$\eta/\rho=\O(1)$ or, equivalently, $\sin\gamma=\O(1)$, implies that
{\it CP is not an approximate symmetry} within the Standard Model.
This is not an obvious fact: after all, the two measured CP violating
quantities, $\epsK$ and $\epspK$, are very small (orders $10^{-3}$
and $10^{-6}$, respectively). The Standard Model accounts for their
smallness by the smallness of the flavor violation, that is the
mixing angles, and not by the smallness of CP violation, that is
a small phase. Indeed, the Standard Model predicts that in some
(yet unmeasured) processes, the CP asymmetry is of order one.

\subsec{The Uniqueness of the Standard Model Picture of CP Violation}
In the previous subsections, we have learnt several features of
CP violation as explained by the Standard Model:
\item{(i)} CP is explicitly broken.
\item{(ii)} $\delta_{\rm KM}$ is the only source of CP violation.
\item{(iii)} CP violation appears only in the charged current interactions 
of quarks.
\item{(iv)} CP violation would vanish in the absence of 
flavor changing interactions.
\item{(v)} CP is not an approximate symmetry ($\delta_{\rm KM}=\O(1)$).

(Non-perturbative corrections to the Standard Model tree-level Lagrangian 
are expected to induce $\theta_{\rm QCD}$, a CP violating parameter. This 
second possible source of CP violation is related to strong interactions
and is flavor diagonal. The bounds on the electric dipole moment of the
neutron imply that $\theta_{\rm QCD}\lsim10^{-9}$. The Standard Model
offers no natural explanation to the smallness of $\theta_{\rm QCD}$.
This puzzle is called `the strong CP problem'. We assume that it is
solved by some type of new physics, such as a Peccei-Quinn symmetry
\ref\PeQui{R.D. Peccei and H.R. Quinn, Phys. Rev. Lett. 38 (1977) 1440;
Phys. Rev. D16 (1977) 1791.},
which sets $\theta_{\rm QCD}$ to zero.)

It is important to realize that (a) none of features (i)-(v) is experimentally
established and that (b) various reasonable extensions of the Standard
Model provide examples where these features do not hold.
In particular, it could be that CP violation in Nature has 
some or all of the following features:
\item{(i)} CP is spontaneously broken.
\item{(ii)} There are many independent sources of CP violation.
\item{(iii)} CP violation appears in lepton interactions and/or in
neutral current interactions and/or in new sectors beyond the SM.
\item{(iv)} CP violation appears also in flavor diagonal interactions.
\item{(v)} CP is an approximate symmetry.

This situation, where the Standard Model has a very unique description
of CP violation and experiments have not yet confirmed this description,
is the basis for the strong interest, experimental and theoretical,
in CP violation. There are two types
of unambiguous tests concerning CP violation in the Standard Model:
First, since there is a single source of CP violation, all observables
are correlated with each other. For example, the CP asymmetries in
$B\ra\psi K_S$ and in $K\ra\pi\nu\bar\nu$ are strongly correlated.
Second, since CP violation is restricted to flavor changing quark
processes, it is predicted to practically vanish in the lepton sector
and in flavor diagonal processes. For example, the transverse
lepton polarization in semileptonic meson decays, CP violation
in $t\bar t$ production, and (assuming $\theta_{\rm QCD}=0$) the electric 
dipole moment of the neutron are all predicted to be orders of magnitude
below the (present and near future) experimental sensitivity.

The experimental investigation of CP violation in $B$ decays will
shed light on some but not all of these questions. In particular,
it will easily test the question of whether CP is an
approximate symmetry: if any $\O(1)$ asymmetry is observed,
for example in the $B\ra\psi K_S$ mode, then we immediately learn
that CP is not an approximate symmetry. It will also test (though
probably in a less definitive way) the question of whether the 
Kobayashi-Maskawa phase is the only source of CP violation.
On the other hand, we will learn little on flavor diagonal
CP violation and on CP violation outside the quark sector.
It is therefore important to search for CP violation in many
different systems.

\newsec{CP Violation in Meson Decays}
In the previous section, we understood how CP violation arises
in the Standard Model. In the next three sections, we would like to understand
the implications of this theory for the phenomenology of CP violation 
in meson decays. Our main focus will be on $B$-meson decays.
To do so, we first present a model independent analysis of CP violation
in meson decays.

There are three different types of CP violation in meson decays:
\item{(i)} CP violation in mixing, which occurs when the two neutral mass
eigenstate admixtures cannot be chosen to be CP-eigenstates;
\item{(ii)} CP violation in decay, which occurs in both charged and neutral
decays, when the amplitude for a decay and its CP-conjugate process
have different magnitudes;
\item{(iii)} CP violation in the interference of decays with and
without mixing, which occurs in decays into final states that are common to
$B^0$ and $\bar B^0$. 


\subsec{Notations and Formalism}
To define the three types of CP violation in meson decays and to discuss 
their theoretical calculation and experimental measurement, we first 
introduce some notations and formalism. We refer specifically to $B$ meson 
mixing and decays, but most of our discussion applies equally well 
to $K$, $B_s$ and $D$ mesons.

Our phase convention for the CP transformation law of
the neutral $B$ mesons is defined by
\eqn\phacon{\CP|{B^0}\rangle=\omega_B|{\bar B^0}\rangle,\ \ \
\CP|{\bar B^0}\rangle=\omega_B^*|{B^0}\rangle,\ \ \ (|\omega_B|=1).}
Physical observables do not depend on the phase factor $\omega_B$.
The time evolution of any linear combination of the neutral $B$-meson flavor 
eigenstates,
\eqn\defab{a|{B^0}\rangle+b|{\bar B^0}\rangle,}
is governed by the Schr\"odinger equation,
\eqn\Schro{i{d\over dt}\pmatrix{a\cr b\cr}=H\pmatrix{a\cr b\cr}
\equiv\left(M-{i\over2}\Gamma\right)\pmatrix{a\cr b\cr},}
for which $M$ and $\Gamma$ are $2\times2$ Hermitian matrices.

The off-diagonal terms in these matrices, $M_{12}$ and $\Gamma_{12}$,
are particularly important in the discussion of mixing and CP violation.
$M_{12}$ is the dispersive part of the transition amplitude from
$B^0$ to $\bar B^0$, while $\Gamma_{12}$ is the absorptive part of that 
amplitude.

The light $B_L$ and heavy $B_H$ mass eigenstates are given by
\eqn\defqp{|{B_{L,H}}\rangle=p|{B^0}\rangle\pm q|{\bar B^0}\rangle.}
The complex coefficients $q$ and $p$ obey the normalization condition
$|q|^2+|p|^2=1$. Note that $\arg(q/p^*)$ is just an overall common
phase for $|B_L\rangle$ and $|B_H\rangle$ and has no physical significance.
The mass difference and the width difference between the
physical states are given by
\eqn\DelmG{
\Delta m\equiv M_H-M_L,\ \ \ \Delta\Gamma\equiv\Gamma_H-\Gamma_L.}
Solving the eigenvalue equation gives
\eqn\eveq{(\Delta m)^2-{1\over4}(\Delta\Gamma)^2=
(4|M_{12}|^2-|\Gamma_{12}|^2),\ \ \ \ \ 
\Delta m\Delta\Gamma=4\Re(M_{12}\Gamma_{12}^*),}
\eqn\solveqp{{q\over p}=-{2M_{12}^*-i\Gamma_{12}^*\over
\Delta m-{i\over2}\Delta\Gamma}=-{\Delta m-{i\over2}\Delta\Gamma
\over 2M_{12}-i\Gamma_{12}}.}
In the $B$ system, $|\Gamma_{12}|\ll|M_{12}|$ (see discussion below),
and then, to leading order in $|\Gamma_{12}/M_{12}|$, eqs. \eveq\ 
and \solveqp\ can be written as
\eqn\eveqB{\Delta m_B=2|M_{12}|,\ \ \
\Delta\Gamma_B=\ 2\Re(M_{12}\Gamma_{12}^*)/|M_{12}|,}
\eqn\solveqpB{{q\over p}=-{M_{12}^*\over|M_{12}|}.}

To discuss CP violation in mixing, it is useful to write 
eq. \solveqp\ to first order in $|\Gamma_{12}/M_{12}|$:
\eqn\solveqpC{{q\over p}=-{M_{12}^*\over|M_{12}|}\left[1-{1\over2}
\Im\left({\Gamma_{12}\over M_{12}}\right)\right].}

To discuss CP violation in decay, we need to consider
decay amplitudes. The CP transformation law for a final state $f$ is
\eqn\phaconf{\CP|{f}\rangle=\omega_f|{\bar f}\rangle,\ \ \
\CP|{\bar f}\rangle=\omega_f^*|{f}\rangle,\ \ \ (|\omega_f|=1).}
For a final CP eigenstate $f=\bar f=f_{\CP}$, the phase factor
$\omega_f$ is replaced by $\eta_{f_{\CP}}=\pm1$, the CP eigenvalue
of the final state.
We define the decay amplitudes $A_f$ and $\bar A_f$ according to
\eqn\defAf{
A_f=\vev{f|{\cal H}_d|B^0},\ \ \ \bar A_f=\vev{f|{\cal H}_d|\bar B^0},}
where ${\cal H}_d$ is the decay Hamiltonian. 

CP relates $A_f$ and $\bar A_{\bar f}$. There are two types of phases that 
may appear in $A_f$ and $\bar A_{\bar f}$. Complex parameters in any 
Lagrangian term that contributes to the
amplitude will appear in complex conjugate form in the CP-conjugate
amplitude. Thus their phases appear in $A_f$ and $\bar A_{\bar f}$ with
opposite signs. In the Standard Model these phases occur only in the CKM
matrix which is part of the electroweak sector of the theory, hence these
are often called ``weak phases''. The weak phase of any single term is
convention dependent. However the difference between the weak phases in
two different terms in $A_f$ is convention independent because the phase
rotations of the initial and final states are the same for every term.
A second type of phase can appear in scattering or decay amplitudes even
when the Lagrangian is real. Such phases do not violate CP and they
appear in $A_f$ and $\bar A_{\bar f}$ with the same sign. Their origin is
the possible contribution from intermediate on-shell states in the
decay process, that is an absorptive part of an amplitude that has
contributions from coupled channels. Usually the dominant
rescattering is due to strong interactions and hence the designation
``strong phases'' for the phase shifts so induced. Again only the
relative strong phases of different terms in a scattering amplitude
have physical content, an overall phase rotation of the entire
amplitude has no physical consequences.
Thus it is useful to write each contribution to $A$ in three parts:
its magnitude $A_i$; its weak phase term $e^{i\phi_i}$; and its strong
phase term $e^{i\delta_i}$. Then, if several amplitudes contribute to
$B\ra f$, we have
\eqn\defAtoA{
\left|{\bar A_{\bar f}\over A_f}\right|=\left|{\sum_i A_i 
e^{i(\delta_i-\phi_i)}\over  \sum_i A_i e^{i(\delta_i+\phi_i)}}\right|.}

To discuss CP violation in the interference of decays with and without 
mixing, we introduce a complex quantity $\lambda_f$ defined by
\eqn\deflam{\lambda_f\ =\ {q\over p}\ {\bar A_f\over A_f}.}

We further define the CP transformation law for the quark fields in the 
Hamiltonian (a careful treatment of CP conventions can be found in 
\ref\BLS{G.C. Branco, L. Lavoura and J.P. Silva, {\it CP Violation}
(Oxford University Press, 1999).}):
\eqn\CPofq{q\ \rightarrow\ \omega_q\bar q,\ \ \ 
\bar q\ \rightarrow\ \omega_q^*q,\ \ \ (|\omega_q|=1).}
The effective Hamiltonian that is relevant to $M_{12}$ is of the form
\eqn\Hbtwo{H^{\Delta b=2}_{\rm eff}\propto
e^{+2i\phi_B}\left[\bar d\gamma^\mu(1-\gamma_5)b\right]^2
+e^{-2i\phi_B}\left[\bar b\gamma^\mu(1-\gamma_5)d\right]^2,}
where $2\phi_B$ is a CP violating (weak) phase. (We use the
Standard Model $V-A$ amplitude, but the results can be generalized
to any Dirac structure.) For the $B$ system, where $|\Gamma_{12}|\ll
|M_{12}|$, this leads to
\eqn\qpforB{q/p=\omega_B\omega_b^*\omega_d e^{-2i\phi_B}.}
(We implicitly assumed that the vacuum insertion approximation
gives the correct sign for $M_{12}$. In general, there is a
sign($B_B$) factor on the right hand side of eq. \qpforB\ 
\ref\GKN{Y. Grossman, B. Kayser and Y. Nir,
 Phys. Lett. B415 (1997) 90, hep-ph/9708398.}.)
To understand the phase structure of decay amplitudes, we take as
an example the $b\rightarrow q\bar qd$ decay ($q=u$ or $c$). 
The decay Hamiltonian is of the form
\eqn\Hdecay{
H_d\propto e^{+i\phi_f}\left[\bar q\gamma^\mu(1-\gamma_5)d\right]
\left[\bar b\gamma_\mu(1-\gamma_5)q\right]
+e^{-i\phi_f}\left[\bar q\gamma^\mu(1-\gamma_5)b\right]
\left[\bar d\gamma_\mu(1-\gamma_5)q\right],}
where $\phi_f$ is the appropriate weak phase. (Again, for simplicity
we use a $V-A$ structure, but the results hold for any Dirac structure.)
Then
\eqn\AbarA{
\bar A_{\bar f}/A_f=\omega_f\omega_B^*\omega_b\omega_d^* e^{-2i\phi_f}.}
Eqs. \qpforB\ and \AbarA\ together imply that for a final CP eigenstate,
\eqn\lamfCP{\lambda_{f_{\CP}}=\eta_{f_{\CP}}e^{-2i(\phi_B+\phi_f)}.}

\subsec{The Three Types of CP Violation in Meson Decays}
{\bf (i) CP violation in mixing:}
\eqn\inmixin{|q/p|\neq1.}
This results from the mass eigenstates being different from the
CP eigenstates, and requires a relative phase between $M_{12}$
and $\Gamma_{12}$. For the neutral $B$ system, this effect could
be observed through the asymmetries in semileptonic decays:
\eqn\mixexa{
a_{\rm SL}={\Gamma(\bar B^0_{\rm phys}(t)\ra\ell^+\nu X)-
\Gamma(B^0_{\rm phys}(t)\ra\ell^-\nu X)\over
\Gamma(\bar B^0_{\rm phys}(t)\ra\ell^+\nu X)+
\Gamma(B^0_{\rm phys}(t)\ra\ell^-\nu X)}.}
In terms of $q$ and $p$,
\eqn\mixter{a_{\rm SL}={1-|q/p|^4\over1+|q/p|^4}.}
CP violation in mixing has been observed in the neutral $K$ system
($\Re\ \epsK\neq0$).

In the neutral $B$ system, the effect is expected to be small, 
$\lsim\O(10^{-2})$. The reason is that, model independently, 
one expects that $a_{\rm SL}\lsim\Delta\Gamma_B/\Delta m_B$.
(We assume here that $\arg(\Gamma_{12}/M_{12})$ is not
particularly close to $\pi/2$.) The difference  in width is produced
by decay channels common to $B^0$ and $\bar B^0$. The branching
ratios for such channels are at or below the level of $10^{-3}$.
Since various channels contribute with differing signs, one expects
that their sum does not exceed the individual level. Hence, we can safely
assume that $\Delta\Gamma_B/\Gamma_B=\O(10^{-2})$. On the other hand, it is
experimentaly known that $\Delta m_B/\Gamma_B\approx0.7$.

To calculate $a_{\rm SL}$, we use \mixter\ and \solveqpC, and get:
\eqn\Absqp{a_{\rm SL}=\Im{\Gamma_{12}\over M_{12}}.}
To predict it in a given model, one needs to calculate $M_{12}$ and 
$\Gamma_{12}$. This involves large hadronic uncertainties, in particular 
in the hadronization models for $\Gamma_{12}$.

{\bf (ii) CP violation in decay:}
\eqn\indecay{|\bar A_{\bar f}/A_f|\neq1.}
This appears as a result of interference among various terms in the
decay amplitude, and will not occur unless at least two terms have
different weak phases and different strong phases. CP asymmetries in
charged $B$ decays,
\eqn\decexa{a_{f^\pm}={\Gamma(B^+\ra f^+)-\Gamma(B^-\ra f^-)\over
\Gamma(B^+\ra f^+)+\Gamma(B^-\ra f^-)},}
are purely an effect of CP violation in decay.
In terms of the decay amplitudes,
\eqn\decter{
a_{f^\pm}={1-|\bar A_{f^-}/A_{f^+}|^2\over1+|\bar A_{f^-}/A_{f^+}|^2}.}
CP violation in decay has been observed in the neutral $K$ system 
($\Re\ \epsK^\prime\neq0$).

To calculate $a_{f^\pm}$, we use \decter\ and \defAtoA. 
For simplicity, we consider decays with contributions from 
two weak phases and with $A_2\ll A_1$. We get:
\eqn\AbsAA{a_{f^\pm}=-2(A_2/A_1)\sin(\delta_2-\delta_1)
\sin(\phi_2-\phi_1).}
The magnitude and strong phase of any amplitude involve long distance
strong interaction physics, and our ability to calculate these from
first principles is limited. Thus quantities that depend only on the weak
phases are much cleaner than those that require knowledge of the
relative magnitudes or strong phases of various amplitude contributions,
such as CP violation in decay.

{\bf (iii) CP violation in the interference between decays
 with and without mixing:}
\eqn\ininter{
|\lambda_{f_{\CP}}|=1,\ \ \Im\ \lambda_{f_{\CP}}\neq0.}
Any $\lambda_{f_{\CP}}\neq\pm1$ is a manifestation of CP violation.
The special case \ininter\ isolates the effects of interest since both
CP violation in decay, eq. \indecay, and in mixing, eq. \inmixin, lead to
$|\lambda_{f_{\CP}}|\neq1$. For the neutral $B$ system, this effect can
be observed by comparing decays into final CP eigenstates of a
time-evolving neutral $B$ state that begins at time zero as $B^0$
to those of the state that begins as $\bar B^0$:
\eqn\intexa{a_{f_{\CP}}={\Gamma(\bar B^0_{\rm phys}(t)\ra f_{\CP})-
\Gamma(B^0_{\rm phys}(t)\ra f_{\CP})\over
\Gamma(\bar B^0_{\rm phys}(t)\ra f_{\CP})+
\Gamma(B^0_{\rm phys}(t)\ra f_{\CP})}.}
This time dependent asymmetry is given (for $|\lambda_{f_{\CP}}|=1$) by
\eqn\intters{a_{f_{\CP}}=-\Im\lambda_{f_{\CP}}\sin(\Delta m_B t).}

CP violation in the interference of decays with and without
mixing has been observed for the neutral $K$ system
($\Im\ \epsK\neq0$). It is expected to be an effect of $\O(1)$
in various $B$ decays.  For such cases, the contribution from CP violation
in mixing is clearly negligible. For decays that are dominated
by a single CP violating phase (for example, $B\ra\psi K_S$ and
$K_L\ra\pi^0\nu\bar\nu$), so that the contribution from CP violation in
decay is also negligible, $a_{f_{\rm CP}}$ is cleanly interpreted
in terms of purely electroweak parameters. Explicitly,
$\Im\lambda_{f_{\CP}}$ gives the difference between the phase of
the $B-\bar B$ mixing amplitude ($2\phi_B$) and twice the phase of the
relevant decay amplitude ($2\phi_f$) (see  eq. \lamfCP):
\eqn\intCKM{
\Im\lambda_{f_{\CP}}=-\eta_{f_{\CP}}\sin[2(\phi_B+\phi_f)].}

A summary of the main properties of the different types of
CP violation in meson decays is given in the table I.
\vskip 1cm

\begintable
Type & Exp. & Theory & Calculation & Uncertainties & Observed in \crthick
mixing & $a_{\rm SL}$ & ${1-\vert q/p\vert^4\over1+\vert q/p\vert^4}$ & 
$\Im{\Gamma_{12}\over M_{12}}$ & $\Gamma_{12}\ \Longrightarrow$ Large 
& $\Re\ \epsK$ \nr
decay & $a_{f^\pm}$  & ${1-\vert\bar A_{f^-}/A_{f^+}\vert^2\over
1+\vert\bar A_{f^-}/A_{f^+}\vert^2}$ & 
$-2{A_2\over A_1}\sin(\delta_2-\delta_1)\sin(\phi_2-\phi_1)$ & 
$\delta_i,A_i\ \Longrightarrow$ Large & $\Re\ \epspK$ \nr
interference & $a_{f_{CP}}$ & $-\Im\lambda_{f_{CP}}$ & 
$\eta_{f_{CP}}\sin[2(\phi_B+\phi_f)]$ & Small & $\Im\ \epsK$ \endtable

\centerline{Table I. The three types of CP violation in meson decays.}

\newsec{CP Violation in $K$ Decays}
The two CP violating observables that have been measured are
related to $K$ meson decays. In this chapter we present these
observables and their significance.

\subsec{Direct and Indirect CP Violation} 
The terms indirect CP violation and direct CP violation
are commonly used in the literature. While various authors use these
terms with different meanings, the most useful definition is the following:

{\bf Indirect CP violation} refers to CP violation in
meson decays where the CP violating phases can all be chosen to
appear in $\Delta F=2$ (mixing) amplitudes.

{\bf Direct CP violation} refers to CP violation in
meson decays where some CP violating phases necessarily
appear in $\Delta F=1$ (decay) amplitudes.

Examining eqs. \inmixin\ and \solveqp, we learn that
CP violation in mixing is a manifestation of indirect CP violation.
Examining eqs. \indecay\ and \defAf, we learn that
CP violation in decay is a manifestation of direct CP violation.
Examining eqs. \ininter\ and \deflam, we learn that
the situation concerning CP violation in the interference of
decays with and without mixing is more subtle. For any single
measurement of $\Im\lambda_f\neq0$, the relevant CP violating phase
can be chosen by convention to reside in the $\Delta F=2$ amplitude
($\phi_f=0$, $\phi_B\neq0$ in the notation of eq. \lamfCP),
and then we would call it indirect CP violation. Consider, however,
the CP asymmetries for two different final CP eigenstates (for the
same decaying meson), $f_a$ and $f_b$. Then, a non-zero difference
between $\Im\lambda_{f_a}$ and $\Im\lambda_{f_b}$ requires
that there exists CP violation in $\Delta F=1$ processes ($\phi_{f_a}-
\phi_{f_b}\neq0$), namely direct CP violation.

Experimentally, both direct and indirect CP violation have been established.
Below we will see that $\epsK$ signifies indirect CP violation
while $\epspK$ signifies direct CP violation. 

Theoretically, most models of CP violation (including the Standard Model)
have predicted that both types of CP violation exist. There is, however,
one class of models, that is {\it superweak models}, that predict
only indirect CP violation. The measurement of $\epspK\neq0$ has
excluded this class of models.

\subsec{The $\epsK$ and $\epspK$ Parameters}
Historically, a different language from the one used by us has been
employed to describe CP violation in $K\ra\pi\pi$ and $K\ra\pi\ell\nu$
decays. In this section we `translate' the language of $\varepsilon_K$
and $\varepsilon_K^\prime$ to our notations. Doing so will make it
easy to understand which type of CP violation is related to each quantity.

The two CP violating quantities measured in neutral $K$ decays are
\eqn\defetaij{
\eta_{00}={\vev{\pi^0\pi^0|\H|K_L}\over\vev{\pi^0\pi^0|\H|K_S}},\ \ \ 
\eta_{+-}={\vev{\pi^+\pi^-|\H|K_L}\over\vev{\pi^+\pi^-|\H|K_S}}.}
Define, for $(ij)=(00)$ or $(+-)$,
\eqn\epsamp{A_{ij}=\vev{\pi^i\pi^j|\H|K^0},\ \ \ 
\bar A_{ij}=\vev{\pi^i\pi^j|\H|\bar K^0},}
\eqn\epslam{\lambda_{ij}=\left({q\over p}\right)_K{\bar A_{ij}\over A_{ij}}.}
Then
\eqn\etapqA{\eta_{00}={1-\lambda_{00}\over1+\lambda_{00}},\ \ \ 
\eta_{+-}={1-\lambda_{+-}\over1+\lambda_{+-}}.}
The $\eta_{00}$ and $\eta_{+-}$ parameters get contributions from 
CP violation in mixing ($|(q/p)|_K\neq1$) and from the interference 
of decays with and without mixing ($\Im\lambda_{ij}\neq0$) 
at $\O(10^{-3})$ and from CP violation in decay 
($|\bar A_{ij}/A_{ij}|\neq1$) at $\O(10^{-6})$.

There are two isospin channels in $K\ra\pi\pi$ leading to final
$(2\pi)_{I=0}$ and $(2\pi)_{I=2}$ states:
\eqn\twois{\eqalign{
\langle\pi^0\pi^0|\ =&\ \sqrt{1\over3}\langle(\pi\pi)_{I=0}|-
\sqrt{2\over3}\langle(\pi\pi)_{I=2}|,\cr
\langle\pi^+\pi^-|\ =&\ \sqrt{2\over3}\langle(\pi\pi)_{I=0}|+
\sqrt{1\over3}\langle(\pi\pi)_{I=2}|.\cr}}
The fact that there are two strong phases allows for CP violation
in decay. The possible effects are, however, small (on top of the 
smallness of the relevant CP violating phases) because the final
$I=0$ state is dominant (this is the $\Delta I=1/2$ rule). Defining
\eqn\defAI{
A_I=\vev{(\pi\pi)_I|\H|K^0},\ \ \ \bar A_I=\vev{(\pi\pi)_I|\H|\bar K^0},}
we have, experimentally, $|A_2/A_0|\approx1/20$.
Instead of $\eta_{00}$ and $\eta_{+-}$ we may define two combinations,
$\epsK$ and $\epspK$, in such a way that the possible effects of
CP violation in decay (mixing) are isolated into $\epspK$ ($\epsK$).

The experimental definition of the $\epsK$ parameter is 
\eqn\defepsex{\epsK\equiv{1\over3}(\eta_{00}+2\eta_{+-}).}
To zeroth order in $A_2/A_0$, we have $\eta_{00}=\eta_{+-}=\epsK$.
However, the specific combination \defepsex\ is chosen in such a 
way that  the following relation holds to {\it first} order in $A_2/A_0$:
\eqn\defepsth{\epsK={1-\lambda_0\over1+\lambda_0},}
where
\eqn\deflamz{
\lambda_0=\left({q\over p}\right)_K\left({\bar A_0\over A_0}\right).}
Since, by definition, only one strong channel contributes to $\lambda_0$,
there is indeed no CP violation in decay in \defepsth.
It is simple to show that $\Re\ \epsK\neq0$ is a manifestation
of CP violation in mixing while $\Im\ \epsK\neq0$ is a manifestation
of CP violation in the interference between decays with and without 
mixing. Since experimentally $\arg\epsK\approx\pi/4$, the two
contributions are comparable. It is also clear that $\epsK\neq0$
is a manifestation of indirect CP violation: it could be described
entirely in terms of a CP violating phase in the $M_{12}$ amplitude.

The experimental value of $\epsK$ is given by \PDG\
\eqn\expeps{|\epsK|=(2.280\pm0.013)\times10^{-3}.}

The experimental definition of the $\epspK$ parameter is 
\eqn\defepspex{\epspK\equiv{1\over3}(\eta_{+-}-\eta_{00}).}
The theoretical expression is
\eqn\defepspth{\epspK\approx{1\over6}(\lambda_{00}-\lambda_{+-}).}
Obviously, any type of CP violation which is independent of the
final state does not contribute to $\epspK$. Consequently,
there is no contribution from CP violation in mixing to \defepspth. 
It is simple to show that $\Re\ \epspK\neq0$ 
is a manifestation of CP violation in decay while $\Im\ \epspK\neq0$ 
is a manifestation of CP violation in the interference between decays 
with and without mixing. Following our explanations in the previous
section, we learn that $\epspK\neq0$ is a manifestation of direct 
CP violation: it requires $\phi_2-\phi_0\neq0$ (where $\phi_I$ is
the CP violating phase in the $A_I$ amplitude defined in \defAI).

The quantity that is actually measured in experiment is
\eqn\epsexpe{1-\left|{\eta_{00}\over\eta_{+-}}\right|^2=6\Re(\epe).}
The average over the experimental measurements of $\epe$ 
\nref\NAto{H. Burkhardt {\it et al.}, NA31 collaboration, 
 Phys. Lett. B206 (1988) 169.}%
\nref\NAton{G.D. Barr {\it et al.}, NA31 collaboration,
 Phys. Lett. B317 (1993) 233.}%
\nref\Esto{L.K. Gibbons {\it et al.}, E731 collaboration,
 Phys. Rev. Lett. 70 (1993) 1203.}%
\nref\KTeV{A. Alavi-Harati {\it et al.}, KTeV collaboration,
 Phys. Rev. Lett. 83 (1999) 22, hep-ex/9905060.}%
\nref\NAfe{V. Fanti {\it et al.}, NA48 collaboration, hep-ex/9909022.}%
\refs{\NAto-\NAfe}\ is given by:
\eqn\expepe{\Re(\epe)=(2.11\pm0.46)\times10^{-3}.}

\subsec{The $\epsK$ Parameter in the Standard Model}
An approximate expression for $\epsK$, that is convenient for calculating
it, is given by
\eqn\appeps{\epsK={e^{i\pi/4}\over\sqrt2}{\Im M_{12}\over\Delta m_K}.}
A few points concerning this expression are worth emphasizing:

(i) Eq. \appeps\ is given in a specific phase convention,
where $A_2$ is real. Within the Standard Model, this is a phase
convention where $V_{ud}V_{us}^*$ is real, a condition fulfilled
in both the standard parametrization of eq. \stapar\ and the Wolfenstein
parametrization of eq. \WCKM.

(ii) The phase of $\pi/4$ is approximate. It is determined by hadronic
parameters and therefore is independent of the electroweak model.
Specifically,
\eqn\argeps{\Delta\Gamma_K\approx-2\Delta m_K\ \Longrightarrow\
\arg(\epsK)\approx{\rm arctan}(-2\Delta m_K/\Delta\Gamma_K)\approx\pi/4.}

(iii) A term of order $2{\Im\ A_0\over\Re\ A_0}{\Re\ M_{12}\over\Im\ M_{12}}
\lsim0.02$ is neglected when \appeps\ is derived.

(iv) There is a large hadronic uncertainty in the calculation of
$M_{12}$ coming from long distance contributions. There are, however, 
good reasons to believe that the long distance contributions are important 
in $\Re\ M_{12}$ (where they could be even comparable to the short distance 
contributions), but negligible in $\Im\ M_{12}$. To avoid this uncertainty, 
one uses $\Im M_{12}/\Delta m_K$, with the experimentally measured value
of $\Delta m_K$, instead of $\Im M_{12}/2\Re\ M_{12}$ with the theoretically
calculated value of $\Re\ M_{12}$.

(v) The matrix element $\vev{\bar K^0|(\bar sd)_{V-A}(\bar sd)_{V-A}|K^0}$ 
is yet another source of hadronic uncertainty. If both $\Im\ M_{12}$ and 
$\Re\ M_{12}$ were dominated by short distance contributions, the matrix 
element would have cancelled in the ratio between them. However, as explained
above, this is not the case.

Within the Standard Model, $\Im\ M_{12}$ is accounted for by
box diagrams. One gets:
\eqn\epsCKM{\epsK=e^{i\pi/4}C_\varepsilon B_K\Im\lambda_t\{\Re\lambda_c
[\eta_1 S_0(x_c)-\eta_3 S_0(x_c,x_t)]-\Re\lambda_t\eta_2S_0(x_t)\},}
where the CKM parameters are $\lambda_i=V_{is}^*V_{id}$, the constant
$C_\varepsilon$ is a well known parameter,
\eqn\Csubeps{C_\varepsilon={G_F^2 f_K^2 m_K m_W^2\over6\sqrt{2}\pi^2
\Delta m_K}=3.8\times10^4,}
the $\eta_i$ are QCD correction factors
\ref\HeNi{S. Herrlich and U. Nierste, Phys. Rev. D52 (1995) 6505,
 hep-ph/9507262; Nucl. Phys. B476 (1996) 27, hep-ph/9604330.},
\eqn\etai{\eta_1=1.38\pm0.20,\ \ \ 
\eta_2=0.57\pm0.01,\ \ \ \eta_3=0.47\pm0.04,}
$S_0$ is a kinematic factor, given approximately by \BuLL\
\eqn\Szero{\eqalign{
S_0&(x_t)=2.4\left({m_t\over170\ GeV}\right)^{1.52},\ \ \ 
S_0(x_c)=x_x=2.6\times10^{-4},\cr
S_0&(x_t,x_c)=x_c\left[\ln{x_t\over x_c}-{3x_t\over4(1-x_t)}
-{3x_t^2\ln x_t\over4(1-x_t)^2}\right]=2.3\times10^{-3},\cr}}
and $B_K$ is the ratio between the matrix element of the four quark
operator and its value in the vacuum insertion approximation
(see \BuLL\ for a precise definition),
\eqn\BsubL{B_K=0.85\pm0.15.}
Note that $\epsK$ is proportional to $\Im\lambda_t$ and, consequently,
to $\eta$. The $\epsK$ constraint on the Wolfenstein parameters
can be written as follows:
\eqn\etaeps{\eta\left[(1-\rho)|V_{cb}|^2\eta_2S_0(x_t)+\eta_3S_0(x_c,x_t)
-\eta_1S_0(x_c)\right]|V_{cb}|^2 B_K=1.24\times10^{-6}.}
Eq. \etaeps\ gives hyperbolae in the $\rho-\eta$ plane. The main
sources of uncertainty are in the $B_K$ parameter and in the $|V_{cb}|^4$
dependence. When the theoretically reasonable range for the first
and the experimentally allowed range for the second are taken into
account, one finds that $\eta\gsim0.2$ is required to explain $\epsK$ 
(see \recon). Hence our statement that CP is not an approximate
symmetry of the Standard Model.

\subsec{The $\epspK$ Parameter in the Standard Model}
A convenient approximate expression for $\epspK$ is given by:
\eqn\appepsp{\epspK={i\over\sqrt2}\left|{A_2\over A_0}\right|
e^{i(\delta_2-\delta_0)}\sin(\phi_2-\phi_0).}
We would like to emphasize a few points:

(i) The approximations used in \appepsp\ are 
$\lambda_0=1$ and $|A_2/A_0|\ll1$.

(ii) The phase of $\epspK$ is determined by hadronic parameters
and, therefore, model independent:
\eqn\argepsp{\arg(\epspK)=\pi/2+\delta_2-\delta_0\approx\pi/4.}
The fact that, accidentally, $\arg(\epsK)\approx\arg(\epspK)$, means that
\eqn\epReAbs{\Re(\epe)\approx\epe.}

(iii) $\Re\ \epspK\neq0$ requires $\delta_2-\delta_0\neq0$, consistent
with our statement that it is a manifestation of CP violation in decay.
$\epspK\neq0$ requires $\phi_2-\phi_0\neq0$, consistent with our statement
that it is a manifestation of direct CP violation.

The calculation of $\epe$ within the Standard Model is complicated
and suffers from large hadronic uncertainties. Let us first try
a very naive order of magnitude estimate. The relevant quark decay
process is $s\ra d\bar uu$. All tree diagrams contribute with the
same weak phase, $\phi_T=\arg(V_{ud}^*V_{us})$. Spectator diagrams
contribute to both $I=0$ and $I=2$ final states. Penguin diagrams
with an intermediate $q=u,c,t$ quarks, contribute with a weak phase
$\phi_P^q=\arg(V_{qd}^*V_{qs})$. Strong penguin contribute to the final
$I=0$ states only. Electroweak penguins contribute also to final $I=2$
states, but we will ignore them for the moment. (The fact that the top
quark is heavy means that the electroweak penguins are important.)
A difference in the weak phase between $A_0$ and $A_2$ is then a result
of the fact that $A_0$ has contributions from penguin diagrams with
intermediate $c$ and $t$ quarks. Consequently, $\epspK$ is suppressed
by the following factors:
\item{a.} $|A_2/A_0|\sim0.045$;
\item{b.} $|A_0^{\rm penguin}/A_0^{\rm tree}|\sim0.05$. 
\item{c.} $\sin\beta_K\sim10^{-3}$.

The last factor appears, however, also in $\epsK$. Therefore, it cancels
in the ratio $\epe$. A very rough order of magnitude estimate is then
$\epe\sim10^{-3}$. Note that $\epe$ is not small because of small 
CP violating parameters but because of hadronic parameters. Actually, 
it is independent of $\sin\delta_{\rm KM}$. (In most calculations one
uses the experimental value of $\epsK$ and the theoretical 
expression for $\epspK$. Then the expression for $\epe$ depends on
$\sin\delta_{\rm KM}$.)
 
The actual calculation of $\epe$ is very complicated.
There are several comparable contributions with differing signs.
The final result can be written in the form
\nref\Burepe{A.J. Buras, hep-ph/9806471.}%
\nref\KNS{Y.-Y. Keum, U. Nierste and A.I. Sanda,
 Phys. Lett. B457 (1999) 157, hep-ph/9903230.}%
\nref\BBGJ{S. Bosch {\it et al.}, hep-ph/9904408.}%
\nref\BuSi{A.J. Buras and L. Silvestrini,
 Nucl. Phys. B546 (1999) 299, hep-ph/9811471.}%
(for recent work, see \refs{\Burepe-\BuSi,\BuLL}\ and references
therein):
\eqn\SMepe{\epe=\Im(V_{td}V_{ts}^*)\left[P^{(1/2)}-P^{(3/2)}\right],}
where we omitted a phase factor using the approximation
$\arg(\epsK)=\arg(\epspK)$, and where
\eqn\defPhalf{\eqalign{
P^{(1/2)}\ =&\ {G_F\Re A_2\over2|\epsK|(\Re A_0)^2}\sum y_i\vev{Q_i}_0
(1-\Omega_{\eta+\eta^\prime}),\cr
P^{(3/2)}\ =&\ {G_F\over2|\epsK|\Re A_0}\sum y_i\vev{Q_i}_2.\cr}}
$Q_i$ are four quark operators, and $y_i$ are the Wilson coefficient
functions. The most important operators are
\eqn\OPEepe{\eqalign{
Q_6\ =&\ (\bar s_\alpha d_\beta)_{V-A}\sum_{q=u,d,s}
(\bar q_\beta q_\alpha)_{V+A},\cr
Q_8\ =&{3\over2}\ (\bar s_\alpha d_\beta)_{V-A}\sum_{q=u,d,s}
e_q(\bar q_\beta q_\alpha)_{V+A}.\cr}}
$P^{(3/2)}$ is dominated by electroweak penguin contributions
while $P^{(1/2)}$ is dominated by QCD penguin contributions.
The latter are suppressed by isospin breaking effects ($m_u\neq m_d$),
parametrized by
\eqn\Ometet{\Omega_{\eta+\eta^\prime}={\Re A_0\over\Re A_2}
{(\Im A_2)_{\rm I.B.}\over\Im A_0}\approx0.25\pm0.08.}
A crude approximation to \SMepe\ which emphasizes the sources of
hadronic uncertainty is given by
\eqn\SMepap{\eqalign{\epe\approx13\ \Im(V_{td}V_{ts}^*)&
\left(110\ MeV\over m_s(2\ GeV)\right)^2
\left({\Lambda^{(4)}_{\overline{\rm MS}}\over340\ MeV}\right)\cr
\times&\left[B_6^{(1/2)}(1-\Omega_{\eta+\eta^\prime})
-0.4B_8^{(3/2)}\left({m_t\over165\ GeV}\right)^{2.5}\right].\cr}}
$B_6^{(1/2)}$ and $B_8^{(3/2)}$ parametrize the hadronic matrix elements:
\eqn\defBi{\eqalign{
\vev{Q_6}_0\equiv B_6^{(1/2)}\vev{Q_6}_0^{\rm (vac)},&\ \ \ 
 B_6^{(1/2)}\approx1.0\pm0.3,\cr
\vev{Q_8}_2\equiv B_8^{(3/2)}\vev{Q_8}_2^{\rm (vac)},&\ \ \ 
 B_8^{(3/2)}\approx0.8\pm0.2.\cr}}

The main sources of uncertainties lie then in the parameters
$m_s$, $B_6^{(1/2)}$, $B_8^{(3/2)}$, $\Omega_{\eta+\eta^\prime}$ and
$\Lambda^{(4)}_{\overline{\rm MS}}$. The importance of these
uncertainties is increased because of the cancellation between the
two contributions in \SMepap.
Taking the above reasonable ranges for the hadronic parameters
(from lattice calculations and $1/N_c$ expansion), one can estimate \BBGJ:
\eqn\SMepeval{\Re(\epe)^{\rm SM}=(7.7^{+6.0}_{-3.5})\times10^{-4}.}
\nref\BEFL{S. Bertolini, J.O. Eeg, M. Fabbrichesi and E.I. Lashin,
 Nucl. Phys. B514 (1998) 93, hep-ph/9706260.}%
\nref\BEF{S. Bertolini, J.O. Eeg and M. Fabbrichesi, hep-ph/9802405.}%
\nref\HKPS{T. Hambye, G.O. Kohler, E.A. Paschos and P.H. Soldan,
 hep-ph/9906434.}%
\nref\BBLM{A.A. Belkov, G. Bohm, A.V. Lanyov and A.A. Moshkin,
 hep-ph/9907335.}%
The fact that the range in \SMepeval\ is lower than the experimentally
allowed range in \expepe\ cannot be taken as evidence
for new physics. With a more conservative treatment of the theoretical
uncertainties, one can stretch the theoretical upper bound to values
consistent with the experimental lower bound \refs{\BEFL-\BBLM,\KNS-\BBGJ}.

\subsec{CP violation in $K\rightarrow\pi\nu\bar\nu$}
CP violation in the rare $K\ra\pi\nu\bar\nu$ decays is very
interesting. It is very different from the CP violation
that has been observed in $K\ra\pi\pi$ decays which is small and 
involves theoretical uncertainties. Similar to the CP asymmetry
in $B\ra\psi K_S$, it is predicted to be large and can be cleanly 
interpreted. Furthermore, observation of the $K_L\ra\pi^0\nu\bar\nu$
decay at the rate predicted by the Standard Model will provide
further evidence that CP violation cannot be attributed to mixing 
($\Delta S=2$) processes only, as in superweak models. 
 
Define the decay amplitudes
\eqn\defApnn{A_{\pi^0\nu\bar\nu}=\vev{\pi^0\nu\bar\nu|\H|K^0},\ \ \
\bar A_{\pi^0\nu\bar\nu}=\vev{\pi^0\nu\bar\nu|\H|\bar K^0},}
and the related $\lambda_{\pi\nu\bar\nu}$ quantity:
\eqn\deflpnn{\lambda_{\pi\nu\bar\nu}=\left({q\over p}\right)_K
{\bar A_{\pi^0\nu\bar\nu}\over A_{\pi^0\nu\bar\nu}}.} 
The decay amplitudes of $K_{L,S}$ into a final $\pi^0\nu\bar\nu$
state are then
\eqn\KLpnn{\vev{\pi^0\nu\bar\nu|\H|\bar K_{L,S}}=pA_{\pi^0\nu\bar\nu}
\mp q\bar A_{\pi^0\nu\bar\nu},}
and the ratio between the corresponding decay rates is
\eqn\KLKSpnn{{\Gamma(K_L\ra \pi^0\nu\bar\nu)\over 
\Gamma(K_S\ra \pi^0\nu\bar\nu)}=
{1+|\lambda_{\pi\nu\bar\nu}|^2-2\Re\lambda_{\pi\nu\bar\nu}\over
1+|\lambda_{\pi\nu\bar\nu}|^2+2\Re\lambda_{\pi\nu\bar\nu}}.}
We learn that the $K_L\ra \pi^0\nu\bar\nu$ decay rate vanishes in
the CP limit ($\lambda_{\pi\nu\bar\nu}=1$), as expected on general
grounds \Litt. (The CP conserving contributions were explicitly 
calculated within the Standard Model
\ref\BuIs{G. Buchalla and G. Isidori, Phys. Lett. B140 (1998) 170,
 hep-ph/9806501.}\
and within its extension with massive neutrinos
\ref\Pere{G. Perez, JHEP 9909 (1999) 019, hep-ph/9907205.},
and found to be negligible.)

Since the effects of CP violation in decay and in mixing are expected
to be negligibly small, $\lambda_{\pi\nu\bar\nu}$ is, to an excellent
approximation, a pure phase. Defining $\theta_K$ to be the relative
phase between the $K-\bar K$ mixing amplitude and the $s\ra d\nu\bar\nu$
decay amplitude, namely $\lambda_{\pi\nu\bar\nu}=e^{2i\theta_K}$, 
we get from \KLKSpnn:
\eqn\KLKSpn{{\Gamma(K_L\ra \pi^0\nu\bar\nu)\over 
\Gamma(K_S\ra \pi^0\nu\bar\nu)}={1-\cos2\theta_K\over1+\cos2\theta_K}=
\tan^2\theta_K.}
Using the isospin relation $A(K^0\ra\pi^0\nu\bar\nu)/
A(K^+\ra\pi^+\nu\bar\nu)=1/\sqrt2$, we get
\eqn\defapnn{a_{\pi\nu\bar\nu}\equiv{\Gamma(K_L\ra \pi^0\nu\bar\nu)\over 
\Gamma(K^+\ra \pi^+\nu\bar\nu)}={1-\cos2\theta_K\over2}=\sin^2\theta_K.}
Note that $a_{\pi\nu\bar\nu}\leq1$, and consequently a measurement
of $\Gamma(K^+\ra \pi^+\nu\bar\nu)$ can be used to set a model independent
upper limit on $\Gamma(K_L\ra \pi^0\nu\bar\nu)$ \GrNi.

Within the Standard Model, the $K\ra\pi\nu\bar\nu$ decays are dominated by 
short distance $Z$-penguins and box diagrams. The branching ratio for $K^+\ra
\pi^+\nu\bar\nu$ can be expressed in terms of $\rho$ and $\eta$ \BuLL:
\eqn\BrKppnnc{BR(K^+\ra\pi^+\nu\bar\nu)=4.11\times10^{-11}
A^4[X(x_t)]^2\left[\eta^2+(\rho_0-\rho)^2\right],}
where
\eqn\defrhpz{\rho_0=1+{P_0(X)\over A^2X(x_t)},}
and $X(x_t)$ and $P_0(X)$ represent the electroweak loop contributions
in NLO for the top quark and for the charm quark, respectively.
The main theoretical uncertainty is related to the strong
dependence of the charm contribution on the renormalization
scale and the QCD scale, $P_0(X)=0.42\pm0.06$. First evidence
for $K^+\ra\pi^+\nu\bar\nu$ was presented recently
\ref\Adler{S. Adler {\it et al.}, E787 Collaboration, 
 Phys. Rev. Lett. 79 (1997) 2204, hep-ex/9708031.}.
The large experimental error does not yet give a useful CKM constraint
and is consistent with the Standard Model prediction.

The branching ratio for the $K_L\ra\pi^0\nu\bar\nu$ decay can be expressed 
in terms of $\eta$ \BuLL:
\eqn\BrKppnnn{BR(K_L\ra\pi^0\nu\bar\nu)=1.80\times10^{-10}
[X(x_t)]^2A^4\eta^2.}
The present experimental bound, $BR(K_L\ra\pi^0\nu\bar\nu)\leq
1.6\times10^{-6}$
\ref\KTeV{J. Adams {\it et al.}, KTeV Collaboration,
 Phys. Lett. B447 (1999) 240, hep-ex/9806007.}\
lies about five orders of magnitude above the Standard Model prediction
\BuLL\ and about two orders of magnitude above the bound that can be deduced
using model independent isospin relations \GrNi\
from the experimental upper bound on the charged mode. 

Note that if the charm contribution to $K^+\ra\pi^+\nu\bar\nu$ were negligible,
so that both the charged and the neutral decays were dominated by
the intermediate top contributions, then $a_{\pi\nu\bar\nu}$ would
simply measure $\sin^2\beta$. While the charm contribution makes
the evaluation of $a_{\pi\nu\bar\nu}$ more complicated, a reasonable
order of magnitude estimate is still given by $\sin^2\beta$.

\newsec{CP Violation in $D\rightarrow K\pi$ Decays}
Within the Standard Model, $D-\bar D$ mixing is expected to be well
below the experimental bound. Furthermore, effects related to CP violation
in $D-\bar D$ mixing are expected to be negligibly small since
this mixing is described to a good approximation by physics
of the first two generations. An experimental observation of
$D-\bar D$ mixing close to the present bound (and, even more convincingly, 
of related CP violation) will then be evidence for New Physics.
The most sensitive experimental searches for $D-\bar D$ mixing
use $D\ra K\pi$ decays
\nref\Esnoa{E791 collaboration,
 Phys. Rev. Lett. 77 (1996) 2384, hep-ex/9606016.}%
\nref\Esnob{E791 collaboration, Phys. Rev. D57 (1998) 13, hep-ex/9608018.}%
\nref\alephD{ALEPH collaboration,
 Phys. Lett. B436 (1998) 211, hep-ex/9811021.}%
\nref\Esnoc{E791 collaboration,
 Phys. Rev. Lett. 83 (1999) 32, hep-ex/9903012.}%
\nref\cleoD{CLEO collaboration, hep-ex/9908040.}%
\refs{\Esnoa-\cleoD}. We now give the formalism of neutral $D$ decays 
into final $K^\pm\pi^\mp$ states. 

\subsec{Formalism}
We define the neutral $D$ mass eigenstates:
\eqn\defqpD{|{D_{1,2}}\rangle=p|{D^0}\rangle\pm q|{\bar D^0}\rangle.}
We define the following four decay amplitudes:
\eqn\defAKpi{\eqalign{A_{K^+\pi^-}=\vev{K^+\pi^-|{\cal H}|D^0},&\ \ \ 
\bar A_{K^+\pi^-}=\vev{K^+\pi^-|{\cal H}|\bar D^0},\cr
A_{K^-\pi^+}=\vev{K^-\pi^+|{\cal H}|D^0},&\ \ \ 
\bar A_{K^-\pi^+}=\vev{K^-\pi^+|{\cal H}|\bar D^0}.\cr}}
We introduce the following two quantities:
\eqn\deflpm{
\lambda_{K^+\pi^-}\ =\ \left({q\over p}\right)_D\ 
{\bar A_{K^+\pi^-}\over A_{K^+\pi^-}},\ \ \ 
\lambda_{K^-\pi^+}\ =\ \left({q\over p}\right)_D\ 
{\bar A_{K^-\pi^+}\over A_{K^-\pi^+}}.}

The following approximations are all experimentally confirmed:
\eqn\Dexp{x\equiv{\Delta m_D\over\Gamma_D}\ll1;\ \ \ 
y\equiv{\Delta \Gamma_D\over2\Gamma_D}\ll1;\ \ \ 
|\lambda_{K^+\pi^-}^{-1}|\ll1;\ \ \ |\lambda_{K^-\pi^+}|\ll1.}
Using these approximations, the decay rates for the doubly Cabibbo
suppressed (DCS) decays are given by
\eqn\DCSmin{\eqalign{
\Gamma&[D^0(t)\ra K^+\pi^-]\ =\ e^{-t}
\left|\bar A_{K^+\pi^-}\right|^2\left|{q/p}\right|^2\times\cr
&\left[\left|\lambda_{K^+\pi^-}^{-1}\right|^2
+\Re(\lambda_{K^+\pi^-}^{-1})y t+\Im(\lambda_{K^+\pi^-}^{-1})x t
+{1\over4}(x^2+y^2)t^2\right],\cr
\Gamma&[\bar D^0(t)\ra K^-\pi^+]\ =\ e^{-t}
\left|A_{K^-\pi^+}\right|^2\left|{p/q}\right|^2\times\cr
&\left[\left|\lambda_{K^-\pi^+}\right|^2
+\Re(\lambda_{K^-\pi^+})y t+\Im(\lambda_{K^-\pi^+})x t
+{1\over4}(x^2+y^2)t^2\right].\cr}}
The time $t$ is given here in units of the $D$-lifetime,
$\tau_D=1/\Gamma_D$. Eqs. \DCSmin\ are valid for times $\lsim\tau_D$.

The decay rates for the Cabibbo favored (CF) modes are given to a good 
approximation by
\eqn\CFmin{\eqalign{
\Gamma[D^0(t)\ra K^-\pi^+]&\ =\ e^{-t}\left|A_{K^-\pi^+}\right|^2,\cr
\Gamma[\bar D^0(t)\ra K^+\pi^-]&\ =\ e^{-t}
\left|\bar A_{K^+\pi^-}\right|^2.\cr}}

\subsec{CP Violation}
We will assume that the CF decays are unaffected by CP violation, that is,
\eqn\noCPCF{\left|A_{K^-\pi^+}\right|=\left|\bar A_{K^+\pi^-}\right|
\equiv \ACF.}
In general, $|q/p|$ is a positive real number and $\lambda_{K^+\pi^-}^{-1}$ 
and $\lambda_{K^-\pi^+}$ are two independent complex numbers. We now
parametrize these quantities in a way that is convenient to separate
the three types of CP violation:
\eqn\genpar{\eqalign{
|q/p|\ =&\ R_m,\cr
\lambda_{K^+\pi^-}^{-1}\ =&\ {R\over R_dR_m} e^{-i(\delta_{K\pi}+\phi_{K\pi})},
\cr
\lambda_{K^-\pi^+}\ =&\ R R_d R_m e^{-i(\delta_{K\pi}-\phi_{K\pi})}.\cr}}
CP violation in mixing, that is violation of $|q/p|=1$, is related to 
$R_m\neq1$. CP violation in decay, that is violation of 
$\left|{A_{K^+\pi^-}\over\bar A_{K^-\pi^+}}\right|=
\left|{\bar A_{K^+\pi^-}\over A_{K^-\pi^+}}\right|=1$, is related to $R_d\neq1$.
CP violation in interference of decays with and without mixing, that is 
violation of ${\Im(\lambda_{K^+\pi^-}^{-1})\over|\lambda_{K^+\pi^-}^{-1}|}=
{\Im(\lambda_{K^-\pi^+})\over|\lambda_{K^-\pi^+}|}$, is related to 
$\phi_{K\pi}\neq0$.

We also define
\eqn\delpri{\eqalign{
x^\prime\ \equiv&\ x\cos\delta_{K\pi} +y\sin\delta_{K\pi},\cr
y^\prime\ \equiv&\ y\cos\delta_{K\pi} -x\sin\delta_{K\pi}.\cr}}

In the language of eqs. \noCPCF, \genpar\ and \delpri, we can rewrite
eq. \DCSmin\ as follows:
\eqn\DCScp{\eqalign{
\Gamma[D^0(t)\ra K^+\pi^-]\ =\ e^{-t}\ACF^2 
&\left[{R^2\over R_d^2}+{RR_m\over R_d}(y^\prime c_\phi-x^\prime s_\phi)t
+{R_m^2\over4}(x^2+y^2)t^2\right],\cr
\Gamma[\bar D^0(t)\ra K^-\pi^+]\ =\ e^{-t}\ACF^2
&\left[R^2 R_d^2+{R R_d\over R_m}(y^\prime c_\phi+x^\prime s_\phi)t
+{1\over4R_m^2}(x^2+y^2)t^2\right],\cr}}
where $s_\phi\equiv\sin\phi_{K\pi}$ and $c_\phi\equiv\cos\phi_{K\pi}$.
Note that the three different mechanisms of CP violation can be
distinguished if the time dependent rates \DCScp\ are measured:
\item{(i)} A different $t^2e^{-t}$ term in the DCS decays of $D^0$ and
$\bar D^0$ is an unambiguous signal of CP violation in mixing.
\item{(ii)} A different $e^{-t}$ term in the DCS decays of $D^0$ and
$\bar D^0$ is an unambiguous signal of CP violation in decay.
\item{(iii)} A measurement of all three terms for each of the two decay rates
can provide an unambiguous signal of CP violation in interference.

In the presence of new physics, the most likely situation is that we
have observable CP violation in interference between decays with and
without mixing, while CP violation in mixing and in decays
\ref\BeNi{S. Bergmann and Y. Nir,  JHEP 9909 (1999) 031, hep-ph/9909391.}\
remain unobservably small. In this case, $R_m=1$, $R_d=1$, but 
$\phi_{K\pi}\neq0$. Furthermore, while $\Delta m_D$ can be enhanced
to the level of present experimental sensitivity, $\Delta\Gamma_D$ is
likely to remain close to the SM prediction. Neglecting $y$, we have
in this case
\eqn\DCSint{\eqalign{
\Gamma[D^0(t)\ra K^+\pi^-]\ =\ e^{-t}\ACF^2 
&\left[R^2-Rx(s_\delta c_\phi+c_\delta s_\phi)t+{x^2\over4}t^2\right],\cr
\Gamma[\bar D^0(t)\ra K^-\pi^+]\ =\ e^{-t}\ACF^2
&\left[R^2-Rx(s_\delta c_\phi-c_\delta s_\phi)t+{x^2\over4}t^2\right].\cr}}
The $te^{-t}$ terms in \DCSint\ are potentially CP violating. There are four 
possibilities concerning these terms
\nref\BSN{G. Blaylock, A. Seiden and Y. Nir,
 Phys. Lett. B355 (1995) 555, hep-ph/9504306.}%
\nref\WolD{L. Wolfenstein,
 Phys. Rev. Lett. 75 (1995) 2460, hep-ph/9505285.}%
\refs{\BSN,\WolD}:
\item{(i)} They vanish: both strong and weak phases play no role.
\item{(ii)} They are equal in magnitude and in sign: weak phases play no role. 
\item{(iii)} They are equal in magnitude but have opposite signs:
strong phases play no role. 
\item{(iv)} They have different magnitudes: both strong and weak phases 
play a role.

\newsec{CP Violation in $B$ Decays in the Standard Model}
\subsec{$|q/p|\neq1$}
As explained in the previous section, in the $B_d$ system we expect 
model independently that $\Gamma_{12}\ll M_{12}$. Within any given model
we can actually calculate the two quantities from quark diagrams. 
Within the Standard Model, $M_{12}$ is given by box diagrams.
For both the $B_d$ and $B_s$ systems, the long distance contributions
are expected to be negligible and the calculation of these diagrams with
a high loop momentum is a very good approximation. 
$\Gamma_{12}$ is calculated from a cut of box diagrams
\ref\BKUS{I.I. Bigi, V.A. Khoze, N.G. Uraltsev and A.I. Sanda, in {\it
 CP Violation}, ed. C. Jarlskog (World Scientific, Singapore, 1992).}. 
Since the cut of a diagram always involves on-shell particles and thus long 
distance physics, the cut of the quark box diagram is 
a poor approximation to $\Gamma_{12}$. However, it does correctly give
the suppression from small electroweak parameters such as the weak
coupling. In other words, though the hadronic uncertainties are large
and could change the result by order $50\%$, the cut in the
box diagram is expected to give a reasonable order of magnitude estimate
of $\Gamma_{12}$. (For $\Gamma_{12}(B_s)$ it has been shown that local
quark-hadron duality holds exactly in the simultaneous limit of small
velocity and large number of colors. We thus expect an uncertainty of
$\O(1/N_C)\sim30\%$
\nref\Alek{R. Aleksan {\it et al.}, Phys. Lett. B316 (1993) 567.}%
\nref\Bene{M. Beneke {\it et al.}, Phys.Lett. B459 (1999) 631, hep-ph/9808385.}%
\refs{\Alek,\Bene}. For $\Gamma_{12}(B_d)$ the small velocity limit
is not as good an approximation but an uncertainty of order 50\% still
seems a reasonable estimate
\ref\WolGot{L. Wolfenstein, Phys. Rev. D57 (1998) 5453.}.) 

Within the Standard Model, $M_{12}$ is dominated by top-mediated box diagrams
\ref\BBL{A.J. Buras, G. Buchalla and M.E. Lautenbacher,
 Rev. Mod. Phys. 68 (1996) 1125, hep-ph/9512380.}:
\eqn\MontwB{M_{12}={G_F^2\over12\pi^2}m_Bm_W^2\eta_BB_Bf_B^2
(V_{tb}V_{td}^*)^2S_0(x_t),}
where $S_0(x_t)$ is given in eq. \Szero, $\eta_B=0.55$ is a QCD correction, 
and $B_Bf_B^2$ parametrizes the hadronic matrix element.
For $\Gamma_{12}$, we have
\nref\Hage{J. Hagelin, Nucl. Phys. B193 (1981) 123.}%
\nref\BSS{A.J. Buras, W. Slominski and H. Steger,
 Nucl. Phys. B245 (1984) 369.}%
\nref\BBD{M. Beneke, G. Buchalla and I. Dunietz,
 Phys. Rev. D54 (1996) 4419, hep-ph/9605259.}%
\refs{\Hage-\BBD}
\eqn\GontwB{\eqalign{\Gamma_{12}\ =&\ -{G_F^2\over24\pi}m_Bm_b^2B_Bf_B^2
(V_{tb}V_{td}^*)^2\cr &\times\left[{5\over3}{m_B^2\over(m_b+m_d)^2}
{B_S\over B_B}(K_2-K_1)+{4\over3}(2K_1+K_2)+8(K_1+K_2)
{m_c^2\over m_b^2}{V_{cb}V_{cd}^*\over V_{tb}V_{td}^*}\right],}}
where $K_1=-0.39$ and $K_2=1.25$ \BBD\ are combinations of Wilson
coefficients and $B_S$ parametrizes the $(S-P)^2$ matrix element.
Note that new physics is not expected to affect
$\Gamma_{12}$ significantly because it usually takes place at a high
energy scale and is relevant to the short distance part only.
Therefore, the Standard Model estimate in eq. \GontwB\ remains
valid model independently. Combining \MontwB\ and \GontwB, one gets  
\eqn\SMGtoM{
{\Gamma_{12}\over M_{12}}=-5.0\times10^{-3}\left(1.4{B_S\over B_B}
+0.24+2.5{m_c^2\over m_b^2}{V_{cb}V_{cd}^*\over V_{tb}V_{td}^*}\right).}
We learn that $|\Gamma_{12}/M_{12}|=\O(m_b^2/m_t^2)$, which confirms our 
model independent order of magnitude estimate, $|\Gamma_{12}/M_{12}|
\lsim10^{-2}$. As concerns the imaginary part of this ratio, we have
\eqn\aSLSM{a_{\rm SL}=\Im{\Gamma_{12}\over M_{12}}=-1.1\times10^{-3}
\sin\beta=-(2-5)\times10^{-4}.}
The strong suppression of $a_{\rm SL}$ compared to $|\Gamma_{12}/M_{12}|$ 
comes from the fact that the leading contribution to $\Gamma_{12}$
has the same phase as $M_{12}$. Consequently, $a_{\rm SL}=\O(m_c^2/m_t^2)$.
The CKM factor, $\Im{V_{cb}V_{cd}^*\over V_{tb}V_{td}^*}=\sin\beta$,
is of order one. In contrast,
for the $B_s$ system, where \SMGtoM\ holds except that $V_{td}$ ($V_{cd}$) is
replaced by $V_{ts}$ ($V_{cs}$), there is an additional suppression from
$\Im{V_{cb}V_{cs}^*\over V_{tb}V_{ts}^*}=\sin\beta_s\sim10^{-2}$ 
(see the corresponding unitarity triangle).

\subsec{$|\bar A_{\bar f}/A_f|\neq1$}
In the previous subsection we estimated the effect of CP violation
in mixing to be of $\O(10^{-3})$ within the Standard Model, and
$\leq\O(|\Gamma_{12}/M_{12}|)\sim10^{-2}$ model independently  
\nref\RaSu{L. Randall and S. Su,
 Nucl. Phys. B540 (1999) 37, hep-ph/9807377.}%
\nref\CaWo{R.N. Cahn and M.P. Worah,
 Phys. Rev. D60 (1999) 076006, hep-ph/9904480.}%
(for recent discussions, see \refs{\RaSu-\CaWo,\BEN}).
In semileptonic decays, CP violation in mixing is the leading effect
and therefore it can be measured through $a_{\rm SL}$. In purely hadronic
$B$ decays, however, CP violation in decay and in the interference of
decays with and without mixing is $\geq \O(10^{-2})$. We can therefore
safely neglect CP violation in mixing in the following discussion and use
\eqn\qpSM{{q\over p}=-{M_{12}^*\over|M_{12}|}=
{V_{tb}^*V_{td}\over V_{tb}V_{td}^*}\omega_B.}
(From here on we omit the convention-dependent quark phases $\omega_q$
defined in eq. \CPofq. Our final expressions for physical
quantities are of course unaffected by such omission.)

A crucial question is then whether CP violation in decay is comparable to 
the CP violation in the interference of decays with and without mixing or 
negligible.  In the first case, we can use the corresponding charged $B$ 
decays to observe effects of CP violation in decay. In the latter case, 
CP asymmetries in neutral $B$ decays are subject to clean theoretical 
interpretation: we will either have precise measurements of CKM parameters 
or be provided with unambiguous evidence for new physics. The question of 
the relative size of CP violation in decay can only be answered
on a channel by channel basis, which is what we do in this section.

Channels that have contributions from tree diagrams only  
depend each on a single CKM combination:
\eqn\onlytree{\eqalign{
A(c\bar u d)\ =&\ T_{c\bar u d} V_{cb}V_{ud}^*,\cr
A(c\bar u s)\ =&\ T_{c\bar u s} V_{cb}V_{us}^*,\cr
A(u\bar c d)\ =&\ T_{u\bar c d} V_{ub}V_{cd}^*,\cr
A(u\bar c s)\ =&\ T_{u\bar c s} V_{ub}V_{cs}^*.\cr}}
The subdivision of tree processes into spectator, exchange and annihilation 
diagrams is unimportant in this respect since they all carry the same weak 
phase. For such modes, $|\bar A/A|=1$. It is possible that $\Im\lambda\neq0$,
but since the final states are not CP eigenstates, a clean theoretical
interpretation is difficult. We do not discuss these modes here any further.
 
Most channels have contributions from both tree- and three types of
penguin-diagrams, the latter classified according to the identity of the 
quark in the loop, as diagrams with different intermediate quarks may have 
both different strong phases and different weak phases 
\ref\BSS{M. Bander, S. Silverman and A. Soni, Phys. Rev. Lett. 43 (1979) 242.}.
Consider first $b\ra q\bar qs$ decays, with $q=c$ or $u$. Using \Unitsb,
we can write the CKM dependence of these amplitudes as follows:
\eqn\ccstype{\eqalign{
A(c\bar c s)\ =&\ (T_{c\bar cs}+P^c_s-P^t_s)V_{cb}V_{cs}^*
+(P^u_s-P^t_s)V_{ub}V^*_{us},\cr 
A(u\bar u s)\ =&\ (P^c_s-P^t_s)V_{cb}V_{cs}^*
+(T_{u\bar us}+P^u_s-P^t_s)V_{ub}V^*_{us},}}
where $T$ stands for the tree amplitude and $P^q$ for a penguin diagram
with an intermediate $q$-quark. Next, consider $b\ra q\bar qd$ decays, 
with $q=c$ or $u$. Using \Unitdb,
we can write the CKM dependence of these amplitudes as follows:
\eqn\ccdtype{\eqalign{
A(c\bar c d)\ =&\ (P^t_d-P^u_d)V_{tb}V_{td}^*
+(T_{c\bar c d}+P^c_d-P^u_d) V_{cb}V_{cd}^*,\cr 
A(u\bar u d)\ =&\ (P^t_d-P^c_d)V_{tb}V_{td}^*
 +(T_{u\bar u d}+P^u_d - P^c_d)V_{ub}V_{ud}^*.\cr}} 
Note that in both \ccstype\ and \ccdtype\ only differences of penguin 
contributions occur, which makes the cancellation of the ultraviolet 
divergences of these diagrams explicit. 

To estimate the size of CP violation in decay for these channels, we need to 
know the ratio of the contribution from the difference between a top and light 
quark strong penguin diagram to the contribution from a tree diagram 
(with the CKM combination factored out):
\eqn\pengtree{r_{PT}={P^t-P^{\rm light}\over T_{q \bar q q^\prime}}
\approx{\alpha_s\over12\pi}\ln{m_t^2\over m_b^2}=\O(0.03).}
However, this estimate does not include the effect of hadronic matrix 
elements, which are the probability factor to produce a particular final 
state particle content from a particular quark content. Since this probability 
differs for different kinematics, color flow and spin structures, it can be 
different for tree and penguin contributions and may partially compensate the 
coupling constant suppression of the penguin term. Recent CLEO results on
$BR(B\ra K\pi)$ and $BR(B\ra\pi\pi)$ 
\ref\CLEOpen{R. Godang {\it et al.}, CLEO Collaboration,
 Phys. Rev. Lett. 80 (1998) 3456, hep-ex/9711010.}\
suggest that the matrix element of penguin operators is indeed enhanced
compared to that of tree operators. The enhancement could be by a factor
of a few, leading to
\eqn\pentre{r_{PT}\sim\lambda^2-\lambda.}
(Note that $r_{PT}$ does not depend on the CKM parameters.
We use powers of the Wolfenstein parameter $\lambda$ to
quantify our estimate for $r_{PT}$ in order to simplify
the comparison between the size of CP violation in decay
and CP violation in the interference between decays with
and without mixing.) Take, for example, the $b\ra c\bar cs$ decays.
Using eqs. \AbsAA\ and \ccstype, the size of CP violation in decay is 
then estimated as follows:
\eqn\ccsdec{1-\left|{\bar A_{c\bar cs}\over A_{c\bar c\bar s}}\right|
\lsim r_{PT}\ \Im{V_{ub}V^*_{us}\over V_{cb}V^*_{cs}}=\O(\lambda^4-\lambda^3).} 
Note that, in the language of eq. \AbsAA, this estimate includes 
$(A_2/A_2)\sin(\phi_2-\phi_1)$ but not $\sin(\delta_2-\delta_1)$.
We only used $\sin(\delta_2-\delta_1)\leq1$ but if, for some reason,
the difference in strong phases is small, the effect of CP violation
in decay will be accordingly suppressed compared to \ccsdec.

Finally, processes involving only down-type quarks have no
contributions from tree diagrams:
\eqn\ssstype{\eqalign{
A(s\bar s s)\ =&\ (P^c_s-P^t_s)V_{cb}V_{cs}^*+(P^u_s-P^t_s)V_{ub}V^*_{us},\cr
A(s\bar s d)\ =&\ (P^t_d-P^u_d)V_{tb}V_{td}^*+(P^c_d-P^u_d)V_{cb}V_{cd}^*.\cr}}
For the estimate of CP violation in decay for these modes, we need
to consider two types of ratios between penguin diagrams:
\eqn\penratios{\eqalign{
r_{P^cP^u}\ =&\ {P^u-P^t\over P^c-P^t}\approx1,\cr
r_{P^lP^t}\ =&\ {P^c-P^u\over P^t-P^{\rm light}}\approx0.1.\cr}}
In the $m_c=m_u$ limit, we would have $r_{P^cP^u}=1$ and $r_{P^lP^t}=0$.
The deviations from these limiting values should then be GIM suppressed
\ref\GIM{S.L. Glashow, J. Iliopoulos and L. Maiani,
 Phys. Rev. D2 (1970) 1285.}.
The estimate of the somewhat surprisingly large $r_{P^lP^t}$ is based on refs.
\nref\Flei{R. Fleischer, Phys. Lett. B341 (1995) 205, hep-ph/9409290.}%
\nref\BuFl{A.J. Buras and R. Fleischer,
 Phys. Lett. B341 (1995) 379, hep-ph/9409244.}%
\refs{\Flei,\BuFl}. We get:
\eqn\sssdec{1-\left|{\bar A_{s\bar ss}\over A_{s\bar s\bar s}}\right|
\lsim r_{P^lP^t}\ \Im{V_{ub}V^*_{us}\over V_{cb}V^*_{cs}}=\O(\lambda^2),}
\eqn\ssddec{1-\left|{\bar A_{s\bar sd}\over A_{s\bar s\bar d}}\right|
\lsim r_{P^uP^c}\ \Im{V_{cb}V^*_{cd}\over V_{tb}V^*_{td}}=\O(0.1).}
As concerns the $b\ra d\bar d s$ and $b\ra d\bar dd$ processes, they mix
strongly through rescattering effects with the tree mediated
$b\ra u\bar u s$ and $b\ra u\bar ud$ decays, respectively. It is difficult
to estimate these soft rescattering effects and we do not consider
these modes here any further. 

We thus classify the $B$ decays described in eqs. \ccstype, \ccdtype\
and \ssstype\ into four classes. Classes (i) and (ii) are
expected to have relatively small CP violation in decay and hence are
particularly interesting for extracting CKM parameters from
interference of decays with and without mixing. In classes (iii) and (iv), 
CP violation in decay could be significant and might be observable
in charged $B$ decays.

(i) Decays dominated by a single term: $b\ra c\bar cs$ and $b\ra
s\bar ss$. The Standard Model predicts very small
CP violation in decay: $\O(\lambda^4-\lambda^3)$ for $b\ra c\bar cs$
and $\O(\lambda^2)$ for $b\ra s\bar ss$.
Any observation of large CP asymmetries in charged $B$ decays
for these channels would be a clue to physics beyond the Standard Model.
The corresponding neutral modes have cleanly predicted relationships
between CKM parameters and the measured asymmetry from interference
between decays with and without mixing. The modes
$B\ra\psi K$ and $B\ra\phi K$ are examples of this class.

(ii) Decays with a small second term: $b\ra c\bar cd$ and $b\ra
u \bar ud$. The expectation that penguin-only contributions are suppressed
compared to tree contributions suggests that these modes will have small
effects of CP violation in decay, of $\O(\lambda^2-\lambda)$, and an
approximate prediction for the relationship between measured asymmetries
in neutral decays and CKM
phases can be made. Examples here are $B\ra\ DD$ and $B\ra\pi\pi$.

(iii) Decays with a suppressed tree contribution: $b\ra u \bar us$.
The tree amplitude is suppressed by small mixing angles, $V_{ub}V_{us}$.
The no-tree term may be comparable or even dominate and give large
interference effects. An example is $B\ra\rho K$.

(iv) Decays with no tree contribution and a small second term: 
$b\ra s\bar s d$. Here the
interference comes from penguin contributions with different charge 2/3
quarks in the loop and gives CP violation in decay that could be as large
as 10\%. An example is $B\ra KK$.

Note that if the penguin enhancement is significant,
then some of the decay modes listed in class (ii) might actually fit
better in class (iii). For example, it is possible that $b\ra u\bar ud$
decays have comparable contributions from tree and penguin amplitudes.
On the other hand, this would also mean that some modes listed in class
(iii) could be dominated by a single penguin term. For such cases
an approximate relationship between measured
asymmetries in neutral decays and CKM phases can be made.

A summary of our discussion in this section is given in the table II.
\vskip 1cm

\begintable
Quark process & Sample $B^\pm$ mode & $\O(1-\vert\bar A/A\vert)$ \crthick
$b\ra c\bar cs$ & $\psi K^\pm$ & $r_{PT}\sin\beta_s\sim\lambda^4-\lambda^3$ \nr
$b\ra s\bar ss$ & $\phi K^\pm$ & $\sin\beta_s\sim\lambda^2$ \nr
$b\ra u\bar ud$ & $\pi^0\pi^\pm$ & $r_{PT}\sin\alpha\sim\lambda^2-\lambda$ \nr
$b\ra c\bar cd$ & $DD^\pm$ & $r_{PT}\sin\gamma\sim\lambda^2-\lambda$ \nr
$b\ra u\bar us$ & $\pi^0K^\pm$ & $r_{PT}^{-1}\sin\beta_s\sim\lambda-1$ \nr
$b\ra s\bar sd$ & $\phi\pi^\pm$ & $r_{PP}\sin\beta\sim0.1$  \endtable

\centerline{Table II. CP violation in $B$ decays.}

\subsec{$\Im\lambda_{f_{\CP}}\neq0$}
Let us first discuss an example of class (i), $B\ra\psi K_S$. A new
ingredient in the analysis is the effect of $K-\bar K$ mixing. For decays
with a single $K_S$ in the final state, $K-\bar K$ mixing is essential
because $B^0\ra K^0$ and $\bar B^0\ra\bar K^0$, and interference is
possible only due to $K-\bar K$ mixing. This adds a factor of
\eqn\qpK{
\left({p\over q}\right)_K={V_{cs}V_{cd}^*\over V_{cs}^*V_{cd}}\omega_K^*}
into $(\bar A/A)$. The quark subprocess in $\bar B^0\ra\psi\bar K^0$ is
$b\ra c\bar cs$ which is dominated by the $W$-mediated tree diagram:
\eqn\ApsiK{
{\bar A_{\psi K_S}\over A_{\psi K_S}}= \eta_{\psi K_S}
\left({V_{cb}V_{cs}^*\over V_{cb}^*V_{cs}}\right)
\left({V_{cs}V_{cd}^*\over V_{cs}^*V_{cd}}\right)\omega_B^*.}
The CP-eigenvalue of the state is $\eta_{\psi K_S} = -1$.
Combining eqs. \qpSM\ and \ApsiK, we find
\eqn\lampsiK{\lambda(B\ra\psi K_S)=-\left({V_{tb}^*V_{td}\over
V_{tb}V_{td}^*}\right)\left({V_{cb}V_{cs}^*\over V_{cb}^*V_{cs}}\right)
\left({V_{cd}^*V_{cs}\over V_{cd}V_{cs}^*}\right)
\ \Longrightarrow\ \Im\lambda_{\psi K_S}=\sin(2\beta).}

We have seen in the previous section that, for $b\ra c\bar cs$ decays,
we have a very small CP violation in decay, $1-|\bar A/A|\sim\lambda^2r_{PT}$.
Consequently, eq. \lampsiK\ is clean of hadronic uncertainties to better than
$\O(10^{-2})$. This means that the measuremnet of $a_{\psi K_S}$ can give
the theoretically cleanest determination of a CKM parameter, even
cleaner than the determination of $|V_{us}|$ from $K\ra\pi\ell\nu$. (If
BR($K_L\ra\pi\nu\bar\nu$) is measured, it will give a comparably clean
determination of $\eta$.)

A second example of a theoretically clean mode in class (i) is
$B\ra\phi K_S$. We showed in the previous section that, for 
$b\ra s\bar ss$ decays, we have small CP violation in decay,
$1-|\bar A/A|=\O(\lambda^2)=\O(0.05)$. We can neglect this effect.  
The analysis is similar to the $\psi K_S$ case, and the asymmetry is 
proportional to $\sin(2\beta)$.

The same quark subprocesses give theoretically clean CP
asymmetries also in $B_s$ decays. These asymmetries are, however,
very small since the relative phase between the mixing amplitude
and the decay amplitudes ($\beta_s$ defined in \bbangles) is very small.

The best known example of class (ii) is $B\ra\pi\pi$. The quark subprocess
is $b\ra u\bar ud$ which is dominated by the $W$-mediated tree diagram.
Neglecting for the moment the second, pure penguin, term we find
\eqn\Apipi{{\bar A_{\pi\pi}\over A_{\pi\pi}}=\eta_{\pi\pi}{V_{ub}V_{ud}^*
\over V_{ub}^*V_{ud}}\omega_B^*.}
The CP eigenvalue for two pions is $+1$.
Combining eqs. \qpSM\ and \Apipi, we get
\eqn\lampipi{
\lambda(B\ra\pi^+\pi^-)=\left({V_{tb}^*V_{td}\over V_{tb}V_{td}^*}\right)
\left({V_{ud}^*V_{ub}\over V_{ud}V_{ub}^*}\right)
\ \Longrightarrow\ \Im\lambda_{\pi\pi}=\sin(2\alpha).}
The pure penguin term in $A(u\bar ud)$ in eq. \ccdtype\ has a weak phase,
$\arg(V_{td}^*V_{tb})$, different from the term with the tree
contribution, so it modifies both $\Im\lambda_{\pi\pi}$ and (if there are
non-trivial strong phases) $|\lambda_{\pi\pi}|$. The recent CLEO results
mentioned above suggest that the penguin contribution
to $B\ra\pi\pi$ channel is significant, probably  $10\%$ or more. This
then introduces CP violation in decay, unless the strong phases
cancel (or are zero, as suggested by factorization arguments).
The resulting hadronic uncertainty can be eliminated using isospin analysis
\ref\GrLo{M. Gronau and D. London, Phys. Rev. Lett. 65 (1990) 3381.}.
This requires a measurement of the rates for the
isospin-related channels $B^+ \ra \pi^+ \pi^0$ and $B^0 \ra \pi^0\pi^0$
as well as the corresponding CP-conjugate processes. The rate for
$\pi^0\pi^0$ is expected to be small and the measurement is difficult,
but even an upper bound on this rate can be used to limit the magnitude
of hadronic uncertainties 
\ref\GrQu{Y. Grossman and H. Quinn,
 Phys. Rev. D58 (1998) 017504, hep-ph/9712306.}.

Related but slightly more complicated channels with the same underlying
quark structure are $B\ra\rho^0\pi^0$ and $B\ra a_1^0\pi^0$. Again an
analysis involving the isospin-related channels can be used to help
eliminate hadronic uncertainties from CP violations in the decays
\nref\LNQS{H.J. Lipkin, Y. Nir, H.R. Quinn and A. Snyder,
 Phys. Rev. D44 (1991) 1454.}%
\nref\SnQu{H.R. Quinn and A. Snyder, Phys. Rev. D48 (1993) 2139.}%
\refs{\LNQS,\SnQu}. Channels such as $\rho\rho$ and $a_1\rho$ could 
in principle also be studied, using angular analysis to determine the 
mixture of CP-even and CP-odd contributions.

The analysis of $B\ra D^+D^-$ proceeds along very similar lines. The quark 
subprocess here is $b\ra c\bar cd$, and so the tree contribution gives
\eqn\lamDD{\lambda(B\ra D^+D^-)=
\eta_{D^+D^-}\left({V_{tb}^*V_{td}\over V_{tb}V_{td}^*}\right)
\left({V_{cd}^*V_{cb}\over V_{cd}V_{cb}^*}\right)
\ \Longrightarrow\ \Im\lambda_{DD}=-\sin(2\beta),}
where we used $\eta_{D^+D^-}=+1$. Again, there are hadronic uncertainties 
due to the pure penguin term in \ccdtype, but they are estimated to be small.

A summary of our results for CP violation in the interference of
decays with and without mixing in $B_q\ra f_{\CP}$ is given in table III.
For each mode, we give the asymmetry that would arise if the dominat
contribution were the {\it only} contribution.
\vskip 1cm

\begintable
Quark process & Sample $B_d$ mode & $\Im\lambda_{B_d\ra f_{\CP}}$ & 
Sample $B_s$ mode & $\Im\lambda_{B_s\ra f_{\CP}}$ \crthick
$b\ra c\bar cs$ & $\psi K_S$ & $\sin2\beta$ & $D_s\overline{D_s}$ & 
$\sin2\beta_s$ \nr
$b\ra s\bar ss$ & $\phi K_S$ & $\sin2\beta$ & $\phi\eta^\prime$ & $0$ \nr
$b\ra u\bar ud$ & $\pi\pi$ & $\sin2\alpha$ & $\pi^0K_S$ & $\sin2\gamma$  \nr
$b\ra c\bar cd$ & $D^+D^-$ & $\sin2\beta$ & $\psi K_S$ & $\sin2\beta_s$ \nr
$b\ra u\bar us$ & $\pi^0K_S$ & $\sin2\beta$ & $\phi \pi^0$ & $0$ \nr
$b\ra s\bar sd$ & $\phi\pi$ & $0$ & $\phi K_S$ & $\sin2\beta$  \endtable

\centerline{Table III. $\Im\lambda(B_q\ra f_{\CP})$.}

In all cases the above discussions have neglected the distinction
between strong penguins and electroweak penguins. The CKM phase
structure of both types of penguins is the same. The only place where
this distinction becomes important is when an isospin argument is used
to remove hadronic uncertainties due to penguin contributions. These
arguments are based on the fact that gluons have isospin zero, and
hence strong penguin processes have definite $\Delta I$. Photons and
$Z$-bosons on the other hand contribute to more than one $\Delta I$
transition and hence cannot be separated from tree terms by isospin
analysis. In most cases electroweak penguins are small, typically no more
than ten percent of the corresponding strong penguins and so their
effects can safely be neglected. However in cases (iii) and (iv), where tree
contributions are small or absent, their effects may need to be considered.
(A full review of the role of electroweak penguins in $B$
decays has been given in ref. 
\ref\ewpenreview{R. Fleischer,
 Int. J. Mod. Phys. A12 (1997) 2459, hep-ph/9612446.}.)

\newsec{CP Violation Can Probe New Physics}
We have argued that the Standard Model picture of CP violation is rather 
unique and highly predictive. We have also stated that reasonable
extensions of the Standard Model have a very different picture of
CP violation. Experimental results are too few to decide between
the various possibilities. But in the near future, we expect many
new measurements of CP violating observables. Our discussion of CP violation
in the presence of new physics is aimed to demonstrate that, indeed,
models of new physics can significantly modify the Standard Model predictions
and that the near future measurements will therefore have a strong impact
on the theoretical understanding of CP violation.

To understand how the Standard Model predictions could be modified
by New Physics, we focus on CP violation in the interference
between decays with and without mixing. As explained above, 
this type of CP violation may give, due to its theoretical cleanliness,
unambiguous evidence for New Physics most easily.

Let us consider five specific CP violating observables.

(i) $\Im\lambda_{\psi K_S}$, the CP asymmetry in $B\rightarrow\psi K_S$.
This measurement will cleanly determine the relative phase between
the $B-\bar B$ mixing amplitude and the $b\ra c\bar cs$ decay amplitude
($\sin2\beta$ in the Standard Model).
The $b\ra c\bar cs$ decay has Standard Model tree contributions and
therefore is very unlikely to be significantly affected by new physics.
On the other hand, the mixing amplitude can be easily modified
by new physics. We parametrize such a modification by a phase $\theta_d$:
\eqn\derthed{2\theta_d=\arg(M_{12}/M_{12}^{\rm SM})\ \ \Longrightarrow\ \ 
\Im\lambda_{\psi K_S}=\sin[2(\beta+\theta_d)].}

(ii) $\Im\lambda_{\phi K_S}$, the CP asymmetry in $B\rightarrow\phi K_S$.
This measurement will cleanly determine the relative phase between
the $B-\bar B$ mixing amplitude and the $b\ra s\bar ss$ decay amplitude.
The $b\ra s\bar ss$ decay has only Standard Model penguin contributions
and therefore is sensitive to new physics. We parametrize the modification
of the decay amplitude by a phase $\theta_A$ 
\ref\GrWo{Y. Grossman and M.P. Worah,
 Phys. Lett. B395 (1997) 241, hep-ph/9612269.}:
\eqn\dertheA{\theta_A=\arg(\bar A_{\phi K_S}/\bar A_{\phi K_S}^{\rm SM})
\ \ \Longrightarrow\ \ \Im\lambda_{\phi K_S}=\sin[2(\beta+\theta_d+\theta_A)].}

(iii) $a_{\pi\nu\bar\nu}$, the CP violating ratio of $K\rightarrow
\pi\nu\bar\nu$ decays, defined in \defapnn.
This measurement will cleanly determine the relative phase between the 
$K-\bar K$ mixing amplitude and the $s\ra d\nu\bar\nu$ decay amplitude.
The experimentally measured small value of $\epsK$ requires that
the phase of the $K-\bar K$ mixing amplitude is not modified from
the Standard Model prediction. (More precisely, it requires that
the phase in the mixing amplitude is very close to the phase in
the $s\rightarrow d\bar uu$ decay amplitude.) On the other hand, the decay,
which in the Standard Model is a loop process with small mixing angles,
can be easily modified by new physics.

(iv) $a_{D\ra K\pi}$, the CP violating quantity in $D\rightarrow 
K^\pm\pi^\mp$ decays (see \DCSmin\ and \DCSint):
\eqn\defaD{a_{D\ra K\pi}={\Im(\lambda_{K^-\pi^+})-
\Im(\lambda_{K^+\pi^-}^{-1})\over |\lambda_{K^-\pi^+}|}\ \ \Longrightarrow
\ \ a_{D\ra K\pi}=2\cos\delta_{K\pi}\sin\phi_{K\pi}.} 
It depends on the relative phase between the $D-\bar D$ mixing amplitude
and the $c\ra d\bar su$ and $c\ra s\bar du$ decay amplitudes. 
The two decay channels are tree level and therefore 
unlikely to be affected by new physics \BeNi. On the other hand, 
the mixing amplitude can be easily modified by new physics \BSN.

(v) $d_N$, the electric dipole moment of the neutron.
We did not discuss this quantity so far because, unlike
CP violation in meson decays, flavor changing couplings
are not necessary for $d_N$. In other words, the CP violation
that induces $d_N$ is {\it flavor diagonal}. It does in general
get contributions from flavor changing physics, but it could
be induced by sectors that are flavor blind. Within the
Standard Model (and ignoring the strong CP angle $\theta_{\rm QCD}$),
the contribution from $\delta_{\rm KM}$ arises at the three loop
level and is at least six orders of magnitude below the experimental bound.
We denote the present 90\% C.L upper bound on $d_N$ by $d_N^{\exp}$.
It is given by 
\ref\harris{P.G. Harris {\it et al.}, Phys. Rev. Lett. 82 (1999) 904.}\
\eqn\dnexp{d_N^{\rm exp}=6.3\times10^{-26}\ e\ {\rm cm}.}

The main features of the observables that we chose are summarized
in Table IV.
\vskip 0.5cm

\begintable
Process & Observable & Mixing & Decay & SM & NP \crthick
$B\ra\psi K_S$ & $\Im\lambda_{\psi K_S}$ & $B-\bar B$ & $b\ra c\bar cs$ &
$\sin2\beta$ & $\sin2(\beta+\theta_d)$ \nr
$B\ra\phi K_S$ & $\Im\lambda_{\phi K_S}$ & $B-\bar B$ & $b\ra s\bar ss$ &
$\sin2\beta$ & $\sin2(\beta+\theta_d+\theta_A)$ \nr
$K\ra\pi\nu\bar\nu$ & $a_{\pi\nu\bar\nu}$ & $K-\bar K$ & $s\ra d\nu\bar\nu$ &
$\sim\sin^2\beta$ & $\sin^2\theta_K$ \nr
$D\ra K\pi$ & $a_{D\ra K\pi}$ & $D-\bar D$ & $c\ra d\bar su$ &
$0$ & $\sim\sin\phi_{K\pi}$ \nr
$d_N$ &  &  &  &
$\lsim10^{-6}d_N^{\exp}$ & FD phases  \endtable

\centerline{Table IV. Features of various CP violating observables.}

The various CP violating observables discussed above are sensitive
then to new physics in the mixing amplitudes for the $B-\bar B$
and $D-\bar D$ systems, in the decay amplitudes for $b\rightarrow s\bar ss$
and $s\ra d\nu\bar\nu$ channels and to flavor diagonal CP violation.
If information about all these processes becomes available
and deviations from the Standard Model predictions are found,
we can ask rather detailed questions about the nature of
the new physics that is responsible to these deviations:
\item{(i)} Is the new physics related to the down sector?
the up sector? both?
\item{(ii)} Is the new physics related to $\Delta B=1$ processes?
$\Delta B=2$? both?
\item{(iii)} Is the new physics related to the third generation?
to all generations?
\item{(iv)} Are the new sources of CP violation flavor changing?
flavor diagonal? both?

It is no wonder then that with such rich information,
flavor and CP violation provide an excellent probe of new physics.

\newsec{Supersymmetry}
A generic supersymmetric extension of the Standard Model contains a host
of new flavor and CP violating parameters. (For recent reviews on 
supersymmetry see refs. 
\nref\RamoTASI{P. Ramond, Lectures given at TASI 94, hep-ph/9412234.}%
\nref\BaggTASI{J.A. Bagger, Lectures given at TASI 95, hep-ph/9604232.}%
\nref\SMart{S. Martin, hep-ph/9709356.}%
\nref\LBH{J. Louis, I. Brunner and S.J. Huber, hep-ph/9811341.}%
\nref\ShSh{Y. Shadmi and Y. Shirman, hep-th/9907225.}%
\refs{\RamoTASI-\ShSh}. The following chapter is based on 
\ref\GNR{Y. Grossman, Y. Nir and R. Rattazzi, in {\it Heavy Flavours II}, eds. 
A.J. Buras and M. Lindner (World Scientific), hep-ph/9701231.}.)
It is an amusing exercise to count the number of parameters.
The supersymmetric part of the Lagrangian depends, in addition to the
three gauge couplings of $G_{\rm SM}$, on the parameters of the superpotential
$W$, which can be written as a function of the scalar matter fields:
\eqn\superp{W=\sum_{i,j}\left(Y^u_{ij}h_u\tilde q_{Li}\tilde u_{Rj}
+Y^d_{ij}h_d\tilde q_{Li}\tilde d_{Rj}
+Y^\ell_{ij}h_d\tilde L_{Li}\tilde \ell_{Rj}\right)+\mu h_u h_d.}
In addition, we have to add soft supersymmetry breaking terms:
\eqn\Lsoft{\eqalign{
\L_{\rm soft}\ =\ -&\left(a^u_{ij}h_u\tilde q_{Li}\tilde u_{Rj}
+a^d_{ij}h_d\tilde q_{Li}\tilde d_{Rj}+a^\ell_{ij}h_d\tilde L_{Li}
\tilde\ell_{Rj}+b h_u h_d+{\rm h.c.}\right)\cr
-&\sum_{\rm all\ scalars}m^{S2}_{ij}A_i\bar A_j-{1\over2}\sum_{(a)=1}^3
\left(\tilde m_{(a)}(\lambda\lambda)_{(a)}+{\rm h.c.}\right).\cr}}
The three Yukawa matrices $Y^f$ depend on 27 real and 27 imaginary
parameters. Similarly, the three $a^f$-matrices depend on 27 real 
and 27 imaginary parameters. The five $m^{S2}$ hermitian $3\times3$ 
mass-squared matrices for sfermions ($S=\tilde Q,\tilde d_R,\tilde u_R,
\tilde L,\tilde\ell_R$) have 30 real 
parameters and 15 phases.  The gauge and Higgs sectors depend on 
\eqn\GaHi{\theta_{\rm QCD},\tilde m_{(1)},\tilde m_{(2)},\tilde m_{(3)},
g_1,g_2,g_3,\mu,b,m_{h_u}^2,m_{h_d}^2,}
that is 11 real and 5 imaginary parameters. Summing over all sectors, we get
95 real and 74 imaginary parameters. If we switch off all the above 
parameters but the gauge couplings, we gain global symmetries: 
\eqn\MSBro{G_{\rm global}^{\rm SUSY}(Y^f,\mu,a^f,b,m^2,\tilde m=0)=
U(3)^5\times U(1)_{\rm PQ}\times U(1)_R,}
where the $U(1)_{\rm PQ}\times U(1)_R$ charge assignments are:
\eqn\PQRcharges{
\matrix{&h_u&h_d&Q\bar u&Q\bar d&L\bar\ell\cr
U(1)_{\rm PQ}&1&1&-1&-1&-1\cr
U(1)_{\rm R}&1&1&1&1&1\cr}.}
Consequently, we can remove at most 15 real and 32 imaginary parameters.
But even when all the couplings are switched on, there is
a global symmtery, that is
\eqn\MSCon{G_{\rm global}^{\rm SUSY}=U(1)_B\times U(1)_L,}
so that 2 of the 32 imaginary parameters cannot be removed.
We are left then with
\eqn\MSpar{124\ =\ \cases{80&real\cr 44&imaginary}\ 
{\rm physical\ parameters}.} 
In particular, there are 43 new CP violating phases! In addition to the single
Kobayashi-Maskawa of the SM, we can put 3 phases in $M_1,M_2,\mu$
(we used the $U(1)_{\rm PQ}$ and $U(1)_{\rm R}$ to remove the phases
from $\mu B^*$ and $M_3$, respectively) and the other 40 phases
appear in the mixing matrices of the fermion-sfermion-gaugino couplings.
(Of the 80 real parameters, there are 11 absolute values of the parameters
in \GaHi, 9 fermion masses, 21 sfermion masses, 3 CKM angles
and 36 SCKM angles.) Supersymmetry provides a nice example to our
statement that reasonable extensions of the Standard Model may have
more than one source of CP violation.

The requirement of consistency with experimental data provides strong 
constraints on many of these parameters. For this reason, the physics of flavor
and CP violation has had a profound impact on supersymmetric model building. 
A discussion of CP violation in this context can hardly avoid addressing the 
flavor problem itself.  Indeed, many of the supersymmetric models that we
analyze below were originally aimed at solving flavor problems.
 
As concerns CP violation, one can distinguish two classes of experimental
constraints. First, bounds on nuclear and atomic electric dipole moments
determine what is usually called the {\it supersymmetric CP problem}.
Second, the physics of neutral mesons and, most importantly, the small
experimental value of $\epsK$ pose the {\it supersymmetric $\epsK$
problem}. In the next two subsections we describe the two problems.
Then we describe various supersymmetric flavor models and
the ways in which they address the supersymmetric CP problem.
 
Before turning to a detailed discussion, we define two scales that
play an important role in supersymmetry: $\Lambda_S$, where
the soft supersymmetry breaking terms are generated, and $\Lambda_F$,
where flavor dynamics takes place. When $\Lambda_F\gg\Lambda_S$, it is
possible that there are no genuinely new sources of flavor and CP
violation. This leads to models with exact universality, which we
discuss in section 8.3. When $\Lambda_F\lsim\Lambda_S$, we do not
expect, in general, that flavor and CP violation are limited to the
Yukawa matrices. One way to suppress CP violation 
would be to assume that, similarly to the Standard Model,
CP violating phases are large, but their effects are screened, possibly
by the same physics that explains the various flavor puzzles.
Such models, with Abelian or non-Abelian horizontal symmetries,
are described in section 8.4. It is also possible that CP violating effects 
are suppressed because squarks are heavy. This scenario is also discussed in 
section 8.4. Another option is to assume that CP is an approximate symmetry 
of the full theory (namely, CP violating phases are all small). We discuss 
this scenario in section 8.5. A brief discussion of the implications of
$\epe$ is included in this subsection. Some concluding comments regarding CP 
violation as a probe of supersymmetric flavor models are given in section 8.6. 

\subsec{The Supersymmetric CP Problem}
One aspect of supersymmetric CP violation involves effects that are
flavor preserving. Then, for simplicity, we describe this aspect in
a supersymmetric model without additional flavor mixings, {\it i.e.} the
minimal supersymmetric standard model (MSSM) with universal sfermion
masses and with the trilinear SUSY-breaking scalar couplings
proportional to the corresponding Yukawa couplings. (The generalization
to the case of non-universal soft terms is straightforward.)  In such a
constrained framework, there are four new phases beyond the two
phases of the Standard Model ($\delta_{\rm KM}$ and $\theta_{\rm QCD}$).
One arises in the bilinear $\mu$-term of the superpotential \superp,
while the other three arise in the soft supersymmetry breaking
parameters of \Lsoft: $\tilde m$ (the gaugino mass), $a$ (the trilinear
scalar coupling) and $b$ (the bilinear scalar coupling).
Only two combinations of the four phases are physical 
\nref\DGH{
M. Dugan, B. Grinstein and L.J. Hall, Nucl. Phys. B255 (1985) 413.}%
\nref\DiTh{S. Dimopoulos ans S. Thomas,
 Nucl. Phys. B465 (1996) 23, hep-ph/9510220.}%
\refs{\DGH,\DiTh}. To see this, note that one could treat the various 
dimensionful parameters in \superp\ and \Lsoft\ as spurions which break the
$U(1)_{\rm PQ}\times U(1)_{\rm R}$ symmetry, thus deriving selection rules:
\eqn\spur{\matrix{&\tilde m&A&b&\mu\cr
U(1)_{\rm PQ}&0&0&-2&-2\cr U(1)_{\rm R}&-2&-2&-2&0&\cr}}
(where we defined $A$ through $a^f=AY^f$).
Physical observables can only depend on combinations of the dimensionful
parameters that are neutral under both $U(1)$'s. There are three such
independent combinations: $\tilde m\mu b^*$, $A\mu b^*$
and $A^* \tilde m$. However, only two of their phases are independent, say
\eqn\phiAB{\phi_A=\arg(A^* \tilde m),\ \ \ \phi_B=\arg(\tilde m\mu b^*).}
In the more general case of non-universal soft terms there is
one independent phase $\phi_{A_{i}}$ for each quark and lepton flavor.
Moreover, complex off-diagonal entries in the sfermion
mass-squared matrices may represent additional sources of CP violation.
 
The most significant effect of $\phi_A$ and $\phi_B$ is their
contribution to electric dipole moments (EDMs). 
The electric dipole moment of a fermion $\psi$ is defined
as the coefficient $d_\psi$ of the operator
\eqn\defdpsi{\L_{d_\psi}=-{i\over2}d_\psi\bar\psi\sigma_{\mu\nu}
\gamma_5\psi F^{\mu\nu}.}
For example, the contribution from one-loop gluino diagrams to the down 
quark EDM is given by 
\nref\BuWy{W. Buchmuller and D. Wyler, Phys. Lett. B121 (1983) 321.}%
\nref\PoWi{J. Polchinski and M. Wise, Phys. Lett. B125 (1983) 393.}%
\refs{\BuWy,\PoWi}:
\eqn\ddsusy{d_d=M_d{e\alpha_3\over 18\pi\tilde m^3}\left(
|A|\sin\phi_A+\tan\beta|\mu|\sin\phi_B\right),}
where we have taken $m^2_Q\sim m^2_D\sim m^2_{\tilde g}\sim\tilde m^2$,
for left- and right-handed squark and gluino masses. We define, as usual,
$\tan\beta = \vev{H_u}/\vev{H_d}$. Similar one-loop diagrams give rise to
chromoelectric dipole moments. The electric and chromoelectric dipole moments 
of the light quarks $(u,d,s)$ are the main source of $d_N$ (the EDM of the 
neutron), giving 
\ref\FPT{W. Fischler, S. Paban and S. Thomas, Phys. Lett. B289 (1992) 373.}\
\eqn\dipole{d_N\sim 2\, \left({100\, GeV\over \tilde m}\right )^2
\sin \phi_{A,B}\times10^{-23}\ e\, {\rm cm},}
where, as above, $\tilde m$ represents the overall SUSY scale. In a generic 
supersymmetric framework, we expect $\tilde m=\O(m_Z)$ and $\sin\phi_{A,B}=
\O(1)$. Then the constraint \dnexp\ is generically violated by about two 
orders of magnitude. This is {\it the Supersymmetric CP Problem} 
\nref\EFN{J. Ellis, S. Ferrara and D. Nanopoulos,
 Phys. Lett. B114 (1982) 231.}%
\nref\Barr{S. Barr, Int. J. Mod. Phys. A8 (1993) 209.}%
\refs{\BuWy-\Barr}.

\nref\MoRa{R.N. Mohapatra and A. Rasin,
 Phys. Rev. D54 (1996) 5835, hep-ph/9604445.}%
Eq. \dipole\ shows what are the possible ways to solve
the supersymmetric CP problem:
\item{(i)} Heavy squarks: $\tilde m\gsim1\ TeV$;
\item{(ii)} Approximate CP (or left-right symmetry \MoRa): 
$\sin\phi_{A,B}\ll1$.

Recently, a third way has been investigated, that is cancellations
between various contributions to the electric dipole moments
\nref\FaOl{T. Falk and K.A. Olive, Phys. Lett. B439 (1998) 71, hep-ph/9806236;
Phys. Lett. B375 (1996) 196, hep-ph/9602299.}%
\nref\IbNa{T. Ibrahim and P. Nath, Phys. Lett. B418 (1998) 98, hep-ph/9707409;
 Phys. Rev. D57 (1998) 478, hep-ph/9708456;
 Phys. Rev. D58 (1998) 111301, hep-ph/9807501.}%
\nref\BGPSS{A. Bartl {\it et al.},
 Phys. Rev. D60 (1999) 073003, hep-ph/9903402.}%
\nref\BaKh{S.M. Barr and S. Khalil, hep-ph/9903425.}%
\nref\BGK{M. Brhlik, G.J. Good and G.L. Kane,
 Phys. Rev. D59 (1999) 115004, hep-ph/9810457.}%
\nref\BEKL{M. Brhlik, L. Everett, G.L. Kane and J. Lykken,
 Phys. Rev. Lett. 83 (1999) 2124, hep-ph/9905215.}%
\nref\PRS{S. Pokorski, J. Rosiek and C.A. Savoy, hep-ph/9906206.}%
\refs{\FaOl-\PRS}. However, there seems to be no symmetry that can guarantee
such a cancellation. This is in contrast to the other two mechanisms
mentioned above that were shown to arise naturally in specific models.
We therefore do not discuss any further this third mechanism.

Finally, we mention that the electric dipole moment of the electron is
also a sensitive probe of flavor diagonal CP phases. The present
experimental bound,
\nref\Comm{E. Commins {\it et al.}, Phys. Rev. A50 (1994) 2960.}%
\nref\Abdu{K. Abdullah {\it et al.}, Phys. Rev. Lett. 65 (1990) 234.}%
\refs{\Comm-\Abdu},
\eqn\deexp{|d_e|\leq4\times10^{-27}\ e\ {\rm cm},}
is also violated by about two orders of magnitue for `natural'
values of supersymmetric parameters.

\subsec{The Supersymmetric $\epsK$ Problem}
The contribution to the CP violating $\epsK$ parameter in the neutral $K$
system is dominated by diagrams involving $Q$ and $\bar d$ squarks in the
same loop 
\nref\DNW{J. Donoghue, H. Nilles and D. Wyler, Phys. Lett. B128 (1983) 55.}%
\nref\GaMa{F. Gabbiani and A. Masiero, Nucl. Phys. B322 (1989) 235.}%
\nref\HKT{J.S. Hagelin, S. Kelley and T. Tanaka,
 Nucl. Phys. B415 (1994) 293, hep-ph/9304218.}%
\nref\GMS{E. Gabrielli, A. Masiero and L. Silvestrini,
 Phys. Lett. B374 (1996) 80, hep-ph/9509379.}%
\nref\GGMS{F. Gabbiani, E. Gabrielli, A. Masiero and L. Silvestrini,
 Nucl. Phys. B477 (1996) 321, hep-ph/9604387.}%
\refs{\DNW-\GGMS}. The corresponding effective four-fermi 
operator involves fermions of both chiralities, so that its matrix elements
are enhanced by $\O(m_K/m_s)^2$ compared to the chirality conserving
operators. For $m_{\tilde g}\simeq m_Q \simeq m_D= \tilde m$ (our results
depend only weakly on this assumption) and focusing on the contribution
from the first two squark families, one gets \GGMS:
\eqn\epsKSusy{
\epsK={5\ \alpha_3^2  \over 162\sqrt2}{f_K^2m_K\over\tilde
m^2\Delta m_K}\left [\left({m_K\over m_s+m_d}\right)^2+{3\over 25}\right]
\Im\left\{{(\delta m_Q^2)_{12}\over m_Q^2}
{(\delta m_D^2)_{12}\over m_D^2}\right\},}
where $(\delta m_{Q,D}^2)_{12}$ are the off diagonal entries in the
squark mass matrices in a basis where the down quark mass matrix
and the gluino couplings are diagonal. These flavor
violating quantities are often written as
$(\delta m_{Q,D}^2)_{12}=V_{11}^{Q,D}\delta m_{Q,D}^2 V_{21}^{Q,D*}$,
where $\delta m_{Q,D}^2$ is the mass splitting among the squarks
and $V^{Q,D}$ are the gluino coupling mixing matrices in the mass
eigenbasis of quarks and squarks. Note that CP would be violated
even if there were two families only 
\ref\NirSusy{Y. Nir, Nucl. Phys. B273 (1986) 567.}.  
Using the experimental value of $\epsK$, we get
\eqn\epsKScon{
{(\Delta m_K\epsK)^{\rm SUSY}\over(\Delta m_K\epsK)^{\rm EXP}}\sim10^7
\left ({300 \ GeV\over\tilde m}\right)^2
\left({m^2_{Q_2}-m^2_{Q_1}\over m_Q^2}\right)
\left({m^2_{D_2}-m^2_{D_1}\over m_D^2}\right)
|K_{12}^{dL}K_{12}^{dR}|\sin\phi,}
where $\phi$ is the CP violating phase. In a generic supersymmetric framework, 
we expect $\tilde m=\O(m_Z)$, $\delta m_{Q,D}^2/m_{Q,D}^2=\O(1)$, 
$K_{ij}^{Q,D}=\O(1)$ and $\sin\phi=\O(1)$. Then the constraint 
\epsKScon\ is generically violated by about seven orders of magnitude. 

Eq. \epsKScon\ also shows what are the possible ways to solve
the supersymmetric $\epsK$ problem:
\item{(i)} Heavy squarks: $\tilde m\gg300\ GeV$;
\item{(ii)} Universality: $\delta m_{Q,D}^2\ll m_{Q,D}^2$;
\item{(iii)} Alignment: $|K_{12}^d|\ll1$;
\item{(iv)} Approximate CP: $\sin\phi\ll1$.

\subsec{Exact Universality}
Both supersymmetric CP problems are solved if, at the scale $\Lambda_S$,
the soft supersymmetry breaking terms are universal and the genuine SUSY
CP phases $\phi_{A,B}$ vanish. Then the Yukawa matrices represent
the only source of flavor and CP violation which is relevant in low
energy physics. This situation can naturally arise when supersymmetry
breaking is mediated by gauge interactions at a scale $\Lambda_S\ll\Lambda_F$ 
\ref\DNNS{M. Dine, A. Nelson, Y. Nir and Y. Shirman,
 Phys. Rev. D53 (1996) 2658, hep-ph/9507378.}. 
In the simplest scenarios, the $A$-terms and the
gaugino masses are generated by the same SUSY and $U(1)_R$ breaking
source (see eq. \spur). Thus, up to very small effects due to the
{\it standard}  Yukawa matrices, $\arg(A)=\arg(m_{\tilde g})$ so that
$\phi_A$ vanishes. In specific models also $\phi_B$ vanishes in a similar way 
\nref\DNeS{M. Dine, A. Nelson and Y. Shirman,
 Phys. Rev. D51 (1994) 1362, hep-ph/9408384.}%
\nref\DNiS{M. Dine, Y. Nir and Y. Shirman,
 Phys. Rev. D55 (1997) 1501, hep-ph/9607397.}%
\refs{\DNeS,\DNiS}. It is also possible that similar boundary
conditions occur when supersymmetry breaking is communicated to the
observable sector up at the Planck scale. 
The situation in this case seems to be less under control from the theoretical 
point of view. Dilaton dominance in SUSY breaking, though, seems a very 
interesting direction to explore 
\nref\KaLo{V. Kaplunovsky and J. Louis,
 Phys. Lett. B306 (1993) 269, hep-th/9303040.}%
\nref\BML{R. Barbieri, J. Louis and M. Moretti, Phys. Lett. B312 (1993) 451, 
hep-ph/9305262; (E) {\it ibid.} B316 (1993) 632.}%
\refs{\KaLo,\BML}.
 
The most important implication of this type of boundary conditions
for soft terms, which we refer to as {\it exact  universality}, 
is the existence of the SUSY analogue of the GIM mechanism 
which operates in the SM. The CP violating phase of the CKM matrix can
feed into the soft terms via Renormalization Group (RG) evolution only
with a strong suppression from light quark masses \DGH.
 
With regard to the supersymmetric CP problem, gluino diagrams
contribute to quark EDMs as in eq. \ddsusy,
but with a highly suppressed effective phase, {\it e.g.}
\eqn\gim{\phi_{A_d}\sim (t_S/16 \pi^2)^4 Y_t^4 Y_c^2 Y_b^2 J.}
Here $t_S=\log (\Lambda_S/M_W)$ arises from the RG evolution from
$\Lambda_S$ to the electroweak scale, the $Y_i$'s are quark Yukawa
couplings (in the mass basis), and $J\simeq 2\times 10^{-5}$ is defined in eq. 
\defJ. A similar contribution comes from chargino diagrams. The 
resulting EDM is $d_N\lsim 10^{-31}\ e\ {\rm cm}$. This maximum can be reached 
only for very large $\tan\beta\sim60$ while, for small $\tan\beta \sim 1$,
$d_N$ is about 5 orders of magnitude smaller. This range of values for
$d_N$ is much below the present ($\sim10^{-25}\ e$ cm) and foreseen
($\sim 10^{-28}\ e$ cm) experimental sensitivities 
\nref\RoStru{A. Romanino and A. Strumia,
 Nucl. Phys. B490 (1997) 3, hep-ph/9610485.}%
\nref\BeVi{S. Bertolini and F. Vissani,
 Phys. Lett. B324 (1994) 164, hep-ph/9311293.}%
\nref\IMSS{T. Inui, Y. Mimura, N. Sakai and T. Sasaki,
 Nucl. Phys. B449 (1995) 49, hep-ph/9503383.}%
\nref\ACW{S.A. Abel, W.N. Cottingham and I.B. Wittingham,
 Phys. Lett. B370 (1996) 106, hep-ph/9511326.}%
\refs{\RoStru-\ACW}.
 
With regard to the supersymmetric $\epsK$ problem, the contribution
to $\epsK$ is proportional to $\Im(V_{td}V_{ts}^*)^2 Y_t^4
(t_S/16 \pi^2)^2$, giving the same GIM suppression as in the SM.
This contribution turns out to be negligibly small 
\nref\BKeK{S. Baek and P. Ko, Phys. Lett. B462 (1999) 95, hep-ph/9904283.}%
\refs{\DGH,\BKeK}. The supersymmetric contribution to
$D-\bar D$ mixing is similarly small and we expect no observable effects.
For the $B_d$ and $B_s$ systems, the largest SUSY contribution
to the mixing comes from box diagrams with intermediate charged Higgs and
the up quarks. It can be up to $\O(0.2)$ of the SM amplitude for
$\Lambda_S=M_{\rm Pl}$ and $\tan\beta = \O(1)$ 
\nref\BBMR{S. Bertolini, F. Borzumati, A. Masiero and G. Ridolfi,
Nucl. Phys. B353 (1991) 591.}%
\nref\GNO{T. Goto, T. Nihei and Y. Okada,
 Phys. Rev. D53 (1996) 5233, hep-ph/9510286.}%
\nref\Nihe{T. Nihei, Prog. Theor. Phys. 98 (1997) 1157, hep-ph/9707336.}%
\nref\BaKo{S. Baek and P. Ko, Phys. Rev. Lett. 83 (1999) 488, hep-ph/9812229.}%
\refs{\BBMR-\BaKo}, and much smaller for large $\tan\beta$. The contribution 
is smaller in models of gauge mediated SUSY breaking where the mass of the 
charged Higgs boson is typically $\gsim 300\ GeV$ \DNNS\ and $t_S\sim5$. The
SUSY contributions to $B_s-\bar B_s$ and $B_d -\bar B_d$ mixing are, to a
good approximation, proportional to $(V_{tb}V_{ts}^*)^2$ and
$(V_{tb}V_{td}^*)^2$, respectively, like in the SM. Then, regardless of
the size of these contributions, the relation $\Delta m_{B_d}/\Delta m_{B_s}
\sim |V_{td}/V_{ts}|^2$ and the CP asymmetries in neutral
$B$ decays into final CP eigenstates are the same as in the SM.

\subsec{Approximate Horizontal Symmetries}
In the class of supersymmetric models with $\Lambda_F\lsim\Lambda_S$, 
the soft masses are generically not universal, and we do not 
expect flavor and CP violation to be limited to the Yukawa matrices.  
Most models where soft terms arise at the Planck scale 
($\Lambda_S\sim M_{\rm Pl}$) belong to this class. It is possible
that, similarly to the Standard Model, CP violating phases are large, 
but their effects are screened, possibly
by the same physics that explains the various flavor puzzles. This
usually requires Abelian or non-Abelian horizontal symmetries.
Two ingredients play a major role here: selection rules that come from
the symmetry and holomorphy of Yukawa and $A$-terms that comes from the
supersymmetry. With Abelian symmetries, the screening mechanism is
provided by {\it alignment} 
\nref\NiSe{Y. Nir and N. Seiberg,
 Phys. Lett. B309 (1993) 337, hep-ph/9304307.}%
\nref\LNSb{M. Leurer, Y. Nir and N. Seiberg,
 Nucl. Phys. B420 (1994) 468, hep-ph/9410320.}%
\nref\GrNiL{Y. Grossman and Y. Nir,
 Nucl. Phys. B448 (1994) 30, hep-ph/9502418.}%
\nref\RaNi{Y. Nir and R. Rattazzi,
 Phys. Lett. B382 (1996) 363, hep-ph/9603233.}%
\refs{\NiSe-\RaNi}, whereby the mixing matrices for gaugino couplings 
have very small mixing angles, particularly for the first two down squark 
generations. With non-Abelian symmetries, the screening mechanism is 
{\it approximate universality}, where squarks of the two families fit into 
an irreducible doublet and are, therefore, approximately degenerate 
\nref\DKL{M. Dine, A. Kagan and R.G. Leigh,
 Phys. Rev. D48 (1993) 4269, hep-ph/9304299.}%
\nref\PoSe{P. Pouliot and N. Seiberg,
 Phys. Lett. B318 (1993) 169, hep-ph/9308363.}%
\nref\HaMu{L.J. Hall and H. Murayama,
 Phys. Rev. Lett. 75 (1995) 3985, hep-ph/9508296.}%
\nref\PoTo{A. Pomarol and D. Tommasini,
 Nucl. Phys. B466 (1996) 3, hep-ph/9507462.}%
\nref\BDH{R. Barbieri, G. Dvali and L.J. Hall,
 Phys. Lett. B377 (1996) 76, hep-ph/9512388.}%
\nref\CHM{C. Carone, L.J. Hall and H. Murayama,
 Phys. Rev. D54 (1996) 2328, hep-ph/9602364.}%
\nref\Zurab{Z.G. Berezhiani, hep-ph/9609342.}%
\nref\Gali{G. Eyal, Phys. Lett. B441 (1998) 191, hep-ph/9807308.}%
\refs{\DKL-\Gali}. In all of these
models, it is difficult to avoid $d_N\gsim10^{-28}$ e cm.
 
As far as the third generation is concerned, the signatures of
Abelian and non-Abelian models are similar. In particular, they allow
observable deviations from the SM predictions for CP asymmetries in
$B$ decays. The recent measurement of $a_{\psi K_S}$ gives first
constraints on these contributions \BEN.
In some cases, non-Abelian models give relations between
CKM parameters and consequently predict strong constraints
on these CP asymmetries.

For the two light generations, only alignment allows interesting effects. 
In particular, it predicts large CP violating effects in $D-\bar D$ mixing 
\refs{\NiSe,\LNSb}. Thus, it allows $a_{D\ra K\pi}=\O(1)$.
 
Finally, it is possible that CP violating effects are suppressed because
squarks are heavy. If the masses of the first and second generations
squarks $m_i$ are larger than the other soft masses, $m_i^2\sim 100\,
\tilde m^2$ then the Supersymmetric CP problem is solved and the $\epsK$
problem is relaxed (but not eliminated) 
\nref\DKS{M. Dine, A. Kagan and S. Samuel,
 Phys. Lett. B243 (1990) 250.}%
\refs{\PoTo,\DKS}. This does not necessarily lead to naturalness problems, 
since these two generations are almost decoupled from the Higgs sector.

Notice though that, with the possible exception of $m_{\tilde b_R}^2$,
third family squark masses cannot naturally be much above $m_Z^2$.
If the relevant phases are of $O(1)$, the main contribution to $d_N$
comes from the third family via the two-loop induced three-gluon operator
\ref\Weintg{S. Weinberg, Phys. Rev. Lett. 63 (1989) 2333.}, 
and it is roughly at the present experimental
bound  when $m_{\tilde t_{L,R}}\sim 100\ GeV$.
 
Models with the first two squark generations heavy have their own
signatures of CP violation in neutral meson mixing 
\ref\CKLN{A.G. Cohen, D.B. Kaplan, F. Lepeintre and A.E. Nelson,
 Phys. Rev. Lett. 78 (1997) 2300, hep-ph/9610252.}.
The mixing angles relevant to $D-\bar D$ mixing are similar, in general,
to those of models of alignment (if alignment is
invoked to explain $\Delta m_K$ with $m^2_{Q,D}\lsim20\ TeV$).
However, since the $\tilde u$ and $\tilde c$ squarks are heavy, the
contribution to $D-\bar D$ mixing is one to two orders
of magnitude below the experimental bound. This may lead to the
interesting situation that $D-\bar D$ mixing will first be observed
through its CP violating part \WolD.
In the neutral $B$ system, $\O(1)$ shifts from the Standard Model
predictions of CP asymmetries in the decays to final CP eigenstates
are possible. This can occur even when squark masses of the third
family are $\sim1\ TeV$ 
\ref\CKNS{A.G. Cohen, D.B. Kaplan and A.E. Nelson,
 Phys. Lett. B388 (1996) 588, hep-ph/9607394.}, 
since now mixing angles can naturally be larger than in the case 
of horizontal symmetries (alignment or approximate universality).

\subsec{Approximate CP Symmetry}
Both supersymmetric CP problems are solved if CP is an approximate
symmetry, broken by a small parameter of order $10^{-3}$. This is another
possible solution to CP problems in the class of supersymmetric
models with $\Lambda_F\lsim\Lambda_S$. (Of course,
some mechanism has also to suppress the real part of the $\Delta S=2$
amplitude by a sufficient amount.) 
 
If CP is an approximate symmetry, we expect also the SM
phase $\delta_{\rm KM}$ to be $\ll 1$. Then the standard box diagrams
cannot account for $\epsK$ which should arise from another
source. In supersymmetry with non-universal soft terms, the source could
be diagrams involving virtual superpartners, mainly squark-gluino box
diagrams. Let us call $(M_{12}^K)^{\rm SUSY}$
the supersymmetric contribution to the $K-\bar K$ mixing amplitude.
Then the requirements $\Re (M_{12}^K)^{\rm SUSY}\lsim\Delta m_K$
and $\Im(M_{12}^K)^{\rm SUSY}\sim\epsK\Delta m_K$ imply that the
generic CP phases are $\geq\O(\epsK)\sim 10^{-3}$.

Of course, $d_N$ constrains the relevant CP violating phases to be
$\lsim10^{-2}$. If all phases are of the same order, then $d_N$ must be
just below or barely compatible with the present experimental bound.
A signal should definitely be found if the accuracy is increased by two
orders of magnitude.
 
The main phenomenological implication of these scenarios is that
CP asymmetries in $B$ meson decays are small, perhaps $\O(\epsK)$, rather 
than ${\cal O}(1)$ as expected in the SM. Also the ratio $a_{\pi\nu\bar\nu}$ 
of eq. \defapnn\ is very small, in contrast to the Standard Model
where it is expected to be of ${\cal O}(\sin^2\beta)$. Explicit models of 
approximate CP were presented in refs. 
\nref\abefre{S.A. Abel and J.M. Frere,
 Phys. Rev. D55 (1997) 1623, hep-ph/9608251.}%
\nref\BaBa{K.S. Babu and S.M. Barr,
 Phys. Rev. Lett. 72 (1994) 2831, hep-ph/9309249.}%
\nref\Eyal{G. Eyal and Y. Nir, Nucl. Phys. B528 (1998) 21, hep-ph/9801411.}%
\nref\BDM{K.S. Babu, B. Dutta and R.N. Mohapatra, hep-ph/9905464.}%
\refs{\abefre-\BDM}.   

The experimntal value of $\epe$ has particularly interesting
implications on models of approximate CP
\ref\EMNS{G. Eyal, A. Masiero, Y. Nir and L. Silvestrini, hep-ph/9908382.}.
In this framework, the standard model cannot account for $\epe$.
A model of approximate CP would then be excluded if it does not provide
sufficiently large contributions from new physics to this parameter.
A generic supersymmetric model where the $a^q$ terms in \Lsoft\ are not 
proportional to the $Y^q$ terms in \superp\ can provide a large contribution 
\ref\MaMu{A. Masiero and H. Murayama,
  Phys. Rev. Lett. 83 (1999) 907, hep-ph/9903363.}\
related to imaginary part of
\eqn\susyepe{a^d_{12}\sim m_s|V_{us}|/\tilde m.}
In models of non-Abelian flavor symmetries, the contribution is typically
not large enough because of cancellation between the $a^d_{12}$ and
$a^d_{21}$ terms
\nref\HMPV{X.G. He, H. Murayama, S. Pakvasa and G. Valencia, hep-ph/9909562.}%
\nref\BCS{R. Barbieri, R. Contino and A. Strumia, hep-ph/9908255.}%
\refs{\HMPV,\BCS}. In models of heavy $\tilde d$ and $\tilde s$ squarks,
the contribution is highly suppressed by the heavy mass scale \EMNS.
Models of alignment can give a contribution that is not much smaller
than the estimate in \susyepe. If, however, the related CP violating
phase is small, then the model can account for $\epe$ only if both the
model parameters and the hadronic parameters assume rather extreme
values \EMNS. We conclude that most existing models of supersymmetry with
approximate CP are excluded (or, at least, strongly disfavored) by
the experimental measurement of $\epe$. (For other recent works on
$\epe$ in the supersymmetric framework, see
\nref\CoIs{G. Colangelo and G. Isidori, JHEP 9809 (1998) 009, hep-ph/9808487.}%
\nref\KKM{S. Khalil, T. Kobayashi and A. Masiero, hep-ph/9903544.}%
\nref\BDM{K.S. Babu, B. Dutta and R.N. Mohapatra, hep-ph/9905464.}%
\nref\BJKP{S. Baek, J.H. Jang, P. Ko and J.H. Park, hep-ph/9907572.}%
\refs{\GMS-\GGMS,\BuSi,\CoIs-\BJKP}.

The fact that the Standard Model and the models of approximate
CP are both viable at present shows that the mechanism of CP violation has 
not really been tested experimentally. The only measured CP violating 
observales, that is $\epsK$ and $\epspK$, are small.
Their smallness could be related to the `accidental' smallness of CP violation
for the first two quark generations, as is the case in the Standard
Model, or to CP being an approximate symmetry, as is the case in the
models discussed here. Future measurements, particularly of processes
where the third generation plays a dominant role (such as $a_{\psi K_S}$
or $a_{\pi\nu\bar\nu}$), will easily distinguish between the two scenarios.
While the Standard Model predicts large CP violating effects
for these processes, approximate CP would suppress them too.
 
\subsec{Some Concluding Remarks}
We can get an intuitive understanding of how the various supersymmetric
flavor models discussed in this chapter solve the supersymmetric flavor
and CP problems by presenting the general form of the squark 
mass-squared matrices for each framework. This is summarized in
Table V. The implications of each flavor model for the various CP violating
observables presented in the previous chapter are given in Table VI.
\vskip 0.3cm

\begintable
Flavor Model & Theory & ${1\over\tilde m^2}\tilde M^2_{LL}\sim$  & 
Physical Parameters \crthick
Exact Universality & GMSB & diag$(a,a,a)$ & 
$\Delta\tilde m^2_{12}\sim m_c^2/m_W^2$ \nr
Approx. Universality & Non-Abelian $H$ & diag$(a,a,b)$ & 
$\Delta\tilde m^2_{12}\sim \sin^2\theta_C$ \nr
Alignment & Abelian $H$ & diag$(a,b,c)$ & 
$(K_L^d)_{12}\ll\sin\theta_C$ \nr
Heavy Squarks & Comp.; Anom. $U(1)$ &
 diag$(A,B,c)$ & $\tilde m^2_{1,2}\sim100\tilde m^2$ \nr
Approximate CP & SCPV &  & $10^{-3}\lsim\phi_{\CP}\ll1$
\endtable

\centerline{Table V. Supersymmetric flavor models.}
\vskip 0.3cm

\begintable
Model & ${d_N\over10^{-25}\ e\ {\rm cm}}$ & $\theta_d$ & $\theta_A$ &
$a_{D\ra K\pi}$ & $a_{K\ra\pi\nu\bar\nu}$ \crthick
Standard Model & $\lsim10^{-6}$ & $0$ & $0$ & $0$ & $\O(1)$ \nr
Exact Universality & $\lsim10^{-6}$ & $0$ & $0$ & $0$ & =SM \nr
Approx. Universality & $\gsim10^{-2}$ & $\O(0.2)$ & $\O(1)$ &
$0$ & $\approx$SM \nr
Alignment & $\gsim10^{-3}$ & $\O(0.2)$ & $\O(1)$ &
$\O(1)$ & $\approx$SM \nr
Heavy Squarks & $\sim10^{-1}$ & $\O(1)$ & $\O(1)$ &
$\O(10^{-2})$ & $\approx$SM \nr 
Approximate CP & $\sim10^{-1}$ & $-\beta$ & $0$ & $\O(10^{-3})$
& $\O(10^{-5})$ \endtable

\centerline{Table VI. Phenomenological implicatons of supersymmetric 
flavor models.}

We would like to emphasize the following points:

(i) For supersymmetry to be established, a direct observation
of supersymmetric particles is necessary. Once it is discovered, then
measurements of CP violating observables will be a very sensitive
probe of its flavor structure and, consequently, of the mechanism
of dynamical supersymmetry breaking.

(ii) It is easy to distinguish between models of exact
universality and models with genuine supersymmetric flavor and
CP violation. This can be done through searches of $d_N$  and
of CP asymmetries in $B$ decays.

(iii) The neutral $D$ system provides a stringent test of alignment.

(iv) The fact that $K\ra\pi\nu\bar\nu$ decays are not affected
by most supersymmetric flavor models 
\nref\NiWo{Y. Nir and M.P. Worah, 
 Phys. Lett. B423 (1998) 319, hep-ph/9711215.}%
\nref\BRS{A.J. Buras, A. Romanino and L. Silvestrini,
 Nucl. Phys. B520 (1998) 3, hep-ph/9712398.}%
\refs{\NiWo,\BRS}\ is actually an advantage. The Standard Model correlation 
between $a_{\pi\nu\bar\nu}$ and $a_{\psi K_S}$ is a much cleaner test than 
a comparison of $a_{\psi K_S}$ to the CKM constraints.

(v) Approximate CP has dramatic effects on all observables.
My guess is that in lectures given a year from now, it will not
appear in the Table as a viable option.

\newsec{Left Right Symmetric Models of Spontaneous CP Violation}
\subsec{The Model}
We consider models with a symmetry $G_{\rm LRS}\times D_{\rm LRS}$,
where $G_{\rm LRS}$ is the gauge group,
\eqn\GLRsym{
G_{\rm LRS}=SU(3)_C\times SU(2)_L\times SU(2)_R\times U(1)_{B-L},}
and $D_{\rm LRS}$ is a discrete group,
\eqn\DLRsym{D_{\rm LRS}=P\times C.}
Various versions of left-right symmetric models differ in
$D_{\rm LRS}$. We are interested here in models where CP
is only spontaneously broken, hence our choice of \DLRsym\
\nref\Chan{D. Chang, Nucl. Phys. B214 (1983) 435.}%
\nref\BFG{G.C. Branco, J.-M. Fr\`ere and J.-M. G\'erard,
 Nucl. Phys. B221 (1983) 317.}%
\nref\HaLe{H. Harari and M. Leurer, Nucl. Phys. B233 (1984) 221.}%
\nref\EGN{G. Ecker, W. Grimus and H. Neufeld,
 Nucl. Phys. B247 (1983) 70.}%
\nref\EGN{G. Ecker and W. Grimus, Nucl. Phys. B258 (1985) 328.}%
\nref\Leur{M. Leurer, Nucl. Phys. B266 (1986) 147.}%
\refs{\Chan-\Leur}.

The fermion representations consist of three generations of
\eqn\LRfrep{Q_{Li}(3,2,1)_{1/3},\ \ \ Q_{Ri}(3,1,2)_{1/3},\ \ \ 
L_{Li}(1,2,1)_{-1},\ \ \ L_{Ri}(1,1,2)_{-1}.}
Under $D_{\rm LRS}$, the fermion fields transform as follows:
\eqn\Dfer{\matrix{
P:\ &Q_L\leftrightarrow Q_R & L_L\leftrightarrow L_R \cr
C:\ &Q_L\leftrightarrow i\sigma_2(Q_R)^*&L_L\leftrightarrow 
i\sigma_2(L_R)^*\cr}}
The scalar representations consist of three multiplets
\ref\MoPa{R.N. Mohapatra and J.C. Pati, Phys. Rev. D11 (1975) 566.},
\eqn\LRsrep{\Delta_R(1,1,3)_2,\ \ \ \Delta_L(1,3,1)_2,\ \ \ 
\Phi(1,2,2)_0.}
Under $D_{\rm LRS}$, the scalar fields transform as follows:
\eqn\Dfer{\matrix{
P:\ &\Delta_L\leftrightarrow \Delta_R & \Phi\leftrightarrow \Phi^\dagger \cr
C:\ &\Delta_L\leftrightarrow (\Delta_R)^*& \Phi\leftrightarrow \Phi^T\cr}}
It is often convenient to write $\Phi$ in a $2\times2$ matrix form:
\eqn\Phiot{\Phi=\pmatrix{\phi_1^0&\phi_1^+\cr \phi_2^-&\phi_2^0\cr}.}

The spontaneous symmetry breaking occurs in two stages,
\eqn\LRSbreak{G_{\rm LRS}\times D_{\rm LRS}\ra G_{\rm SM}\ra
SU(3)_C\times U(1)_{\rm EM},}
due to the VEVs of the neutral members of the scalar fields:
\eqn\LRVEVs{\vev{\Delta_R}=\pmatrix{0\cr0\cr v_Re^{i\beta}\cr},\ \ \ 
\vev{\Delta_L}=\pmatrix{0\cr0\cr v_L\cr},\ \ \
\vev{\Phi}=\pmatrix{k_1&0\cr0&k_2e^{i\alpha}\cr}.}
(In general, all four VEVs are complex. There is, however, freedom of
rotations by $U(1)_{B-L}$ for $\Delta_L$ and $\Delta_R$ and by
$U(1)_{T_{3L}}\times U(1)_{T_{3R}}$ for $\phi_1$ and $\phi_2$, so
that only two phases are physical.) These VEVs are assumed to
satisfy the hierarchy
\eqn\VEVhie{v_R \gg k_1,k_2 \gg v_L.}
The first stage of symmetry breaking in \LRSbreak\ takes place at
the scale $v_R$ and the second at $k=\sqrt{k_1^2+k_2^2}$.

\subsec{Flavor Parameters}
The quark Yukawa couplings have the following form:
\eqn\Lyuk{{\cal L}_{\rm Yuk}=f\overline{Q_L}\Phi Q_R+
h\overline{Q_L}\widetilde\Phi Q_R+{\rm h.c.},}
where $\widetilde\Phi=\tau_2\Phi^*\tau_2$. As a consequence of $D_{\rm LRS}$,
the Yukawa matrices $f$ and $h$ are {\it symmetric and real}: $P$ requires 
that they are hermitian, $C$ requires that they are symmetric, and CP requires 
that they are real. The resulting mass matrices,
\eqn\Qmass{\eqalign{M_u\ =&\ fk_1+hk_2e^{-i\alpha},\cr
M_d\ =&\ hk_1+fk_2e^{i\alpha},\cr}}
are complex symmetric matrices.

How many independent physical flavor parameters (and, in particular, phases)
does this model have? We have two symmetric and complex mass matrices,
that is twelve real and twelve imaginary Yukawa parameters.
If we set $h=f=0$, we gain a  global $U(3)$ symmetry ($D_{\rm LRS}$ 
does not allow independent $U(3)_L$ and $U(3)_R$ rotations).
This means that we can remove three real and six imaginary parameters.
When $h$ and $f$ are different from zero, there is no global symmetry
($U(1)_{B-L}$ is part of the gauge symmetry). We conclude
that there are nine real and six imaginary flavor parameters.
Six of the real parameters are the six quark masses. To identify
the other flavor parameters, note that the symmetric mass matrices
can be diagonalized by a unitary transformation of the form
\eqn\diamQ{V_u M_u V_u^T=M_u^{\rm diag},\ \ \ V_d M_d V_d^T=M_d^{\rm diag}.}
Consequently, the mixing matrices $V_L$ and $V_R$ describing,
respectively, the $W_L$ and $W_R$ interactions, 
\eqn\LCC{{\cal L}_{CC}={g\over\sqrt2}\left(
W^+_{L\mu}\overline{u_L}V_L\gamma^\mu d_L+
W^+_{R\mu}\overline{u_R}V_R\gamma^\mu d_R\right)+{\rm h.c.},}
are related:
\eqn\relVLR{V_L=P^u V_R^* P^d,}
where $P^u$ and $P^d$ are diagonal phase matrices. The three
real parameters are then the three mixing angles,
which are equal in $V_L$ and $V_R$. The six phases can be
arranged in various ways. A convenient choice is to have
a single phase in $V_L$, which is then just $\delta_{\rm KM}$
of $V_{\rm CKM}$ of the Standard Model, and five phases in $V_R$.
(It is also possible to have $V_L=V_R^*$ with six phases in each.)    

\subsec{What is the Low Energy Effective Theory of the LRS Model?}
It is interesting to ask what is the low enrgy effective theory
below the scale $v_R$. It is straightforward to show that the
light fields are precisely those of the Standard Model: the fermions
are chiral under $SU(2)_L$ except for the right-handed neutrinos in $L_{Ri}$
which acquire Majorana masses at the scale $v_R$ due to their
coupling to $\Delta_R$. (The left-handed neutrinos acquire very light masses
from both the see-saw mechanism and their direct coupling to $\Delta_L$.)
Only the one Higgs doublet related to $G_{\rm SM}$ breaking, that is 
$k_1\phi_1+k_2e^{-i\alpha}\phi_2$, remains light. The question is then 
whether the left-right symmetry constrains Standard Model parameters.

To answer this question, we first argue that phenomenological constraints 
forbid $r\equiv k_2/k_1={\cal O}(1)$. (More precisely, it is $r\sin\alpha$ 
which is constrained to be very small.)
Consider eqs. \Qmass. They lead to the following equations:
\eqn\rMuMd{\eqalign{
M_u re^{i\alpha}-M_d\ =&\ k_1h(r^2-1),\cr
M_u-M_d re^{-i\alpha}\ =&\ k_1f(1-r^2).\cr}}
The right hand side of these equations is real. Then, the
imaginary part of the left-hand side should vanish. Let us
put all quark masses to zero, except for $m_t$ and $m_b$.
We take then $(M_u)_{33}=m_te^{i\theta_t}$ and 
$(M_d)_{33}=m_be^{i\theta_b}$. We get:
\eqn\imrMuMd{\eqalign{r m_t\sin(\theta_t+\alpha)-m_b\sin\theta_b=&0,\cr
m_t\sin\theta_t-m_b\sin(\theta_b-\alpha)=&0.\cr}}
The second equation implies that $\theta_t\lsim m_b/m_t$. Then,
the first equation gives
\ref\BBRK{G. Barenboim, J. Bernabeu and M. Raidal,
 Nucl. Phys. B478 (1996) 527, hep-ph/9608450.}
\eqn\rsia{r\sin\alpha\leq m_b/m_t.}
(Note that the model is symmetric under $r\ra1/r$ and $\alpha\ra-\alpha$.
Therefore, $r\sin\alpha\geq m_t/m_b$ is acceptable and physically 
equivalent.)

The only source of CP violation in the quark mass matrices is the phase
$\alpha$. (The phase $\beta$ in $\vev{\Delta_R}$ does not affect quark
masses, though it may affect neutrino masses.) Moreover, if one of the
$\vev{\phi_i^0}$ vanished, then again there would be no CP violation
in the quark mass matrices. As a consequence of these two facts,
all CP violating phases in the mixing matrices $V_L$ and $V_R$ are
proportional to $r\sin\alpha$. Hence the importance of \rsia. 
In particular, for the Kobayashi-Maskawa phase, one finds \BBRK\
\eqn\dKMLR{\delta_{\rm KM}\sim r\sin\alpha(m_c/m_s)\leq{\cal O}(0.1).}
We learn that the low energy effective theory of the left-right symmetric
model is the Standard Model with a small value for $\delta_{\rm KM}$.

Phenomenologically, it is difficult, though not impossible, to account
for $\epsK$ with just the Standard Model contribution and a small KM phase.
There are then two possibilities:
\item{(i)} The left-right symmetry is broken at a very high scale.
The low energy theory is to a good approximation just the Standard Model.
CP is, however, an approximate symmetry in the kaon sector. 
The hadronic parameters playing
a role in the calculation of $\epsK$ have to assume rather extreme values.
\item{(ii)} The left-right symmetry is broken at low enough scale
so that there are significant new contributions to various rare processes.
In particular, box diagrams with intermediate $W_R$-boson and tree diagrams
with a heavy neutral Higgs dominate $\epsK$. This sets up an upper bound
on the scale $v_R$, of order tens of $TeV$.
  
\subsec{Phenomenology of CP Violation}
The smallness of $r\sin\alpha$ does not necessarily mean that CP is an 
approximate symmetry in the quark sector; the phases in the mixing matrices 
depend, in addition to $r\sin\alpha$, on quark mass ratios, some of which are
large. An explicit calculation shows that the six phases actually
divide to two groups: the KM phase and the three phases that appear
in $V_R$ in a two generation model (usually denoted by $\delta_1$,
$\delta_2$ and $\gamma$) are all small \BBRK, while the two extra phases that
appear in the three generation $V_R$ (denoted by $\sigma_1$, $\sigma_2$)
are not
\ref\BBRB{G. Barenboim, J. Bernabeu and M. Raidal,
 Nucl. Phys. B511 (1998) 577, hep-ph/9702337.}:
\eqn\LRoom{\eqalign{
\delta,\delta_1,\delta_2,\gamma\ \propto&\ r\sin\alpha(m_c/m_s)
\leq{\cal O}(0.1),\cr
\sigma_1,\sigma_2\ \propto&\ r\sin\alpha(m_t/m_b)
\leq{\cal O}(1).\cr}}

In $\epsK$, it is mainly $\delta_1$ and $\delta_2$ which play a role.
(We here assume that the hadronic parameters are close to their present
theoretical estimate and therefore $\epsK$ cannot be explained in
this framework by the Standard Model contribution alone.)
Assuming that the $W_L-W_R$ box diagram gives the dominant contribution, 
one is led to conclude that \BBRK\
\eqn\upperWR{M(W_R)\lsim20\ TeV}
is favored. Note that CP conserving processes provide a lower bound
\nref\BBSlr{G. Beall, M. Bander and A. Soni, Phys. Rev. Lett. 48 (1982) 848.}%
\nref\BBPR{G. Barenboim, J. Bernabeu, J. Prades and M. Raidal,
 Phys. Rev. D55 (1997) 4213, hep-ph/9611347.}%
\refs{\BBSlr,\BBPR},
\eqn\upperWR{M(W_R)\gsim1.6\ TeV.}
The favored range for $M(W_R)$ is then very constrained in this framework.

Taking into account this upper bound and the fact that the $\sigma_i$
phases are enhanced by a factor of about 10 compared to the $\delta_i$
phases, one finds that the left-right symmetric contributions compete with
or even dominate over the Standard Model contributions to $B-\bar B$
mixing and to $B_s-\bar B_s$ mixing
\nref\EcGr{G. Ecker and W. Grimus, Z. Phys. C30 (1986) 293.}%
\nref\LoWy{D. London and D. Wyler, Phys. Lett. B232 (1989) 503.}%
\nref\BBMR{G. Barenboim, J. Bernabeu, J. Matias and M. Raidal,
 hep-ph/9901265.}%
\refs{\BBRB,\EcGr-\BBMR}.  This means that CP asymmetries in
$B$ or $B_s$ decays into final CP eigenstates could be substantially
different from the Standard Model prediction. Moreover, the phases
in the left-right symmetric contributions to $B-\bar B$ and $B_s-\bar B_s$ 
mixing are closely related, predicting correlations between the deviations. 
The CP asymmetry in semileptonic $B$
decays could also be significantly enhanced
\ref\BBRB{G. Barenboim, Phys. Lett. B443 (1998) 317, hep-ph/9810325.}.
The recent measurement of $a_{\psi K_S}$ gives first constraints on
$\sigma_1$ leading to new bounds on $a_{\rm SL}$ \BEN. 

Finally, LRS models could enhance the electric dipole moments of
the neutron and of the electron
\nref\BeSo{G. Beall and A. Soni, Phys. Rev. Lett. 47 (1981) 552.}%
\nref\HMP{X.G. He, H.J. McKellar and S. Pakvasa,
 Phys. Rev. Lett. 61 (1988) 1267.}%
\nref\CLY{D. Chang, C.S. Li and T.C. Yuan, Phys. Rev. D42 (1990) 867.}%
\nref\BBde{G. Barenboim and J. Bernabeu,
 Z. Phys. C73 (1997) 321, hep-ph/9603379.}%
\refs{\BeSo-\BBde}.

\newsec{Multi-Scalar Models}
The Standard Model has a single scalar field, $\phi(1,2)_{1/2}$,
that is responsible for the spontaneous symmetry breaking, $SU(2)_L
\times U(1)_Y\ra U(1)_{\rm EM}$. Within the framework of the
Standard Model, the complex Yukawa couplings of the scalar doublet 
to fermions are the only source of flavor physics and of CP violation.
However, in the mass basis, the interactions of the Higgs particle
are flavor diagonal and CP conserving.

There are several good reasons for the interest in multi-scalar models 
in the context of flavor and CP violation:
\item{a.} If there exist additional scalars and, in particular, 
$SU(2)_L$-doublets, then not only there are new sources of CP violation, 
but also the Yukawa interactions in the mass basis as well as the scalar
self-interactions may violate CP. 
\item{b.} CP violation in scalar interactions has very different features
from the $W$-mediated CP violation of the Standard Model. For example, it
could lead to observable {\it flavor diagonal} CP violation in top physics
or in electric dipole moments, or it could induce transverse lepton 
polarization in semileptonic meson decays.
\item{c.} With more than a single scalar doublet, CP violation could be 
spontaneous.

Indeed, there is no good reason to assume that the Standard Model doublet is 
the only scalar in Nature. Most extensions of the Standard Model predict
that there exist additional scalars. For example, models with an extended
gauge symmetry (such as GUTs and left-right symmetric models) need extra
scalars to break the symmetry down to $G_{\rm SM}$; Supersymmetry requires
that there exists a scalar partner to each Standard Model fermion.
However, scalar masses are generically not protected by a symmetry. 
Consequently, in models where the low energy effective theory is the
Standard Model, we expect in general that the only light scalar is
the Standard Model doublet.

The study of multi-scalar models is then best motivated in
the following cases:
\item{(i)} The scale of new physics is not very high above
the electroweak scale. One has to remember, however, that in such
cases there is more to the new physics than just extending the
scalar sector.
\item{(ii)} The scalar is related to the spontaneous breaking of
a global symmetry. In some cases, a discrete symmetry is enough
to make a scalar light.

We will discuss scalar $SU(2)_L$-doublets and -singlets only.
There are two reasons for that. First, the VEVs of higher multiplets 
need to be very small in order to avoid large deviations from the 
experimentally successful relation $\rho=1$. Second, higher multiplets  
do not couple to the known fermions. (The only exception is an 
$SU(2)_L$-triplet that can couple to the left-handed leptons.)

\subsec{Multi Higgs Doublet Models}
The most popular extension of the Higgs sector is the multi Higgs 
doublet model (MHDM) and, in particular, the two Higgs doublet model 
(2HDM). These models have, in addition to the Kobayashi-Maskawa phase
of the quark mixing matrix, several new sources of CP violation
\ref\WoWu{Y.L. Wu and L. Wolfenstein,
 Phys. Rev. Lett. 73 (1994) 1762, hep-ph/9409421.}:
\item{(i)} A complex mixing matrix for charged scalars
\ref\WeCH{S. Weinberg, Phys. Rev. Lett. 37 (1976) 657.}.
\item{(ii)} Mixing of CP-even and CP-odd neutral scalars 
\ref\TDLee{T.D. Lee, Phys. Rev. D8 (1973) 1226.}.
\item{(iii)} CP-odd Yukawa couplings (in the quark mass basis).
\item{(iv)} Complex quartic scalar couplings.

The CP violation that is relevant to near future experiments
always involves fermions. Therefore, we will only discuss the 
new sources (i), (ii) and (iii).

A generic MHDM, with all dimensionful parameters at the electroweak
scale and all dimensionless parameters of order one, leads to severe
phenomenological problems. In particular, some of the physical scalars
have flavor changing (and CP violating) couplings at tree level,
violating bounds on rare processes such as $\Delta m_K$
and $\epsK$ by several orders of magnitude. There are three possible
solutions to these problems:

(I) {\it Natural flavor conservation (NFC)}
\ref\GlWe{S.L. Glashow and S. Weinberg, Phys. Rev. D15 (1977) 1958.}: 
only a single scalar doublet
couples to each fermion sector. 2HDM where the same (a different) scalar 
doublet couples to the up and the down quarks are called type I (II).
The absence of flavor changing and/or CP violating Yukawa interactions
in this case is based on the same mechanism as within the Standard Model.

(II) {\it Approximate flavor symmetries (AFS)}
\ref\ChSh{T.P. Cheng and M. Sher, Phys. Rev. D35 (1987) 3484.}: 
it is quite likely that the
smallness and hierarchy in the fermion masses and mixing angles are
related to an approximate flavor symmetry, broken by a small parameter.
If so, then it is unavoidable that the Yukawa couplings of all scalar
doublets are affected by the selection rules related to the flavor
symmetry. In such a case, couplings to the light generations and,
in particular, off-diagonal couplings, are suppressed.

(III) {\it Heavy scalars}
\ref\GeNa{H. Georgi and D.V. Nanopoulos, Phys. Lett. 82B (1979) 95.}: 
all dimensionful parameters that are
not constrained by the requirement that the spontaneous breaking
of $G_{\rm SM}$ occurs at the electroweak scale are actually
higher than this scale, $\Lambda_{\rm NP}\gg\Lambda_{\rm EW}$. Then 
all the new sources of flavor and CP violation in the scalar sector
are suppressed by ${\cal O}(\Lambda_{\rm EW}^2/\Lambda_{\rm NP}^2)$.
 
In table VII we summarize the implications of the various multi-scalar 
models for CP violation. Note that, if we impose NFC, 
spontaneous CP violation (SCPV) \TDLee\
is impossible in 2HDM \WeCH\ and (since the combination of SCPV and NFC
leads to $\delta_{\rm KM}=0$
\ref\Bran{G.C. Branco, Phys. Rev. Lett. 44 (1980) 504.})
is phenomenologically excluded in MHDM
\nref\KrPo{P. Krawczyk and S. Pokorski, Nucl. Phys. B364 (1991) 11.}%
\nref\GrNis{Y. Grossman and Y. Nir,
 Phys. Lett. B313 (1993) 126, hep-ph/9306292.}%
\refs{\KrPo,\GrNis}. Explicit models of spontaneous CP violation have 
been constructed within the frameworks of approximate NFC
\ref\LiWo{J. Liu and L. Wolfenstein, Nucl. Phys. B289 (1987) 1.},
approximate flavor symmetries \refs{\RaNi,\Eyal}\ and heavy scalars \BBde.
In the supersymmetric framework, one has to add at least two scalar
singlets to allow for spontaneous CP violation
\ref\MaRa{M. Masip and A. Rasin,
 Phys. Rev. D58 (1998) 035007, hep-ph/9803271.}.
Entries marked with `$*$' mean that the number of scalar doublets
should be larger than 2 (that is, the answer is `No' in 2HDM).
\vskip 1cm

\begintable
Framework & Model (Example) & SCPV & (i) & (ii) & (iii) \crthick
NFC & MSSM & Excluded & Yes$^*$ & Yes$^*$ & No \nr
AFS & Horizontal Sym. & Yes & Yes$^*$  & Yes & $\O(m_q/m_Z)$ \nr
Heavy & LRS & Yes & $\O({\Lambda_{\rm EW}^2\over\Lambda_{\rm NP}^2})^*$ & 
$\O({\Lambda_{\rm EW}^2\over\Lambda_{\rm NP}^2})$  
& $\O({\Lambda_{\rm EW}^2\over\Lambda_{\rm NP}^2})$  
\endtable

\centerline{Table VII. Multi Higgs Doublet Models}

\subsec{(i) Charged Scalar Exchange}
We investigate a multi Higgs doublet model (with $n\ge3$ doublets) with
NFC and assume that a different doublet couples to each of the the down, 
up and lepton sectors:
\eqn\LCSi{-\L_Y=
-{\phi_1^+\over v_1}\overline{U}V_{\rm CKM}M_d^{\rm diag}P_R \, D
+{\phi_2^+\over v_2}\overline{U}M_u^{\rm diag}V_{\rm CKM} P_L \,D
-{\phi_3^+\over v_3}\overline{\nu}M_\ell P_R \, \ell + {\rm h.c.},}
where $P_{L,R}=(1\mp\gamma_5)/2$.
We denote the physical charged scalars by $H_i^+$ ($i=1,2,\ldots,n-1$),
and the would-be Goldstone boson (eaten by the $W^+$) by $H_n^+$. We
define $K$ to be the matrix that rotates the charged scalars from the
interaction- to the mass-eigenbasis. Then the Yukawa Lagrangian in the
mass basis (for both fermions and scalars) is
\eqn\LCSm{\L_Y={G_F^{1/2}\over2^{1/4}}\sum_{i=1}^{n-1}
\{H_i^+\overline{U}[Y_i M_u^{\rm diag}V_{\rm CKM} P_L 
+X_i V_{\rm CKM}M_d^{\rm diag}P_R ]D  +
H_i^+\overline{\nu} [Z_i M_\ell P_R ]\ell  \}+{\rm h.c.},}
where
\eqn\XYdef{X_i=-{K_{i1}^*\over K_{n1}^*},\ \ \
Y_i=-{K_{i2}^*\over K_{n2}^*},\ \ \
Z_i=-{K_{i3}^*\over K_{n3}^*}.}
CP violation in the charged scalar sector comes from phases in $K$.
CP violating effects are largest when
the lightest charged scalar is much lighter than the heavier ones
\nref\WeiL{S. Weinberg, Phys. Rev. D42 (1990) 860.}%
\nref\Lav{L. Lavoura, Int. J. Mod. Phys. A8 (1993) 375.}%
\refs{\WeiL,\Lav}. Here we assume that all but the lightest charged
scalar ($H_1^+$) effectively decouple from the fermions.
Then, CP violating observables depend on three parameters:
\eqn\threeCPV{\eqalign{
{\Im(XY^*)\over m_H^2} \equiv\ &\ {\Im(X_1Y_1^*) \over m_{H_1}^2}\
\approx\ \sum_{i=1}^{n-1} {\Im(X_iY_i^*) \over m_{H_i}^2},\cr
{\Im(XZ^*)\over m_H^2} \equiv\ &\ {\Im(X_1Z_1^*) \over m_{H_1}^2}\
\approx\ \sum_{i=1}^{n-1} {\Im(X_iZ_i^*) \over m_{H_i}^2},\cr
{\Im(YZ^*)\over m_H^2} \equiv\ &\ {\Im(Y_1Z_1^*) \over m_{H_1}^2}\
\approx\ \sum_{i=1}^{n-1} {\Im(Y_iZ_i^*) \over m_{H_i}^2}.\cr }}
$\Im(XY^*)$ induces CP violation in the quarks sector, while $\Im(XZ^*)$
and $\Im(YZ^*)$ give CP violation that is observable in
semi-leptonic processes.
 
There is an interesting question of whether charged
scalar exchange could be the {\it only} source of CP violation. In other
words, we would like to know whether a model of extended scalar sector
with spontaneous CP violation and NFC is viable. In these models, 
$\delta_{\rm KM}=0$ and $\epsK$ has to be accounted for by charged
Higgs exchange. This requires very large long distance
contributions. The CP violating coupling should fulfill
\nref\BiSa{I.I. Bigi and A.I. Sanda, Phys. Rev. Lett. 58 (1987) 1605.}%
\nref\Che{H.Y. Cheng, Phys. Rev. D42 (1990) 2329.}%
\refs{\BiSa-\Che}
\eqn\epsllim{\Im(XY^*)\geq\O(40).} 
However, the upper bounds on $d_N$ \KrPo\ and on BR$(b\ra s\gamma)$
\GrNis\ require
\eqn\epsulim{\Im(XY^*)\leq\O(1).} 
We conclude that models of SCPV and NFC are excluded. It is, of course, 
still a viable possibility that CP is explicitly broken, in which case 
both quark and Higgs mixings provide CP violation.

The bound \epsulim\ implies that the charged Higgs contribution
to $B-\bar B$ mixing is numerically small and would modify the Standard
Model predictions for CP asymmetries in $B$ decays by no more
than $\O(0.02)$ \GrNis. On the other hand, the contribution
to $d_N$ can still be close to the experimental upper bound.

The lepton transverse polarization cannot be generated by vector or
axial-vector interactions only 
\nref\Miriam{M. Leurer, Phys. Rev. Lett. 62 (1989) 1967.}%
\nref\CFK{P. Castoldi, J.M. Fr\`ere and G. Kane,
 Phys. Rev. D39 (1989) 2633.}%
\refs{\Miriam,\CFK}, so it is particularly
suited for searching for CP violating scalar contributions.
As triple-vector correlation is odd under time-reversal, the experimental
observation of such correlation would signal T and -- assuming CPT
symmetry -- CP violation. (It is possible to get non-vanishing
T$-$odd observables even without CP violation (see e.g.
\ref\FSIphase{M.B. Gavela {\it et al.}, Phys. Rev. D39 (1989) 1870.}).
Such ``fake" asymmetries can arise 
from final state interactions (FSI). They can be removed by comparing the 
measurements in two CP conjugate channels.) The muon transverse polarization 
in $K\to\pi\,\mu\,\nu$ decays and the tau transverse polarization in 
$B\ra X\tau\nu$ are examples of such observables.
The lepton transverse polarization, $P_\perp$, in semileptonic decays
is defined as the lepton polarization component along the normal vector
of the decay plane,
\eqn\defpper{P_\perp = {\vec s_{\ell} \cdot (\vec p_{\ell} \times
\vec p_X) \over |\vec p_{\ell} \times \vec p_X|}\,,}
where $\vec s_{\ell}$ is the lepton spin three-vector and $\vec p_{\ell}$
($\vec p_X$) is the three-momentum of the lepton (hadron).
Experimentally, it is useful to define the integrated CP violating asymmetry
\eqn\defacp{a_{CP} \equiv \langle P_\perp \rangle =
{\Gamma^+ - \Gamma^- \over \Gamma^+ + \Gamma^-},}
where $\Gamma^+$ ($\Gamma^-$) is the rate of finding the lepton spin
parallel (anti-parallel) to the normal vector of the decay plane.
A non-zero $a_{CP}$ can arise in our model from the interference between the 
$W$-mediated and the $H^+$-mediated tree diagrams. For example, in
the semi-taonic bottom quark decay, the asymmetry is given by
$a_{CP} = C_{ps} {\Im(XZ^*) \over m_H^2}$ and could be as large as 0.3
\nref\Eilam{D. Atwood, G. Eilam and A. Soni,
 Phys. Rev. Lett. 71 (1993) 492, hep-ph/9303268.}%
\nref\GrLi{Y. Grossman and Z. Ligeti,
 Phys. Lett. B332 (1994) 373, hep-ph/9403376;
 Phys. Lett. B347 (1995) 399, hep-ph/9409418.}%
\nref\GrHaNi{Y. Grossman, H.E. Haber and Y. Nir,
 Phys. Lett. B357 (1995) 630, hep-ph/9507213.}%
(see {\it e.g.} \refs{\Eilam-\GrHaNi}).

\subsec{(ii) Effects of CP-even and CP-odd Scalar Mixing in Top Physics}
 It is possible that the neutral scalars are mixtures
of CP-even and CP-odd scalar fields
\nref\AST{C.H. Albright, J. Smith, and S.H.H. Tye,
 Phys. Rev D21 (1980) 711.}%
\nref\BrRe{G.C. Branco and M.N. Rebelo, Phys. Lett. B160 (1985) 117.}%
\nref\MePo{A. M\'endez and A. Pomarol, Phys. Lett. B272 (1991) 313.}%
\nref\PoVe{A. Pomarol and R. Vega, Nucl. Phys. B413 (1993) 3, hep-ph/9305272.}%
\refs{\TDLee,\AST-\PoVe,\WeiL-\Lav}. Such a scalar couples to
both scalar and pseudoscalar currents:
\eqn\Lyuxmix{\L_Y=H_i\bar f(a_i^f+ib_i^f\gamma_5)f,}
where $H_i$ is the physical Higgs boson and $a_i^f,b_i^f$ are functions
of mixing angles in the matrix that diagonalizes the neutral scalar
mass matrix. (Specifically, they are proportional to the components
of, respectively, $\Re\phi_u$ and $\Im\phi_u$ in $H_i$.)
CP violation in processes involving fermions is
proportional to $a_i^fb_i^{f*}$. The natural place to look for
manifestations of this type of CP violation is top physics,
where the large Yukawa couplings allow large asymmetries (see e.g.
\ref\Pes{C.E. Schmidt and M.E. Peskin, Phys. Rev. Lett. 69 (1992) 410.}).
Note that unlike our discussion above, the asymmetries
here have nothing to do with FCNC processes. Actually, in models with
NFC (even if softly broken \BrRe), the effects discussed here contribute
negligibly to $\epsK$ and to CP asymmetries in $B$ decays. On the
other hand, two loop diagrams with intermediate neutral scalar and top
quark can induce a CP violating three gluon operator
\nref\Weintg{S. Weinberg, Phys. Rev. Lett. 63 (1989) 2333.}%
\nref\Dicu{D. Dicus, Phys. Rev. D41 (1990) 999.}%
\refs{\Weintg,\Dicu} that would give $d_N$ close to the experimental bound
\nref\GuWy{J.F. Gunion and D. Wyler, Phys. Lett. B248 (1990) 170.}%
\nref\BLY{E. Braaten, C.S. Li and T.C. Yuan,
 Phys. Rev. Lett. 64 (1990) 1709.}%
\nref\RGPV{A. De R\'ujula, M.B. Gavela, O. P\`ene and F.J. Vegas,
 Phys. Lett. B245 (1990) 640.}%
\refs{\Dicu-\RGPV}.

\subsec{(iii) Flavor Changing Neutral Scalar Exchange}
Natural flavor conservation needs not be exact in models of extended
scalar sector
\nref\LiWoB{J. Liu and L. Wolfenstein, Phys. Lett. B197 (1987) 536.}%
\nref\HaNiS{H. Haber and Y. Nir, Nucl. Phys. B335 (1990) 363.}%
\refs{\LiWo,\BrRe,\LiWoB,\HaNiS}. In particular, it is quite likely that 
the existence of the additional scalars is related to flavor symmetries 
that explain the smallness and hierarchy in the Yukawa couplings. 
In this case, the new flavor changing couplings of these scalars are 
suppressed by the same selection rules as those that are responsible 
to the smallness
of fermion masses and mixing, and there is no need to impose NFC
\nref\JoRi{A.S. Joshipura and S.D. Rindani,
 Phys. Lett. B260 (1991) 149.}%
\nref\AHR{A. Antaramian, L.J. Hall and A. Rasin,
 Phys. Rev. Lett. 69 (1992) 1871, hep-ph/9206205.}%
\nref\HaWeS{L.J. Hall and S. Weinberg,
 Phys. Rev. D48 (1993) 979, hep-ph/9303241.}%
\nref\BGL{G.C. Branco, W. Grimus and L. Lavoura,
 Phys. Lett. B380 (1996) 119, hep-ph/9601383.}%
\nref\ARS{D. Atwood, L. Reina and A. Soni,
 Phys. Rev. D54 (1996) 3269, hep-ph/9603210;
 Phys. Rev. D55 (1997) 3156, hep-ph/9609279.}%
\refs{\ChSh,\JoRi-\ARS,\WoWu}. An explicit framework, with Abelian horizontal
symmetries, was presented in
\nref\LNS{M. Leurer, Y. Nir and N. Seiberg,
 Nucl. Phys. B398 (1993) 319, hep-ph/9212278.}%
\refs{\LNS,\LNSb,\GNR}. (For another related study, see
\ref\ACHM{N. Arkani-Hamed, C.D. Carone, L.J. Hall and H. Murayama,
 Phys. Rev. D54 (1996) 7032, hep-ph/9607298.}.)
We explain the general idea using these models. We emphasize
that in this example the scalar with flavor changing couplings is
a Standard Model singlet, and not an extra doublet, but the idea
that these couplings are suppressed by approximate horizontal symmetries
works in the same way.

The simplest model of ref. \LNS\ extends the SM by supersymmetry and by
an Abelian horizontal symmetry $\H=U(1)$ (or $Z_N$). The symmetry $\H$ is
broken by a VEV of a single scalar $S$ that is a singlet of the SM gauge 
group. Consequently, Yukawa couplings that violate $\H$ arise only from 
nonrenormalizable terms and are therefore suppressed. Explicitly,
the quark Yukawa terms have the form
\eqn\Yukeff{\L_Y=
X^d_{ij}\left({S\over M}\right)^{n^d_{ij}}Q_i\bar d_j\phi_d+
X^u_{ij}\left({S\over M}\right)^{n^u_{ij}}Q_i\bar u_j\phi_u,}
where $M$ is some high energy scale and $n^q_{ij}$ is the horizontal charge
of the combination $Q_i\bar q_j\phi_q$ (in units of the charge of $S$).
The terms \Yukeff\ lead to quark masses and mixing as well as to
flavor changing couplings, $Z^q_{ij}$, for the scalar $S$. The magnitude 
of the latter is then  related to that of the
effective Yukawa couplings $Y^q_{ij}$:
\eqn\relZY{Z^q_{ij}\sim {M^q_{ij}\over\vev{S}}.}
Since the order of magnitude of each entry in the quark mass matrices
is fixed in these models in terms of quark masses and mixing, the
$Z^q_{ij}$ couplings can be estimated in terms of these physical
parameters and the scale $\vev{S}$. For example, these couplings contribute
to $K-\bar K$ mixing proportionally to
\eqn\FCNS{Z^d_{12}Z^{d*}_{21}\sim{m_d m_s\over\vev{S}^2}.}
 
With arbitrary phase factors in the various $Z^q_{ij}$ couplings,
the contributions to neutral meson mixing are, in general, CP violating.
In particular, there will be a contribution to $\epsK$ from
$\Im(Z^d_{12}Z^{d*}_{21})$. Requiring that the $S$-mediated tree level
contribution does not exceed the experimental value of $\epsK$ gives,
for $\O(1)$ phases,
\eqn\bouFCNS{M_S\vev{S}\gsim1.8\ TeV^2.}
We learn that (for $M_S\sim\vev{S}$) the mass of the $S$-scalar
could be as low as 1.5\ TeV, some 4 orders of magnitude below the
bound corresponding to $\O(1)$ flavor changing couplings.
 
The flavor changing couplings of the $S$-scalar lead also to
a tree level contribution to $B-\bar B$ mixing proportional to
\eqn\FCNSB{Z^d_{13}Z^{d*}_{31}\sim{m_d m_b\over\vev{S}^2}.}
This means that, for phases of order 1, the neutral scalar exchange
accounts for at most a few percent of $B-\bar B$ mixing. This cannot
be signaled in $\Delta m_B$ (because of the hadronic uncertainties
in the calculation) but could be signalled (if $\vev{S}$ is at the
lower bound) in CP asymmetries in $B^0$ decays.
 
Finally, the contribution to $D-\bar D$ mixing, proportional to
\eqn\FCNSC{Z^u_{12}Z^{u*}_{21}\sim{m_u m_c\over\vev{S}^2},}
is below a percent of the current experimental bound. This is unlikely
to be discovered in near-future experiments, even if the new phases
maximize the interference effects in the $D^0\ra K^-\pi^+$ decay.
 
To summarize, models with horizontal symmetries naturally suppress
flavor changing couplings of extra scalars. There is no need to
invoke NFC even for new scalars at the TeV scale. Furthermore,
the magnitude of the flavor changing couplings is related to the
observed fermion parameters. Typically, contributions from
neutral scalars with flavor changing couplings could dominate
$\epsK$. If they do, then a signal at the few percent level
in CP asymmetries in neutral $B$ decays is quite likely

\subsec{The Superweak Scenario} 
The original {\it superweak} scenario
\ref\SuperWeak{L. Wolfenstein, Phys. Rev. Lett. 13 (1964) 562.}\
stated that CP violation appears in a new
$\Delta S=2$ interaction while there is no CP violation in the SM
$\Delta S=1$ transitions. Consequently, the only large observable
CP violating effect is $\epsK$, while $\epe\sim10^{-8}$ and EDMs are
negligibly small. At present, the term ``superweak CP violation"
has been used for many different types of models. There are several 
reasons for this situation:
 
(i) The work of ref. \SuperWeak\ was concerned only with
CP violation in $K$ decays. In extending the idea to other mesons,
one may interpret the idea in various ways. On one side, it is possible
that the superweak interaction is significant only in $K-\bar K$
mixing and (apart from the relaxation of the $\epsK$-bounds on the
CKM parameters) has no effect on mixing of heavier mesons.
On the other extreme, one may take the superweak scenario to imply
that CP violation comes from $\Delta F=2$ processes only for all mesons.
The most common use of the term `superweak' refers to the latter
option, namely that there is no direct CP violation.
 
(ii) The scenario proposed in \SuperWeak\ did not employ any specific
model. It was actually proposed even before the formulation of the
Standard Model. To extend the idea to, for example, the neutral $B$
system, a model is required. Various models give very different
predictions for CP asymmetries in $B$ decays.
 
If one extends the superweak scenario to the $B$ system by assuming that
there is CP violation in $\Delta B=2$ but not in $\Delta B=1$
transitions, the prediction for CP asymmetries in $B$ decays into final
CP eigenstates is that they are equal for all final states
\nref\WinSW{B. Winstein, Phys. Rev. Lett. 68 (1992) 1271.}%
\nref\SWSW{J.M. Soares and L. Wolfenstein, Phys. Rev. D46 (1992) 256.}%
\nref\WWSW{B. Winstein and L. Wolfenstein, Rev. Mod. Phys. 65 (1993) 1113.}%
\refs{\WinSW-\WWSW}. Whether these asymmetries are all small or could be
large is model dependent. In addition, the asymmetries in charged $B$
decays vanish.

CP violation via neutral scalar exchange is the most commonly
studied realization of the superweak idea. In particular,
if the complex $Z_{ij}^q$ couplings presented in the previous section
were the only source of CP violation, then this model would be
superweak. The smallness of the $Z_{ij}^q$ couplings would make
the contribution from neutral Higgs mediated diagrams negligible
compared to the Standard Model diagrams in $\Delta S=1$ processes, 
but the fact that they contribute to mixing at tree level would allow 
them to dominate the $\Delta S=2$ processes. 
Various models (or scenarios) that realize the main features of the
superweak idea can be found in refs.
\nref\GeNa{J.-M. Gerard and T. Nakada, Phys. Lett. B261 (1991) 474.}%
\nref\LavSW{L. Lavoura, Int. J. Mod. Phys. A9 (1994) 1873.}%
\refs{\GeNa-\LavSW,\LiWo,\LiWoB}.
As mentioned above, there is
a considerable variation in their predictions for $\epe$, $d_N$ and
other quantities. If we take the term `superweak CP violation' to imply
that there is only indirect CP violation, or at least that there is
no direct CP violation in $K$ decays, then $\epe\neq0$ is inconsistent
with this scenario which is therefore excluded.

\newsec{Extensions of the Fermion Sector: Down Singlet Quarks}
The fermion sector of the Standard Model is described in eq. \SMrep.
It can be extended by either a fourth, sequential generation or
by non-sequential fermions, namely `exotic' representations,
different from those of \SMrep. (The four generation model 
can only be viable if it is further extended to evade bounds
related to the neutrino sector
\ref\HaNi{H. Harari and Y. Nir, Nucl. Phys. B292 (1987) 251.}\
and to electroweak precision data \PDG.)
Our discussion in this chapter is focused
on non-sequential fermions and their implications on CP asymmetries
in neutral $B$ decays and on $K_L\to\pi\nu\bar\nu$.
 
\subsec{The Theoretical Framework}
We consider a model with extra quarks in vector-like
representations of the Standard Model gauge group,
\eqn\extra{d_4(3,1)_{-1/3}\ +\ \bar d_4(\bar 3,1)_{+1/3}.}
Such (three pairs of) quark representations appear, for example, in $E_6$
GUTs. The mass matrix in the down sector, $M^d$, is now $4\times4$.
(Note that the $M^d_{4i}$ entries do not violate $G_{\rm SM}$ and
are, therefore, bare mass terms.)

How many independent CP violating parameters are there in 
$M^d$ and $M^u$? Since $M^d$ ($M^u$) is $4\times4$ ($3\times3$) and complex, 
there are 25 real and 25 imaginary parameters in these matrices. 
If we switch off the mass matrices, there is a global
symmetry added to the model, 
\eqn\GglobE{G_{\rm global}^{{\rm extra}\ d}(M^{d,u}=0)=
U(3)_Q\times U(4)_{\bar d}\times U(3)_{\bar u}\times U(1)_{d_4}.}
One can remove, at most, 12 real and 23 imaginary parameters. However, the 
model with the quark mass matrices switched on has still a global symmetry of
$U(1)_B$, so one of the imaginary parameters cannot be removed.
We conclude that there are 16 flavor parameters: 13 real ones,
that is seven masses and six mixing angles, and 3 phases. 
These three phases are independent sources of CP violation.

\subsec{$Z$-Mediated FCNC} 
The most important feature of this model for our purposes is that
it allows CP violating $Z$-mediated Flavor Changing Neutral Currents
(FCNC). To understand how these FCNC arise, it is convenient to work
in a basis where the up sector interaction eigenstates are identified
with the mass eigenstates. The down sector interaction eigenstates
are then related to the mass eigenstates by a $4\times4$ unitary
matrix $K$. Charged current interactions are described by
\eqn\Zcci{\L^W_{\rm int}\ =\ {g\over\sqrt2}
(W_\mu^+V_{ij}\bar u_{iL}\gamma^\mu d_{jL}+{\rm h.c.}).}
The charged current mixing matrix $V$ is a $3\times4$ sub-matrix of $K$:
\eqn\ZVinK{V_{ij}=K_{ij}\ \ {\rm for}\ \ i=1,2,3;\ j=1,2,3,4.}
The $V$ matrix is parameterized, as anticipated above, by six real angles 
and three phases, instead of three angles and one phase in the original 
CKM matrix. All three phases may affect CP asymmetries in $B^0$
decays. Neutral current interactions are described by
\eqn\Zcci{\eqalign{
\L_{\rm int}^Z\ =&\ {g\over\cos\theta_W}Z_\mu
(J^{\mu3}-\sin^2\theta_WJ^{\mu}_{\rm EM}),\cr
J^{\mu3}\ =&\ -{1\over2}U_{pq}\bar d_{pL}\gamma^\mu d_{qL}
+{1\over2}\delta_{ij}\bar u_{iL}\gamma^\mu u_{jL}.\cr}}
The neutral current mixing matrix for the down sector is $U=V^\dagger V$.
As $V$ is not unitary, $U\neq{\bf 1}$. In particular, its non-diagonal
elements do not vanish:
\eqn\ZUpq{U_{pq}=-K_{4p}^*K_{4q}\ \ {\rm for}\ \ p\neq q.}
The three elements which are most relevant to our study are
\eqn\Zrele{\eqalign{
U_{ds}\ =&\ V_{ud}^*V_{us}+V_{cd}^*V_{cs}+V_{td}^*V_{ts},\cr
U_{db}\ =&\ V_{ud}^*V_{ub}+V_{cd}^*V_{cb}+V_{td}^*V_{tb},\cr
U_{sb}\ =&\ V_{us}^*V_{ub}+V_{cs}^*V_{cb}+V_{ts}^*V_{tb}.\cr}}
The fact that, in contrast to the Standard Model, the various $U_{pq}$
do not necessarily vanish, allows FCNC at tree level. This may
substantially modify the predictions for CP asymmetries.
 
The flavor changing couplings of the $Z$ contribute to various FCNC
processes. Relevant constraints arise from semileptonic FCNC $B$ decays:
\eqn\ZBll{{\Gamma(B\ra\ell^+\ell^- X)_Z\over\Gamma(B\ra \ell^+\nu X)}=
\left[(1/2-\sin^2\theta_W)^2+\sin^4\theta_W\right]
{|U_{db}|^2+|U_{sb}|^2\over|V_{ub}|^2+F_{\rm ps}|V_{cb}|^2},}
where $F_{\rm ps}\sim0.5$ is a phase space factor.
The experimental upper bound on $\Gamma(B\ra\ell^+\ell^- X)$ gives
\eqn\ZUqb{\left|{U_{db}\over V_{cb}}\right|\leq0.04,\ \ \
\left|{U_{sb}\over V_{cb}}\right|\leq0.04.}
Additional constraints come from neutral $B$ mixing:
\eqn\ZdelmB{(\Delta m_B)_Z={\sqrt2G_FB_Bf_B^2m_B\eta_B\over3}|U_{db}|^2.}
Using $\sqrt{B_B}f_B\gsim0.16\ GeV$, we get
\eqn\ZdmB{|U_{db}|\lsim9\times10^{-4}.} 
As concerns $\Delta m_{B_s}$, only lower bounds exist
and consequently there is no analog bound on $|U_{sb}|$.

Bounds on $U_{ds}$ can be derived from the measurements
of BR($K_L\ra\mu^+\mu^-$), BR$(K^+\ra\pi^+\nu\bar\nu)$, $\epsK$
and $\epe$ yielding, respectively (for recent derivations, see
\refs{\CoIs,\BuSi}), 
\eqn\sdreb{\eqalign{
|\Re(U_{ds})| \lsim&\ 10^{-5},\cr
|U_{ds}| \leq&\  3\times 10^{-5},\cr
|\Re(U_{ds}) \, \Im(U_{ds})| \lsim&\ 1.3 \times 10^{-9},\cr
|\Im(U_{ds})|\lsim&\ 10^{-5}.\cr}}
(Note that the combination of bounds from BR($K_L\ra\mu^+\mu^-$)
and from the recently improved $\epe$ is stronger than the 
bounds from BR$(K^+\ra\pi^+\nu\bar\nu)$ and $\epsK$ \BuSi. 
The latter bounds are, however, subject to smaller hadronic uncertainties.)

\subsec{CP Asymmetries in $B$ Decays}
The most interesting effects in this model
concern CP asymmetries in neutral $B$ decays into final CP eigenstates
\nref\NiSiZ{Y. Nir and D. Silverman, Phys. Rev. D42 (1990) 1477.}%
\nref\SilvZ{D. Silverman, Phys. Rev. D45 (1992) 1800;
 Int. J. Mod. Phys. A11 (1996) 2253, hep-ph/9504387}%
\nref\BMPRZ{G.C. Branco, T. Morozumi, P.A. Parada and M.N. Rebelo,
 Phys. Rev. D48 (1993) 1167.}%
\nref\SilvZ{W-S. Choong and D. Silverman, Phys. Rev. D49 (1994) 2322.}%
\nref\BBPZ{V. Barger, M.S. Berger and R.J.N. Phillips,
 Phys. Rev. D52 (1995) 1663, hep-ph/9503204.}%
\nref\GrLo{M. Gronau and D. London,
 Phys. Rev. D55 (1997) 2845, hep-ph/9608430 (1996).}%
\nref\BaBo{G. Barenboim and F.J. Botella,
 Phys. Lett. B433 (1998) 385, hep-ph/9708209.}%
\nref\BBBV{G. Barenboim, F.J. Botella, G.C. Branco and O. Vives,
 Phys. Lett. B422 (1998) 277, hep-ph/9709369.}%
\refs{\NiSiZ-\BBBV,\ENcp}. We describe these effects in detail as they
illustrate the type of new ingredients that are likely to affect
CP asymmetries in neutral $B$ decays and the way in which the SM
predictions might be modified. (If there exist light up quarks in
exotic representations, they may introduce similar, interesting
effects in neutral $D$ decays \BSN.)
 
If the $U_{qb}$ elements are not much smaller than the bounds \ZUqb\
and \ZdmB, they will affect several aspects of physics related to CP 
asymmetries in $B$ decays.
 
(i) {\it Neutral $B$ mixing}:
The experimentally measured value of $\Delta m_{B_d}$ (and the lower bound on
$\Delta m_{B_s}$) can be explained by Standard Model processes, namely box
diagrams with intermediate top quarks. Still, the uncertainties in
the theoretical calculations, such as the values of $f_B$ and $V_{td}$
(and the absence of an upper bound on $\Delta m_{B_s}$) allow a situation 
where SM processes do not give the dominant contributions to either or both
of $\Delta m_{B_d}$ and $\Delta m_{B_s}$ \BEN.
The ratio between the $Z$-mediated tree diagram and the Standard
Model box diagram is given by  ($q=d,s$)
\eqn\ZtoSM{{(\Delta m_{B_q})_{\rm tree}\over(\Delta m_{B_q})_{\rm box}}=
{2\sqrt2\pi^2\over G_F m_W^2 S_0(x_t)}
\left|{U_{qb}\over V_{tq}V_{tb}^*}\right|^2
\approx150\left|{U_{qb}\over V_{tq}V_{tb}^*}\right|^2\lsim
\cases{5&$q=d$\cr 0.25&$q=s$\cr}.}
(The last inequality is derived under the assumption that the violation
of CKM unitarity is not strong. The bound on $(\Delta m_{B_d})_{\rm tree}/
(\Delta m_{B_d})_{\rm box}$ is higher if $|V_{td}V_{tb}^*|<0.005$ holds.)
{}From \ZUqb\ and \ZtoSM\ we learn that the $Z$-mediated
tree diagram could give the dominant contribution to $\Delta m_{B_d}$
but at most $25\%$ of $\Delta m_{B_s}$.
 
(ii) {\it Unitarity of the $3\times3$ CKM matrix}:
Within the SM, unitarity of the three generation CKM matrix gives:
\eqn\three{\eqalign{
\U_{ds}\equiv\ &\ V_{ud}^*V_{us}+V_{cd}^*V_{cs}+V_{td}^*V_{ts}=0,\cr
\U_{db}\equiv\ &\ V_{ud}^*V_{ub}+V_{cd}^*V_{cb}+V_{td}^*V_{tb}=0,\cr
\U_{sb}\equiv\ &\ V_{us}^*V_{ub}+V_{cs}^*V_{cb}+V_{ts}^*V_{tb}=0.\cr}}
Eq. \Zrele, however, implies that now \three\ is replaced by
\eqn\reprel{\U_{ds}=U_{ds},\ \ \ \U_{db}=U_{db},\ \ \ \U_{sb}=U_{sb}.}
A measure of the violation of \three\ is given by
\eqn\viothr{
\left|{U_{ds}\over V_{ud}V_{us}^*}\right|\,\lsim \, 5 \times 10^{-4},\ \ \
\left|{U_{db}\over V_{td}V_{tb}^*}\right|\,\lsim \,0.18,\ \ \
\left|{U_{sb}\over V_{ts}V_{tb}^*}\right|\,\lsim \,0.04.}
The bound on $|U_{db}/(V_{td}V_{tb}^*)|$ is even weaker if $|V_{td}|$ is
lower than the three generation unitarity bound. We learn that the first 
of the SM relations in \three\ is
practically maintained, while the third can be violated by at most 4\%.
However, the $\U_{db}=0$ constraint may be violated by $\O(0.2)$ effects.
The Standard Model unitarity triangle should be replaced by a unitarity
{\it quadrangle}. After the recent measurement of $a_{\psi K_S}$ \CDFBpK,
not only the magnitude of $U_{db}$ but also the
phases $\bar\alpha$ and $\bar\beta$,
\eqn\defbeal{\bar\alpha\equiv\arg\left({V_{ud}V_{ub}^*\over U_{db}^*}
\right),\ \ \ \
\bar\beta\equiv\arg\left({U_{db}^*\over V_{cd}V_{cb}^*}\right),}
are constrained \ENcp, but the constraints are not very strong.
 
(iii) {$Z$-mediated $B$ decays}:
Our main interest in this chapter is in hadronic $B^0$ decays to
CP eigenstates, where the quark sub-process is $\bar b\ra\bar u_i
u_i\bar d_j$, with $u_i=u,c$ and $d_j=d,s$. These decays get
new contributions from $Z$-mediated tree diagrams, in addition
to the standard $W$-mediated ones. The ratio between the amplitudes is
\eqn\ratamp{{A_Z\over A_W}=\left[{1\over2}-{2\over3}\sin^2\theta_W
\right]\left|{U_{jb}^*\over V_{ij}V_{ib}^*}\right|.}
We find that the $Z$ contributions can be safely neglected in 
$\bar b\ra \bar cc\bar s$ ($\lsim0.013$) and $\bar b\ra\bar cc\bar d$ 
($\lsim0.03$). On the other hand, it may be significant in
$\bar b\ra\bar uu\bar d$ ($\lsim0.12$), and in processes with
no SM tree contributions, e.g. $\bar b\ra\bar ss\bar s$, that may have
comparable contributions from penguin and $Z$-mediated tree diagrams.
 
(iv) {\it New contributions to $\Gamma_{12}(B_q)$}:
The difference in width comes from modes that are common to $B_q$ and
$\bar B_q$. As discussed above, there are new contributions to such modes
from $Z$-mediated FCNC. However, while the new contributions to $M_{12}$
are from tree level diagrams, {\it i.e.} $\O(g^2)$, those to $\Gamma_{12}$
are still coming from a box-diagram, {\it i.e.} $\O(g^4)$. Consequently,
no significant enhancement
of the SM value of $\Gamma_{12}$ is expected, and the relation
$\Gamma_{12}\ll M_{12}$ is maintained. (The new contribution could
significantly modify the leptonic asymmetry in neutral $B$ decays
\nref\BPM{G.C. Branco, P.A. Parada, T. Morozumi and M.N. Rebelo,
 Phys. Lett. B306 (1993) 398.}%
\refs{\BPM,\ENcp}\ though the asymmetry remains small.)

The fact that $M_{12}(B^0)$ could be dominated by the $Z$-mediated
FCNC together with the fact that this new amplitude depends
on new CP violating phases means that large deviations from the
Standard Model predictions for CP asymmetries are possible.
As $\Gamma_{12}\ll M_{12}$ is maintained, future measurements
of certain modes will still be subject to a clean theoretical
interpretation in terms of the extended electroweak sector parameters.
 
Let us assume that, indeed, $M_{12}$ is dominated by the new
physics. (Generalization to the case that the new contribution is
comparable to (but not necessarily dominant over) the Standard Model
one is straightforward \refs{\BMPRZ,\BaBo}.) Then
\eqn\Zmix{\left({p\over q}\right)_B\approx{U_{db}^*\over U_{db}}.}
We argued above that $b\ra c\bar cs$ is still dominated by the
$W$ mediated diagram. Furthermore, the first unitarity constraint in
\three\ is practically maintained. Then it is straightforward to evaluate
the CP asymmetry in $B\ra\psi K_S$. We find that it simply measures
an angle of the unitarity quadrangle:
\eqn\ZpsiKS{a_{CP}(B\ra\psi K_S)=-\sin2\bar\beta.}
The new contribution to $b\ra c\bar cd$ is  $\O(3\%)$, which is somewhat 
smaller than the SM penguins. So we still have, to a good approximation, 
(taking into account CP-parities)
\eqn\Zsd{a_{CP}(B\ra\psi K_S)\approx-a_{CP}(B\ra DD).}
Care has to be taken regarding $b\ra u\bar ud$ decays. Here, direct
CP violation may be large \GrLo\ and prevent a clean theoretical
interpretation of the asymmetry. Only if the asymmetry is large, so
that the shift from the $Z$-mediated contribution to the decay is small,
we get
\eqn\Zpipi{a_{CP}(B\ra\pi\pi)=-\sin2\bar\alpha.}
The important point about the modification of the SM
predictions is then not that the angles $\alpha,\beta$ and $\gamma$
may have very different values from those predicted by the SM, but
rather that the CP asymmetries do not measure these angles anymore.
As the experimental constraints on $\bar\alpha$ and $\bar\beta$ are still
rather weak \ENcp, a large range is possible for each of the asymmetries.
This model demonstrates that there exist extensions of the SM
where dramatic deviations from its predictions for CP asymmetries
in $B$ decays are not unlikely.
 
Another interesting point concerns $B_s$ decays. If $B_s-\bar B_s$
mixing as well as the $b\ra c\bar cs$ decay are dominated by the SM
diagrams, we have, similar to the SM,
\eqn\Bspsiphi{a_{CP}(B_s\ra\psi\phi)\approx0.}
As shown in ref. \NiSi,
this is a sufficient condition for the angles extracted from
$B\ra\psi K_S$, $B\ra\pi\pi$ and the relative phase between the 
$B_s-\bar B_s$ mixing amplitude and the $b\ra u\bar ud$ decay
amplitude (if it can be deduced from experiment) to sum up to $\pi$
(up to possible effects of direct CP violation). This happens in
spite of the fact that the first two asymmetries do not correspond
to $\beta$ and $\alpha$ of the unitarity triangle.
 
\subsec{The $K_L \to \pi \nu \bar \nu$ Decay}
In chapter 7 we argued that the only potentially significant new 
contribution to $a_{\pi\nu\nu}$ can come from the decay amplitude. 
$Z$-mediated FCNC provide an explicit example of New Physics that may
modify the SM prediction for $a_{\pi\nu\nu}$ of eq. \defapnn.
Assuming that the $Z$-mediated tree diagram dominates
$K\ra\pi\nu\bar\nu$, we get \NiSiZ\
\eqn\Zkpnn{\sin\theta_K=\Im U_{ds}/|U_{ds}|.}
 
Bounds on the relevant couplings were given in eq. \sdreb\ above.
We learn that large effects are possible. When $|\Re(U_{ds})|$ and
$|\Im(U_{ds})|$ are close to their upper bounds, the branching ratios
$BR(K^+ \to \pi^+ \nu \bar \nu)$ and $BR(K_L \to \pi^0 \nu \bar \nu)$
are both $O(10^{-10})$ and $a_{\pi\nu\nu}=\O(1)$.
Furthermore, as in this case the SM contribution is small,
the measurement of $BR(K^+ \to \pi^+ \nu \bar \nu)$ approximately determines
$|U_{ds}|$, and with the additional measurement of
$BR(K_L \to \pi^0 \nu \bar \nu)$, $\arg(U_{ds})$ is 
approximately determined as well.

\newsec{Conclusions}
Experiments have not yet probed in a significant way the
mechanism of CP violation. There is a large number of open questions 
concerning CP violation. Here are some examples:

$\bullet$
{\bf Why are the measured parameters, $\epsK$ and $\epspK$, small?}

The answer in the Standard Model is that CP violation is screened
in processes that are dominated by the first two quark generations
by small mixing angles. We have seen examples of new physics, that is
supersymmetry with approximate CP, where the reason is the smallness 
of all CP violating phases.

Observing CP asymmetries of order one, as expected in processes that
involve the first and third generations such as $B\ra\psi K_S$,
$K\to\pi\nu\bar\nu$ or even in charged $B$ decays, $B^\pm\ra K^\pm\pi^0$,
will exclude the approximate CP scenario.

$\bullet$
{\bf What is the number of independent CP violating phases?}

The answer in the Standard Model is {\it one}, the Kobayashi-Maskawa
phase. We have encountered models with a larger number, {\it e.g.}
forty four in the supersymmetric standard model.

If the pattern predicted by the Standard Model, {\it e.g.} small
CP asymmetries in $B_s\ra\psi\phi$ and in $D\ra K\pi$, 
a strong correlation between CP violation in $B\ra\psi K_S$ and 
in $K_L\ra\pi\nu\bar\nu$, equal asymmetries in $B\ra\psi K_S$
and in $B\ra\phi K_S$, etc., is inconsistent with measurements, 
then probably there are several independent phases.

$\bullet$
{\bf Why is CP violated?}

The answer in the Standard Model is explicit breaking by complex 
Yukawa couplings. In left-right-symmetric models, the Lagrangian 
can be CP symmetric and the breaking is spontaneous.

It will be difficult to answer this question by experimental
measurements, unless the correlations predicted by a specific
model of spontaneous CP violation will be experimentally confirmed.

$\bullet$
{\bf Is CP violation restricted to flavor changing interactions?}

This is indeed the case in the Standard Model. But in many of its
extensions, such as supersymmetry, there is flavor diagonal  CP violation.

Observation of an electric dipole moment or of CP violation in 
$t\bar t$ production will provide strong hints for flavor diagonal
CP violation.

$\bullet$
{\bf Is CP violation restricted to quark interactions?}

This is the case in the Standard Model but not if neutrinos have masses.

Observation of CP asymmetries in neutrino oscillation experiments
will be a direct evidence of CP violation in the lepton sector.

$\bullet$
{\bf Is CP violation restricted to the weak interactions?}

In the Standard Model, CP violation appears in charged current
(that is, $W$-mediated) weak interactions only. In multi-scalar
models, it appears in scalar interactions. In supersymmetry,
it appears in strong interactions.

Observation of transverse lepton polarization in meson decays
will provide evidence for CP violation in interactions
that are not mediated by vector bosons.

There are more questions that we can ask and answers that we will
learn in the near future. But the list above is enough to demonstrate
how unique the Standard Model picture of CP violation is, how
sensitive is CP violation to new physics, and how important are
present and future experiments that will search for CP violation.

\vskip 1 cm
\centerline{\bf Acknowledgements}
I thank Gabriela Barenboim, Sven Bergmann, Galit Eyal, Yuval Grossman, 
Stephane Plaszczynski, Helen Quinn, Marie-Helene Schune and Joao Silva 
for useful discussions. 
Y.N. is supported by a DOE grant DE-FG02-90ER40542,
by the Ambrose Monell Foundation,  
by the Israel Science Foundation founded by the
Israel Academy of Sciences and Humanities, 
and by the Minerva Foundation (Munich).

\listrefs
\end